\documentclass[12pt]{article}
\usepackage{graphicx} % Required for inserting images
\usepackage{natbib}
\usepackage{amsmath}
\usepackage{amssymb}
\usepackage[linesnumbered,ruled,vlined]{algorithm2e}
\usepackage[a4paper,margin=1in]{geometry}
\usepackage{comment}
\usepackage{booktabs}
\usepackage{float}
\usepackage{xcolor}
\usepackage{subcaption}
\usepackage[margin=1in]{geometry}
\newcommand{\notered}[1]{{\color{red}{#1}}}
\usepackage{gensymb}
\usepackage{longtable}
\usepackage{placeins} % in preamble
\usepackage{amsmath} 
\usepackage{bm}
\usepackage{bbm}
\usepackage{setspace}
\usepackage{multirow}
\usepackage{diagbox}
\usepackage{hyperref}

\title{A Statistical Framework for Spatial Boundary Estimation and Change Detection: Application to the Sahel–Sahara Climate Transition}
\author{Stephen Tivenan, Indranil Sahoo\thanks{Corresponding author: sahooi@vcu.edu}, Yanjun Qian \\ \\
Department of Statistical Sciences and Operations Research, \\
Virginia Commonwealth University, Richmond, VA, USA.}
\date{}

\begin{document}

\maketitle

\begin{abstract}
Spatial boundaries, such as ecological transitions or climatic regime interfaces, capture steep environmental gradients, and shifts in their structure can signal emerging environmental changes. Quantifying uncertainty in spatial boundary locations and formally testing for temporal shifts remains challenging, especially when boundaries are derived from noisy, gridded environmental data. We present a unified framework that combines heteroskedastic Gaussian process (GP) regression with a scaled Maximum Absolute Difference (MAD) Global Envelope Test (GET) to estimate spatial boundary curves and assess whether they evolve over time. The heteroskedastic GP provides a flexible probabilistic reconstruction of boundary lines, capturing spatially varying mean structure and location-specific variability, while the test offers a rigorous hypothesis-testing tool for detecting departures from expected boundary behaviors. Simulation studies show that the proposed method achieves the correct size under the null and high power for detecting local boundary shifts. Applying our framework to the Sahel–Sahara transition zone, using annual K\"oppen–Trewartha climate classifications from 1960–1989, we find no statistically significant decade-scale changes in the arid/semi-arid or semi-arid/non-arid interfaces. However, the method successfully identifies localized boundary shifts during the extreme drought years of 1983–1984, consistent with climate studies documenting regional anomalies in these interfaces during that period.

\end{abstract}

\section{Introduction}

Spatial boundaries such as ecological transition zones, climatic regime interfaces, and other regions of rapid change in environmental fields play a central role in describing ecosystem structure, land–atmosphere interactions, and climate-driven shifts in terrestrial systems. Changes in the location or morphology of these boundaries often provide early indicators of environmental stress, altered climatic patterns, and longer-term ecological reorganization, and they are particularly relevant in regions vulnerable to desertification and drought, such as the Sahel–Sahara transition zone. From a statistical perspective, it is natural to formalize such boundaries as function-valued objects embedded in spatial process models and to study how these functions evolve under temporal forcing. However, environmental boundaries are typically inferred from noisy, remotely sensed, or reanalysis-based products with nontrivial discrepancies, and their evolution is driven by complex, partially observed dynamical systems. This combination of measurement error, spatial dependence, and temporal variability makes it challenging to both estimate boundaries and rigorously assess whether they have shifted over time.

A substantial methodological literature addresses boundary detection via `wombling', which uses gradients of an underlying spatial surface to identify points of rapid change. The original formulation by \cite{WombleWilliamH.1951DS} introduced the notion of `wombling boundaries' as curves along which the spatial gradient is unusually large. Building on this idea, \cite{Banerjee01122006} developed Bayesian wombling for point-referenced data using Gaussian process models to assess curvilinear gradients along prespecified paths. \cite{lu2005bayesian} extended wombling to areal data via hierarchical conditional autoregressive models, with later work on multivariate areal wombling and adjacency-based formulations. These model-based wombling approaches have been used in ecological and environmental applications, including detection of species distribution boundaries and genetic discontinuities \citep{fitzpatrick2010ecological}. \cite{gelfand2015bayesian} provided a broader overview of Bayesian wombling in spatial maps, emphasizing gradient-based boundary inference within fully probabilistic frameworks. More recently, \cite{halder2024bayesian} proposed Bayesian spatial curvature processes to model and infer rapid changes and directional curvature on spatial response surfaces, strengthening the theoretical connection between Gaussian process modeling and detection of sharp transitions. Parallel developments in spatial testing have introduced global envelope tests for functional and spatial statistics \citep{myllymaki2017global}, which provide Monte Carlo–based inference with graphical envelopes for a broad class of functionals derived from spatial processes.

In a complementary line of work, change detection for remote sensing imagery has focused primarily on pixel- or object-level changes between multi-temporal images. Early digital change detection relied on image differencing, thresholding, regression, and principal component analysis comparisons \citep{singh1989review}. Recent approaches increasingly use deep learning architectures for change detection, including convolution and encoder–decoder networks for land-cover mapping, as well as attention-based and transformer models designed for complex, heterogeneous scenes \citep{Cheng_etal_2024}. In addition, Siamese network frameworks have demonstrated superior precision and accuracy in boundary extraction across a range of remote-sensing datasets \citep{lei2022boundary}. In parallel, a variety of boundary-detection frameworks have been applied to land-based datasets, including encoder–decoder convolution neural network models that integrate local and temporal features to detect urbanization changes in Hong Kong \citep{IECG2020-08544} and vegetation-based climate classifications used to monitor highland swamp boundaries \citep{jenkins2010high}. However, these approaches typically involve black-box algorithms rather than model-based frameworks, and focus on classification accuracy, detection rates, or segmentation quality, rather than on formal statistical inference for function-valued boundaries with quantified uncertainty.

Despite this rich literature, there remains a methodological gap at the interface of boundary estimation, uncertainty quantification, and temporal change assessment. Wombling methods are well suited for detecting regions of rapid change and, in some cases, for characterizing the posterior distribution of gradient-based boundary sets, but they rarely target inference on differences between boundary curves at distinct time points, nor do they typically incorporate explicit heteroskedastic structure in the underlying spatial process. Remote sensing change-detection algorithms, in turn, operate at the level of pixel-wise labels or image segments and do not usually provide function-based boundary reconstructions with associated uncertainty, and they do not involve a principled test based on observed differences in boundaries. Global envelope tests offer a general framework for testing hypotheses using functional or spatial data, but have not, to our knowledge, been systematically integrated with Gaussian process–based boundary reconstructions to yield a coherent procedure for boundary shift detection. This lack of joint boundary estimation and formal change testing limits our ability to distinguish signal from noise in observed boundary movements. In applications such as the Sahel–Sahara climate transition, where projected desert climate expansion and shifts of arid, semi-arid interfaces are of central concern for risk, adaptation, and regional planning \citep{nicholson1994recent, salgado2004sahel, tian2023northward} our technique can provide a useful tool for others.

In this paper, we propose a unified statistical framework that combines heteroskedastic Gaussian process (GP) regression with a scaled Maximum Absolute Deviate (MAD) Global Envelope Test (GET) to estimate and compare spatial boundary curves over time. We treat the boundary as a functional spatial process and use heteroskedastic GPs to model both the mean boundary and its spatially varying variance, while representing temporal structure through Fourier basis functions. This yields a fully probabilistic reconstruction of boundary location at any given time, together with prediction intervals that reflect both spatial dependence and heterogeneous noise. Building on the global envelope methodology, we then construct a scaled MAD statistic and its corresponding global envelopes to test whether two boundary curves, or an observed boundary and its model-based prediction, are statistically different under the fitted GP model. Through simulation studies, we demonstrate that the proposed test attains the correct size under the null hypothesis of no change, and achieves good power for detecting local boundary shifts. We then apply the framework to the Saharan/Sahelian climate transition zones by extracting annual boundaries from the K\"oppen–Trewartha arid climate classifications \citep{Koppen1918, Trewartha1966} for 1960–1989, and show how our method can be used to assess whether decades with documented drought conditions and the extreme drought year of 1984 exhibit statistically significant shifts in the arid/semi-arid and semi/arid–non-arid climate interfaces.

The remainder of the paper is organized as follows. Section \ref{sec:data_desc} describes the K\"oppen– Trewartha arid climate classification dataset. Section \ref{sec:methods} presents the proposed methodology, including the heteroskedastic Gaussian process model and the scaled MAD Global Envelope Test. Section \ref{sec:sim} reports simulation studies designed to assess the properties of the testing framework. Section \ref{sec:results} applies the method to the Saharan/Sahelian climate boundaries from 1960–1989, and Section \ref{sec:disc} concludes with a discussion of the methodological implications and potential future research.

\section{GLDAS Climate Classification Data} \label{sec:data_desc}

The Sahara–Sahel region in northern Africa underwent marked climatic variability during the period of 1960 to 1989, including prolonged drought episodes in the 1970s and 1980s \citep{NicholsonSharonE.2018RotA}. These extreme events triggered widespread ecological degradation, agricultural and pastoral system collapse, and substantial socio-economic stress across the region \citep{hess-18-3635-2014, glantz1987drought}. In response, international agencies initiated major desertification-monitoring and mitigation efforts, including the Desertification Conventions held in 1977 and 1992, initiated by the United Nations, and the Long-Term Ecological Monitoring Network, such as R\'eseau d'observatoires de surveillance \'ecologique \`a long terme (ROSELT)  established by the Sahara and Sahel Observatory (OSS) \citep{kassas1995desertification, VogtJ.V.2011Maao}. With ongoing concerns about the potential northward expansion of arid conditions and the vulnerability of dryland populations \citep{HuangJianping2020Gdvt}, this three-decade window offers a scientifically meaningful period for evaluating how these climatic events may have influenced the region’s dry-climate boundaries.

To investigate long-term changes in dry-climate boundaries across northern Africa, we utilize annual maps derived from the K\"oppen–Trewartha climate classification (KTC). The KTC system provides a physically interpretable categorization of global climate zones based on temperature and precipitation data, and remains widely used in climatology and environmental sciences for identifying coherent climate regions. Originally proposed by K\"oppen in the early twentieth century \citep{Koppen1918} and later refined by Trewartha to improve partitioning of arid and semi-arid environments \citep{Trewartha1966}, the KTC classification provides a structured and deterministic rule-based framework that yields climate zones from gridded meteorological data. 

In this study, we focus exclusively on the dry-climate subdivision of the KTC system, which assigns each location to one of three categories relevant to our analysis, namely, arid (desert), semi-arid (steppe), or non-arid. The classification is determined using Patton’s precipitation threshold \citep{patton1962note},
\begin{equation*} R = 2.3T - 0.64P_W + 41, \end{equation*}
where $T$ denotes the mean annual temperature in degrees Celsius ($\degree$C), and $P_W$ represents the percentage of total annual precipitation occurring during the winter months (October - March, in the Northern Hemisphere). Based on this threshold, the classification rule for dry climates is as follows:
\begin{equation*} \text{Classification} = \begin{cases} \text{Arid}, & \text{if } P < R/2, \\ \text{Semi-Arid}, & \text{if } R/2 \leq P < R, \\ \text{Non-Arid}, & \text{if } P \geq R, \end{cases} \end{equation*}
where $P$ denotes the mean annual precipitation. 

\begin{figure}[!ht]
  \centering
  % First row
  \begin{subfigure}[b]{0.3\textwidth}
    \includegraphics[width=\textwidth]{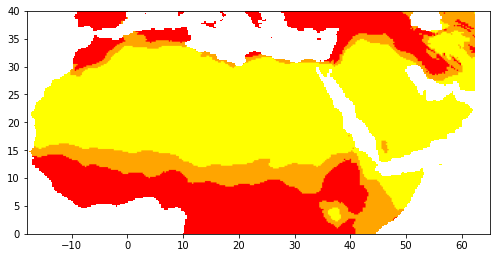}
    \caption{1965}
  \end{subfigure}
  \hfill
  \begin{subfigure}[b]{0.3\textwidth}
    \includegraphics[width=\textwidth]{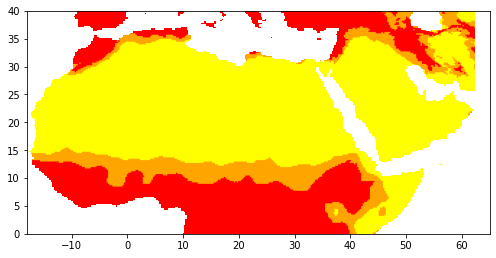}
    \caption{1970}
  \end{subfigure}
  \hfill
  \begin{subfigure}[b]{0.3\textwidth}
    \includegraphics[width=\textwidth]{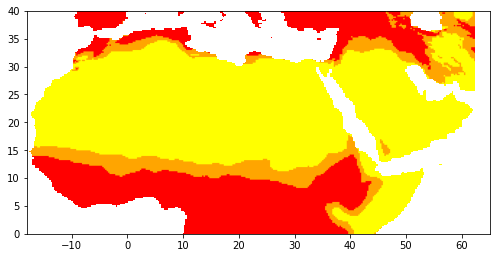}
    \caption{1975}
  \end{subfigure}

  \vspace{0.5cm}
 
  % Second row
  \begin{subfigure}[b]{0.3\textwidth}
    \includegraphics[width=\textwidth]{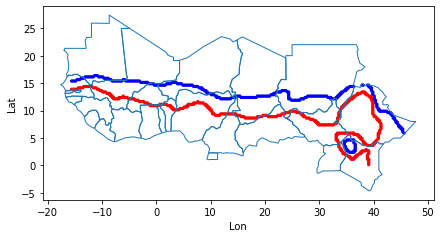}
    \caption{1965}
  \end{subfigure}
  \hfill
  \begin{subfigure}[b]{0.3\textwidth}
    \includegraphics[width=\textwidth]{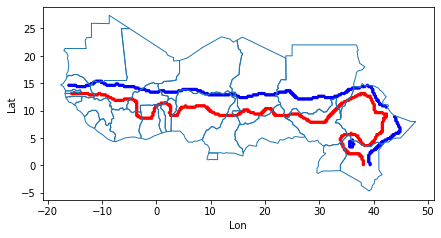}
    \caption{1970}
  \end{subfigure}
  \hfill
  \begin{subfigure}[b]{0.3\textwidth}
    \includegraphics[width=\textwidth]{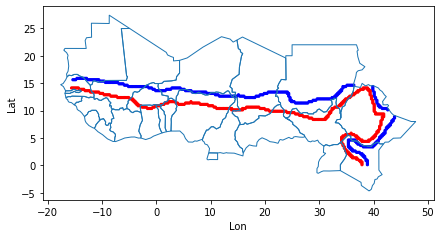}
    \caption{1975}
  \end{subfigure}

  \caption{\textit{Top row}: K\"oppen Trewartha dry-climate classification maps for the years 1965, 1970, and 1975. The yellow, orange, and red pixels correspond to arid, semi-arid, and non-arid regions, respectively. \textit{Bottom row}: Extracted boundary points for each corresponding year. The blue points denote the arid/semi-arid interface, while red points indicate the semi-arid/non-arid interface.} \label{fig:data_plots}
\end{figure}

To compute annual KTC classifications for each year from 1960 to 1989, we use temperature and precipitation fields from the GLDAS Noah Land Surface Model Version 2.0, which is a global land data assimilation system (GLDAS) that provides data on various climatic and temporal variables from 1948 to 2014 \citep{beaudoing_rodell_2020}. GLDAS provides globally consistent monthly estimates of near-surface air temperature (measured in Kelvin) and precipitation rate (measured in $\textrm{kg}~\textrm{m}^{-2}~\textrm{s}^{-1}$) at a 
0.25$\degree$ spatial resolution. We convert temperature values to degrees Celsius and aggregate precipitation to annual totals using standard hydrological unit conversions. These annual summaries are combined to generate discrete KTC dry-climate labels at each grid cell for all 30 years in the study period, from 0$\degree$N to 40$\degree$N latitudes and from -20$\degree$W to 60$\degree$E longitudes. More details on the collection and processing of pixel-level data can be found in \cite{tivenan26}. To illustrate the discrete climate classification maps used in our analysis, the top row of Figure \ref{fig:data_plots} displays the K\"oppen–Trewartha dry-climate classifications for three representative years (1965, 1970, and 1975). In these maps, yellow, orange, and red pixels correspond to arid (desert), semi-arid (steppe), and non-arid regions, respectively, highlighting the spatial structure of the dry-climate regimes across the Sahel–Sahara domain.

We obtain the annual climate boundaries used in our statistical analysis by extracting the interfaces separating arid, semi-arid, and non-arid zones directly from the discrete K\"oppen–Trewartha (KTC) classification maps. Since our modeling framework requires boundary curves (analogous to wombling applications, where the gradient field is used to identify contours of rapid change), we first convert each classified raster into a set of boundary points along transition lines between adjacent climate classes using the Canny edge detection algorithm \citep{opencv_library}.

First, the image is smoothed using a $5 \times 5$ Gaussian filter to suppress small-scale noise and avoid detecting spurious edges. Second, intensity gradients are approximated by applying the following Sobel convolution kernels $G_x$ and $G_y$, in the horizontal and vertical directions respectively,
\begin{equation*}
G_x = \begin{bmatrix}
1 & 0 & -1\\
2 & 0 & -2\\
1 & 0 & -1 \end{bmatrix}
\hspace{1cm}
G_y = \begin{bmatrix}
1 & 2 & 1\\
0 & 0 & 0\\
-1 & -2 & -1 \end{bmatrix}.
\end{equation*}
This yields for each pixel an estimated gradient magnitude and orientation, given by
\begin{equation*}
G = \sqrt{(G_x)^2+(G_y)^2}
\hspace{1 cm}
\theta=\arctan{\frac{G_y}{G_x}}.
\end{equation*}
Non-maximum suppression is then applied along the gradient direction to retain only local maxima, thereby identifying pixels most likely to correspond to genuine edges. Finally, hysteresis thresholding is used to identify a connected set of consistent edges by retaining strong gradients and adding weak gradients only if they are adjacent to confirmed edge pixels. In the context of the KTC raster, this procedure isolates locations where the climate class changes between adjacent grid cells, thereby producing a clean set of boundary points between arid and semi-arid zones, and between semi-arid and non-arid zones.

Let
$\mathcal{D}_t = \lbrace (x_{ti}, y_{ti}): ~i = 1, \ldots, n_t\rbrace
$
denote the set of extracted longitude–latitude boundary points for year $t$, where $x_{ti}$ denotes the longitude where the boundary is observed, $y_{ti}$ denotes the latitude of the boundary at that longitude, and $n_t$ denotes the number of points extracted along the boundary for year $t$. Across the period of 1960 to 1989, the arid/semi-arid and semi-arid/non-arid interface contains, on average, approximately 256 and 201 points per year, respectively. To ensure that the extracted boundaries represent transitions between climate zones over the continent only, we exclude all points within 10 km of the African coastline. For the arid/semi-arid interface, we further restrict boundary points to latitudes above 9$\degree$N latitudes, while for the semi-arid/non-arid interface we retain points west of 30$\degree$E longitudes to avoid unrelated transitions in the Horn of Africa. To visualize these extracted boundaries, the bottom row of Figure~\ref{fig:data_plots} displays the boundary points $(x_{ti}, y_{ti})$ for three representative years, namely, 1965, 1970, and 1975. The blue points trace the interface between the arid and semi-arid zones, while the red points trace the interface between the semi-arid and non-arid zones.

% Starting at classification maps from the previous paper, we want to mathematical model the boundary after the classification between the boundary. And test statistically whether the boundary has changed overtime, previously we just classified the points and we are only focused on the line

% How do you model the line and how do detect the changes overtime
% We want to get the points, data that we are starting with preprocessing step, canny Edge detection,

% Preprocessing Data Description Canny Edge Detection
% to estimate spatial mean boundary line $f$ and it's uncertainties given the set of boundary points \( \{x_t, y_t(x_t)\}_{t=1960}^{1989} \).

\section{Statistical Methodology} \label{sec:methods}

This section outlines the statistical framework used to (i) estimate annual spatial boundary curves from extracted boundary points using heteroskedastic Gaussian process (GP) regression, (ii) obtain uncertainty quantification through predictive distributions, and (iii) formally test for temporal changes in boundaries using the scaled Maximum Absolute Deviate (MAD) Global Envelope Test (GET). Throughout, we treat the latitude values $y_{ti}$ as spatial responses observed at longitudes $x_{ti}$, and define a latent boundary function $f_t(x)$ to be estimated for each year $t$.

\subsection{Heteroskedastic Gaussian process Regression}

For each year $t \in \lbrace 1960, \ldots, 1989 \rbrace$, we observe a set of boundary points
$\mathcal{D}_t = \lbrace (x_{ti}, y_{ti}): ~i = 1, \ldots, n_t\rbrace
$. These observations are viewed as noisy realizations of an underlying spatial boundary function $f_t(x)$. Because the extracted boundary points vary in smoothness and noise level across longitude, we model the data using a heteroskedastic Gaussian process following \citet{Binois_2018}. 

Although our goal is to estimate the spatial boundary curve for each year, we first account for low-frequency temporal variations across the period of 1960 to 1989. To avoid confounding temporal structure with spatial shape, we regress observed latitudes on a set of centered Fourier basis functions. Let
$$
t_{\text{center}} = t - \bar{t}, \qquad \bar{t} = \frac{1}{30} \sum_{s = 1960}^{1989} s.
$$
For each year we define the vector of temporal Fourier regressors
$$
\bm{r}_t^\top = \left\{ \sin\left( \frac{2\pi \cdot j \cdot t_{\text{center}} }{T_{\text{period}}} \right),  ~\cos\left( \frac{2\pi \cdot j \cdot t_{\text{center}} }{T_{\text{period}}} \right), ~~j = 1, 2\right\}
$$
evaluated at periods $T_{\text{period}}= \{3, 6, 9, 12, 15, 18\}$. These six periods and four harmonics yield $q = 24$ temporal basis functions. For each boundary, the latitudes $y_{ti}$ are regressed on $\bm{r}_t^\top$ with regression coefficients $\beta$.

The model for the observed latitude $y_{ti}$ at longitude $x_{ti}$ is then given by
\begin{align*}
y_{ti} &= \bm{r}_t^\top \boldsymbol{\beta} + f_t(x_{ti}) + \varepsilon_{ti}, \label{eq:model-y}\\
f_t(x) &\sim \mbox{GP}\left(\mu_f(x), k_f(x,x';\bm{\theta}_f)\right),\\
\varepsilon_{ti} &\sim \mathcal{N}\left(0,\sigma_t^2(x_{ti})\right),
\end{align*}  
that is, for each year $t$, the latent boundary function $f_t(x)$ is modeled as a Gaussian process with mean function $\mu_f(x)$ and covariance kernel $k_f(x, x'; \bm{\theta}_f)\big)$ and $\varepsilon_{ti}$ denotes the heteroskedastic noise term at $x_{ti}$, which is assumed to follow a normal distribution with mean 0 and spatially varying variance $\sigma^2(x_{ti})$. To model heteroskedasticity, the log-variance of the observational noise follows a second GP \citep{Binois_2018} with mean function $\mu_g(x)$ and covariance kernel $k_g(x, x'; \bm{\theta}_g)$ as follows,
$$
\log \sigma_t^2(x) = g_t(x), \qquad g_t(x) \sim \mbox{GP}\left(\mu_g(x), k_g(x,x';\bm{\theta}_g) \right).
$$
This joint specification allows spatially varying noise levels and avoids the instability of estimating variances at individual points.  The covariance kernels for both $f_t$ and $g_t$ are the Mat\'ern covariance function with smoothness parameter $\nu = 3/2$, that is, 
\[ k_p(x, x'; \bm{\theta}_p) = \kappa_p^2 \left(1 + \frac{\sqrt{3}d}{\ell_p}\right) \exp\left(-\frac{\sqrt{3}d}{\ell_p}\right) \]
for $p \in \lbrace f, g \rbrace$, $\bm{\theta}_p=(\kappa_p^2, \ell_p)$, where $\kappa_p^2$ is the process variance for process $p_t$, $\ell_p > 0$ denotes the length-scale controlling the rate of spatial decay in correlation, and $d = ||x - x'||$. 

Let $\mathbf{x}_t = (x_{t1}, \ldots, x_{tn_t})^\top$ and $\mathbf{y}_t = (y_{t1}, \ldots, y_{tn_t})^\top$ denote the vectors of observed longitudes and latitudes for year $t$, respectively. Then, following \cite{goldberg1997regression, Binois_2018}, the covariance matrix of the observational vector $\mathbf{y}_t$ is given by
\[
\mathrm{Cov}(\mathbf{y}_t)
=
\kappa_f^2\big(C_f + \Lambda_t\big),
\]
where $C_f$ is the $n_t \times n_t$ correlation matrix of the latent Gaussian process $f_t$ and $\Lambda_t 
= \mathrm{diag}(\lambda_{t1},\ldots,\lambda_{tn_t})$, with $\lambda_{ti} = \sigma_t^2(x_{ti})/\kappa_f^2$. Finally, let 
$\mathbf{R}_t$ be the $n_t \times q$ design matrix with each row equal to $\bm{r}_t^\top$. Then, the corresponding log-likelihood for year $t$ is
\begin{align*}
\log L_t( \bm{\theta}_f, \bm{\theta}_g)
=&
-\frac{n_t}{2}\log(2\pi)
-\frac{1}{2}\log\big|\kappa_f^2\big(C_f + \Lambda_t)\big|\\
&-\frac{1}{2}
\big(\mathbf{y}_t - \mathbf{R}_t \beta\big)^\top
\big[\kappa_f^2\big(C_f + \Lambda_t)\big]^{-1}
\big(\mathbf{y}_t - \mathbf{R}_t \beta\big).
\end{align*}
Using $\big|\kappa_f^2\big(C_f + \Lambda_t)\big| = \kappa_f^{2n_t}\big|\big(C_f + \Lambda_t)\big|$ and $\big[\kappa_f^2\big(C_f + \Lambda_t)\big]^{-1} = (\kappa_f^2)^{-1}\big(C_f + \Lambda_t)^{-1}$ we have, 
\begin{align*}
\log L_t( \bm{\theta}_f, \bm{\theta}_g)
=
&-\frac{n_t}{2}\log(2\pi) - \frac{n_t}{2}\log(\kappa_f^2)
-\frac{1}{2}\log\big|\big(C_f + \Lambda_t)\big|\\
&-\frac{1}{2\kappa_f^2}
\big(\mathbf{y}_t - \mathbf{R}_t \beta\big)^\top
\big(C_f + \Lambda_t)^{-1}
\big(\mathbf{y}_t - \mathbf{R}_t \beta\big).
\end{align*}
Now, maximizing the log-likelihood with respect to $\kappa_f^2$ (first-order condition) gives the plug-in estimator
$$
\widehat{\kappa}_f^2 = \frac{1}{n_t} \big(\mathbf{y}_t - \mathbf{R}_t \beta\big)^\top
\big(C_f + \Lambda_t)^{-1}
\big(\mathbf{y}_t - \mathbf{R}_t \beta\big),
$$
and substituting $\widehat{\kappa}_f^2$ into the log-likelihood gives the conditional log-likelihood 
\begin{equation} \label{eq:loglik}
\log L_t(\bm{\theta}_f,\bm{\theta}_g \mid \widehat{\kappa}_f^2)
=
-\frac{n_t}{2}\log(2\pi)
-\frac{n_t}{2}\log \widehat{\kappa}_f^2
-\frac{1}{2}\log|(C_f+\Lambda_t)|
-\frac{n_t}{2}.
\end{equation} 
The conditional log-likelihood in \eqref{eq:loglik} is then maximized with respect to $(\bm{\theta}_f,\bm{\theta}_g)$ using the L-BFGS-B optimization algorithm \citep{zhu1997algorithm} to obtain the remaining hyperparameter estimates.

\subsection{Boundary Curve Estimation and Predictive Uncertainty}

Using the fitted hyperparameters $( \widehat{\boldsymbol{\theta}}_f, \widehat{\boldsymbol{\theta}}_g)$, and 
$\widehat{\kappa}_f^2$, we obtain predictions of the latent boundary function $f_{t_p}(x)$ at a new year $t_p$. 
Let $\bm{x}^* = (x^*_1,\dots,x^*_m)$ be a fixed prediction grid spanning the observed longitude range, $x_j^* \in \mathcal{X} = \Bigl[\min_{t,i}\{x_{ti}\}, \; \max_{t,i}\{x_{ti}\}\Bigr], ~j = 1, \ldots, m$. For the prediction locations $\bm{x}^*$, we define $C_* = C_f(\bm{x}_t, \bm{x}^*)$ as the $n_t \times m$ cross correlation matrix between the observed and prediction locations and $C_{**}$ as the $m \times m$ correlation matrix among the prediction locations. All correlation matrices below are evaluated at the maximum likelihood estimates of the hyperparameters, and hats are omitted for notational simplicity. 

The predictive mean and covariance of the latent boundary function $f_{t_p}(x)$ at locations $\bm{x}^*$ are given by
\begin{equation}
\mathbb{E}\!\left[f_{t_p}(\bm{x}^*) \mid \mathbf{x}_t, \mathbf{y}_t\right]
=
C_*^\top (C_f + \Lambda_t)^{-1} (\mathbf{y}_t - \mathbf{R}_t \beta),
\label{eq:pred_mean_boundary_cf}
\end{equation}
\begin{equation}
\mathrm{Cov}\!\left(f_{t_p}(\bm{x}^*)\right)
=
\widehat{\kappa}_f^2
\Big[
C_{**} - C_*^\top (C_f + \Lambda_t)^{-1} C_*
\Big].
\label{eq:pred_cov_boundary_cf}
\end{equation}
Equation \eqref{eq:pred_mean_boundary_cf} provides the estimated mean boundary curve at year $t_p$, whereas \eqref{eq:pred_cov_boundary_cf} yields its associated confidence bands. To obtain predictions for the observed (noisy) boundary $y_{t_p}(x^*)$, we add back the temporal mean structure and the input-dependent noise variance. The predicted mean boundary is then given by
\begin{equation}
\mathbb{E}\!\left[y_{t_p}(\bm{x}^*) \mid \mathbf{x}_t, \mathbf{y}_t\right]
=
\mathbf{R}_{t_p}^\top \beta
+
\mathbb{E}\!\left[f_{t_p}(\bm{x}^*) \mid \mathbf{x}_t, \mathbf{y}_t\right] 
= \mathbf{R}_{t_p}^\top \beta
+ C_*^\top (C_f + \Lambda_t)^{-1} (\mathbf{y}_t - \mathbf{R}_t \beta),
\label{eq:pred_mean_noisy_cf}
\end{equation}
and the predictive covariance is
\begin{equation}
\mathrm{Cov}\!\left(y_{t_p}(\bm{x}^*)\right)
=
\mathrm{Cov}\!\left(f_{t_p}(\bm{x}^*)\right)
+
\widehat{\kappa}_f^2 \Lambda_{t_p}^* 
= \widehat{\kappa}_f^2
\Big[
C_{**} - C_*^\top (C_f + \Lambda_t)^{-1} C_*
 + \Lambda_{t_p}^*\Big],
\label{eq:pred_cov_noisy_cf}
\end{equation}
where $\Lambda_{t_p}^* = \mathrm{diag}(\lambda^*_{1},\dots,\lambda^*_{m})$ contains the scaled noise variances at the prediction locations, with $\lambda^*_{j} = \sigma^2(x^*_j)/\widehat{\kappa}_f^2, ~~j = 1,\dots,m$. 

In \eqref{eq:pred_mean_noisy_cf}–\eqref{eq:pred_cov_noisy_cf}, 
$\mathbb{E}[y_{t_p}(\bm{x}^*)]$ represents the predicted boundary line at year $t_p$ (including the temporal trend), and 
$\mathrm{Cov}(y_{t_p}(\bm{x}^*))$ provides prediction bands that reflect both spatial uncertainty in the latent boundary and location-specific heteroskedastic noise. These estimated curves and their corresponding uncertainty quantification form the basis for the scaled MAD global envelope tests described in the next subsection. In practice, although the heteroskedastic GP model predictions are specified for a single year $t$, 
we typically aggregate information across multiple years to construct smoothed mean boundary curves for different time 
periods (e.g., decades) or reference intervals. These aggregated mean boundaries and their predictive distributions are then used as inputs to the global scaled Maximum Absolute Difference (MAD) envelope test
\citep{myllymaki2017global} to formally assess temporal changes in the boundary.

%\notered{Qian: Sections 3.1 and 3.2 are very well-written, and I can follow the descriptions nicely. The only concern is that I could not find the estimation for the two mean functions $\mu_f$ and $\mu_g$. I remember that they are fitted using a B-spline. Should we add a few sentences for that?}

\subsection{Global Scaled MAD Envelope Test}

% Mention a comparison between two different mean line residuals 

The global scaled Maximum Absolute Difference (MAD) envelope test 
\citep{myllymaki2017global} is a functional hypothesis test that compares an observed functional statistic $T_{\text{obs}}(x)$,  defined over a spatial domain $x \in \mathcal{X}$, to a collection of simulated functions 
$T_1(x),\dots,T_M(x)$ generated under a null model. A MAD-type test statistic is computed from each  $T_i(x)$ and from $T_{\text{obs}}(x)$; if the observed statistic is extreme relative to its simulated 
counterparts, the null hypothesis is rejected. The same simulated functions are also used to construct a global envelope, yielding an interpretable graphical summary of where the observed function deviates 
from the null.

In our application, the functional statistic $T(x)$ is based on boundary curves estimated from the heteroskedastic GP model. We consider two types of hypotheses:

\medskip
\noindent
\emph{Case 1: Comparing two mean boundary curves from different time periods.}
Let $\widehat{y}^{(A)}(x)$ and $\widehat{y}^{(B)}(x)$ denote the GP-predicted mean boundary curves obtained from two disjoint sets of years $A$ and $B$, respectively, evaluated on a common longitude grid $x \in \mathcal{X}$. The goal is to test whether the underlying latent boundary functions associated with periods $A$ and $B$ differ. The observed functional statistic is the pointwise difference
\[
T_{\mathrm{obs}}(x) = 
\widehat{y}^{(A)}(x) - \widehat{y}^{(B)}(x).
\]

To construct the null model, we generate an ensemble from the estimated latent boundary process corresponding to one of the periods (say, period $A$). Let $f^{(A)}_i(x)$ and $f^{(A)}_j(x)$ denote two independent draws from the predictive distribution of the latent GP boundary for period $A$ (i.e., GP realizations based on the fitted mean function $\widehat{\mu}_f(x)$ and covariance function $\widehat{k}_f(x,x')$). The null ensemble is defined by
\[
T_i(x) = f^{(A)}_i(x) - f^{(A)}_j(x), 
\qquad i = 1,\ldots,M.
\]
These realizations $\{T_i(x)\}$ may be viewed as i.i.d.\ samples from an underlying population distribution $T(x)$ with mean zero, representing the natural variability in boundary differences when no temporal change has occurred.

The statistical hypotheses can be written directly in terms of the underlying latent 
boundary mean functions:
\[
\begin{aligned}
H_0: &\quad \mu_f^{(A)}(x) = \mu_f^{(B)}(x)
      \qquad\text{for all } x \in \mathcal{X},\\
H_a: &\quad \mu_f^{(A)}(x) \neq \mu_f^{(B)}(x)
      \qquad\text{for some } x \in \mathcal{X}.
\end{aligned}
\]
Equivalently, in terms of the functional differences,
\[
H_0:~~ T_{\mathrm{obs}}(x) \stackrel{d}{=} T(x) 
\qquad\text{for all } x \in \mathcal{X}, 
\qquad
H_a:~~ T_{\mathrm{obs}}(x) \not\stackrel{d}{=} T(x),
\]
meaning that under $H_0$ the observed difference between the two predicted mean boundaries is consistent with the natural variability in the reference period's boundary model, whereas under $H_a$ the difference exceeds that natural variability, indicating a boundary shift.

\medskip
\noindent
\emph{Case 2: Comparing an observed annual boundary to a mean boundary curve from a reference period.}

In many applications, interest lies in assessing whether the boundary observed in a particular year $t$ differs from what would be expected under the mean boundary structure of a reference period (for example, comparing the boundary for year $t = 1984$ to the mean boundary over the preceding years $A = \{1960,\ldots,1980\}$, say). Let $\widehat{y}^{(A)}(x)$ denote the GP-predicted mean boundary curve obtained by fitting 
the GP model to period $A$, evaluated on the longitude grid $x \in \mathcal{X}$ at which boundary points in year $t$ are observed. Let $y_t(x)$ denote the set of observed boundary latitudes at the corresponding longitude locations for year $t$. Our goal is to test whether the underlying latent boundary function for year $t$ deviates systematically from that of period $A$. The observed functional statistic is the pointwise difference
\[
T_{\mathrm{obs}}(x) = \widehat{y}^{(A)}(x) - y_t(x), 
\qquad x \in \mathcal{X}.
\]

To generate the null model, we again draw from the predictive distribution of the latent boundary function for period $A$. Let $f^{(A)}_i(x)$ and $f^{(A)}_j(x)$ denote two independent draws from the latent GP associated with the reference period, based on the fitted mean function $\widehat{\mu}_f(x)$ and covariance function $\widehat{k}_f(x,x')$. The null ensemble is constructed as
\[
T_i(x) = f^{(A)}_i(x) - f^{(A)}_j(x), 
\qquad i = 1,\ldots,M,
\]
so that $\{T_i(x)\}$ are i.i.d.\ draws from a distribution $T(x)$ with mean zero representing the natural variability in differences between boundary curves arising from the same underlying climate regime.

The hypotheses may be written in terms of the latent boundary mean function for the 
reference period and the latent boundary function for the specific year $t$:
\[
\begin{aligned}
H_0: &\quad \mu_f^{(A)}(x) = \mu_f^{(t)}(x)
      \qquad\text{for all } x \in \mathcal{X},\\
H_a: &\quad \mu_f^{(A)}(x) \neq \mu_f^{(t)}(x)
      \qquad\text{for some } x \in \mathcal{X}.
\end{aligned}
\]
Equivalently, in terms of the functional differences,
\[
H_0:~~ T_{\mathrm{obs}}(x) \stackrel{d}{=} T(x)
\qquad\text{for } x \in \mathcal{X},
\qquad
H_a:~~ T_{\mathrm{obs}}(x) \not\stackrel{d}{=} T(x),
\]
meaning that under $H_0$ the observed deviation of the year $t$ boundary from the reference mean is consistent with natural variability within period $A$, while under $H_a$ the deviation is too large to be explained by this variability, indicating a year-specific shift in the boundary.
\begin{comment}
\notered{Qian: Our observed $T_{\text{obs}}$ were calculated based on $\hat{y}$ but the references $T_i$ was calculated based on $f$. They need to follow the same distribution if the null is true. The definition of $\hat{y}$ is not very clear to me. Is it randomly generated or an expectation, like $\mu_f$? How to set $t$ in $\bm{r}_t^\top \boldsymbol{\beta}$ for $\hat{y}$?

I found that the details for obtaining $\hat{y}$ were introduced in Section 5.1. We may add a reference here. }
\end{comment}
%\subsection{Scaled MAD Global Envelope Test}

\medskip
\noindent 
\emph{Testing of hypothesis framework}.  Although Cases~1 and~2 differ in how the observed statistic $T_{\mathrm{obs}}(x)$ is constructed, the evaluation of the test statistic follows the same scaled Maximum Absolute Difference (MAD) Global Envelope Test (GET) framework of \cite{myllymaki2017global}. In both cases, we use the ensemble 
$\{T_i(x)\}_{i=1}^{M}$ generated under $H_0$ to characterize the null distribution of functional deviations in boundary differences.

For each simulated difference curve $T_i(x)$, we compute its pointwise 
standardization,
\[
Z_i(x) = \frac{T_i(x) - \mu_{\text{en}}(x)}{\sigma_{\text{en}}(x)}, \qquad i=1,\ldots,M,
\]
where
\[
\mu_{\text{en}}(x) = \frac{1}{M}\sum_{i=1}^{M} T_i(x), 
\qquad
\sigma_{\text{en}}(x) = \sqrt{\frac{1}{M-1}\sum_{i=1}^{M} \left(T_i(x)-\mu_{\text{en}}(x)\right)^2},
\]
denote the pointwise sample mean and standard deviation of the null ensemble, respectively. The simulated MAD statistic for iteration $i$ is then
\[
R_i = \max_{x \in \mathcal{X}} |Z_i(x)|.
\]

Similarly, the observed difference is standardized using the same $\mu_{\text{en}}(x)$ and $\sigma_{\text{en}}(x)$ to obtain the observed test statistic as follows,
\[
Z_{\mathrm{obs}}(x) 
= \frac{T_{\mathrm{obs}}(x) - \mu_{\text{en}}(x)}{\sigma_{\text{en}}(x)},
\qquad
R_{\mathrm{obs}} 
= \max_{x \in \mathcal{X}} |Z_{\mathrm{obs}}(x)|.
\]
The Monte Carlo $p$--value is then computed as
\[
p = \frac{1}{M} \sum_{i=1}^{M} \mathbbm{1}\left(R_{\mathrm{obs}} \ge R_i\right),
\]
and the null hypothesis is rejected at significance level $\alpha$ whenever $p < \alpha$, indicating that the observed difference exceeds the natural variability captured by the null ensemble.

We also construct a $100(1-\alpha)\%$ global envelope for the test to visualize departures from the null.  Let $\{R_{(1)} \le \cdots \le R_{(M)}\}$ be the ordered simulated MAD values and 
define $R_{(\alpha)} = R_{(M - \lfloor \alpha M \rfloor)}$.  
The global envelope is then given by
\[
\phi_{\mathrm{env}}(x) = 
\left\{
\mu_{\text{en}}(x) - R_{(\alpha)}\,\sigma_{\text{en}}(x),\;
\mu_{\text{en}}(x) + R_{(\alpha)}\,\sigma_{\text{en}}(x)
\right\}.
\]
These joint envelopes have also been studied in \cite{crainiceanu2012bootstrap, cui2022fast}. Plotting $T_{\mathrm{obs}}(x)$ together with this envelope allows us to identify the specific longitude ranges where the observed boundary deviates significantly from the null model. In addition, if $T_{\mathrm{obs}}(x)$ exceeds the envelope for any $x$, 
the test rejects $H_0$, signaling a statistically significant boundary shift.

\section{Simulation Studies} \label{sec:sim}

The primary goal of the simulation study is to evaluate the finite-sample performance of the scaled MAD global envelope test for detecting changes in spatial boundaries. In particular, we focus on (i) whether the test attains the nominal Type~I error rate under the null hypothesis of no boundary shift and (ii) its power under controlled alternative scenarios. We do not aim to assess the accuracy of Gaussian process boundary estimation itself, since GP regression is already well established in the spatial statistics literature. 

\begin{figure}[!ht]
    \centering
    \includegraphics[width=0.5\textwidth]{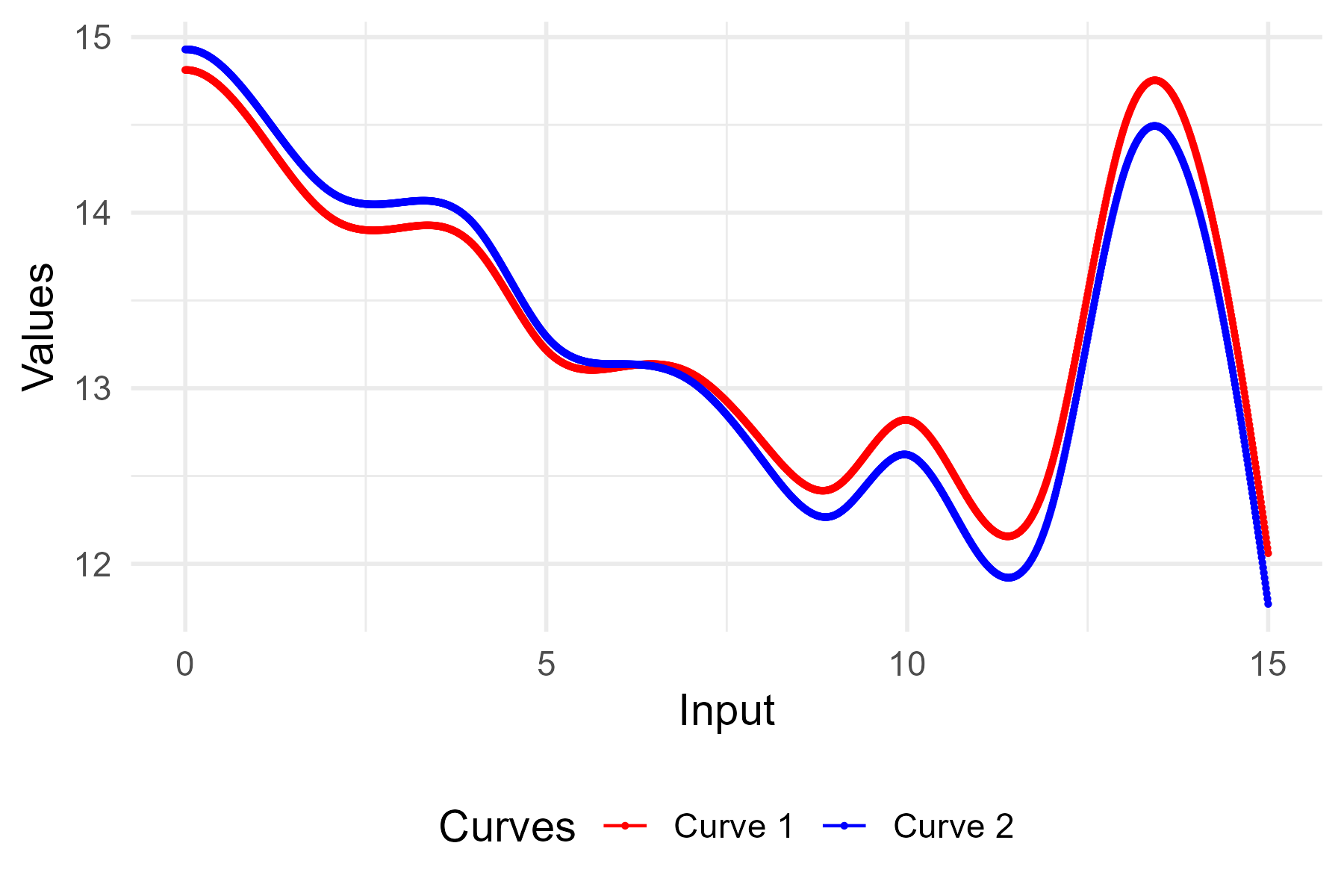}
    \caption{Two independent boundary curves generated from the same B-spline mean function $f_0(x)$ under the null hypothesis of no boundary shift.}
    \label{fig:alternative_curves}
\end{figure}

We construct a ``true'' mean boundary curve $f_0(x)$ on a one-dimensional spatial domain $x \in \mathcal{X}$ using cubic B-splines, explicitly shaping it to resemble the Sahel–Sahara boundary in the real data while retaining flexible and interpretable structural control. Let $B_{j,k}(x)$ denote the B-spline basis functions of degree $k$ associated with a non-decreasing knot vector $\{t_j\}$. Then the mean boundary is defined as
\[
f_0(x) = \sum_{j=0}^{n} \beta_j B_{j,k}(x) + C_0,
\]
where $\beta_j$ are spline coefficients, $k = 3$ for cubic splines, $n + 1$ is the number of basis functions, and $C_0$ is a constant chosen to align the resulting curve with the latitude range of the Sahel-Sahara boundary in the real data. The basis functions 
satisfy the standard B-spline recursion,
\begin{align*}
B_{j,0}(x) = &
\begin{cases}
1, & \text{if } t_j \leq x < t_{j+1},\\
0, & \text{otherwise},
\end{cases}
\\
B_{j,k}(x) = &
\frac{x - t_j}{t_{j+k} - t_j} B_{j,k-1}(x)
+
\frac{t_{j+k+1} - x}{t_{j+k+1} - t_{j+1}} B_{j+1,k-1}(x).
\end{align*}
We fix the knots at 12 equi-spaced knot locations and choose a $16 \times 1$ coefficient vector $\boldsymbol{\beta}^\top := \{0, -1,  -1, -1, -2, -2, -2, -2.5, -1, -2, -2, -3,  -3, 1,  -1, -3 \}^\top$ so that $f_0(x)$ visually resembles the Sahel-Sahara transition boundary (see Figure~\ref{fig:alternative_curves}). The constant term $C_0$ is set to $15$ to match the approximate latitude range. 

To mimic spatially correlated measurement variability around this mean boundary, we generate independent realizations
\[
f_r(x) = f_0(x) + \varepsilon_r(x), 
\qquad r = 1,\dots,R,
\]
where $\varepsilon_r(x)$ is a zero-mean Gaussian process with a squared exponential covariance function
\[
\varepsilon_r(x) \sim \mathcal{GP}\big(\bm{0}, k_e(x,x';\boldsymbol{\theta}_e)\big), 
\qquad k_e(x,x';\boldsymbol{\theta}_e)
=
\sigma_\varepsilon^2
\exp\!\left\{-\frac{(x-x')^2}{2\ell_\varepsilon^2}\right\},
\]
with length-scale $\ell_\varepsilon = 5$ and variance $\sigma_\varepsilon^2 = 0.01$. This construction yields a family of smooth boundary curves with local variability around the common mean $f_0(x)$, analogous to the spatial fluctuations in the estimated climate classification interfaces. 

The use of B-splines is motivated by their flexibility, local shape control, and interpretability. They also have a long history of use in interface and curve modeling in spatial and image analysis \citep{Rethore2013, Pan2011, Liu2024}. Figure~\ref{fig:alternative_curves} illustrates two independent realizations (fitted under the same mean boundary $f_0(x)$) generated under the null hypothesis of no boundary shift.

\subsection{Empirical Size under the Null Hypothesis}

To evaluate whether the global scaled MAD test maintains the correct Type I error rate in our setting, we conduct a simulation study under the null hypothesis of no difference between two boundary curves. Under $H_0$, both curves being compared arise from the same latent climatic boundary model, meaning that any observed differences should reflect only natural stochastic variability rather than a systematic shift in the boundary. This simulation setting corresponds conceptually to Case~1 in Section~\ref{sec:methods}. To assess the size of the test, we perform the test on varying number of spatial grid points, namely, $n = \{200, 500, 1000, 1500\}$ and varying number of Monte Carlo simulations, $N_{\mathrm{sim}} = \{200, 500, 1000, 1500\}$.

For each grid size and within each iteration, we generate an ensemble of $M = 2500$ independent null difference curves from a mean zero Gaussian process with the above-mentioned squared-exponential covariance function as follows
\[
T_k(x) = f_{r_k}(x) - f_{s_k}(x),
\qquad k = 1,\dots,M = 2500,
\]
where $(r_k,s_k)$ denotes indexes two independent realizations of $f_0(x)$. The choice of 2500 follows the recommendation of \citet{myllymaki2017global}, who suggest that 100–200 curves are typically sufficient for stable estimation. However, we use a larger number of ensemble members as a conservative choice. For each iteration, we perform the scaled MAD test as described in Section \ref{sec:methods}, and compute the corresponding $p$-value, denoted by $p_k$. Iteration $k$ is classified as a rejection when $p_k < 0.05$. The empirical size of the test for each grid size and simulation size combination is then calculated as $p_I = \frac{1}{N_{\mathrm{sim}}} \sum_{k = 1}^{N_{\mathrm{sim}}} \mathbbm{1}\{ p_k < 0.05 \}$.  Table~\ref{tab:size} reports the estimated size (in percent) across all combinations of the number of iterations $N_{\mathrm{sim}}$ and number of spatial locations.

\begin{table}[!ht]
\centering
\caption{Empirical size (\%) of the MAD global envelope test under the null hypothesis for varying numbers of Monte Carlo iterations and spatial locations.}
\label{tab:size}
\begin{tabular}{c|cccc}
\toprule
\multirow{2}{*}{\textbf{\# Locations}} 
& \multicolumn{4}{c}{\textbf{\# Iterations}} \\
\cline{2-5}
& 200 & 500 & 1000 & 1500 \\
\midrule
200  & 5.5 & 6.0 & 4.8 & 4.6 \\
500  & 6.5 & 5.4 & 5.6 & 5.3 \\
1000 & 4.5 & 6.2 & 5.9 & 4.5 \\
1500 & 7.5 & 6.0 & 5.0 & 5.4 \\
\bottomrule
\end{tabular}
\end{table}

\begin{figure}[!ht]
    \centering
    \includegraphics[width=0.5\textwidth]{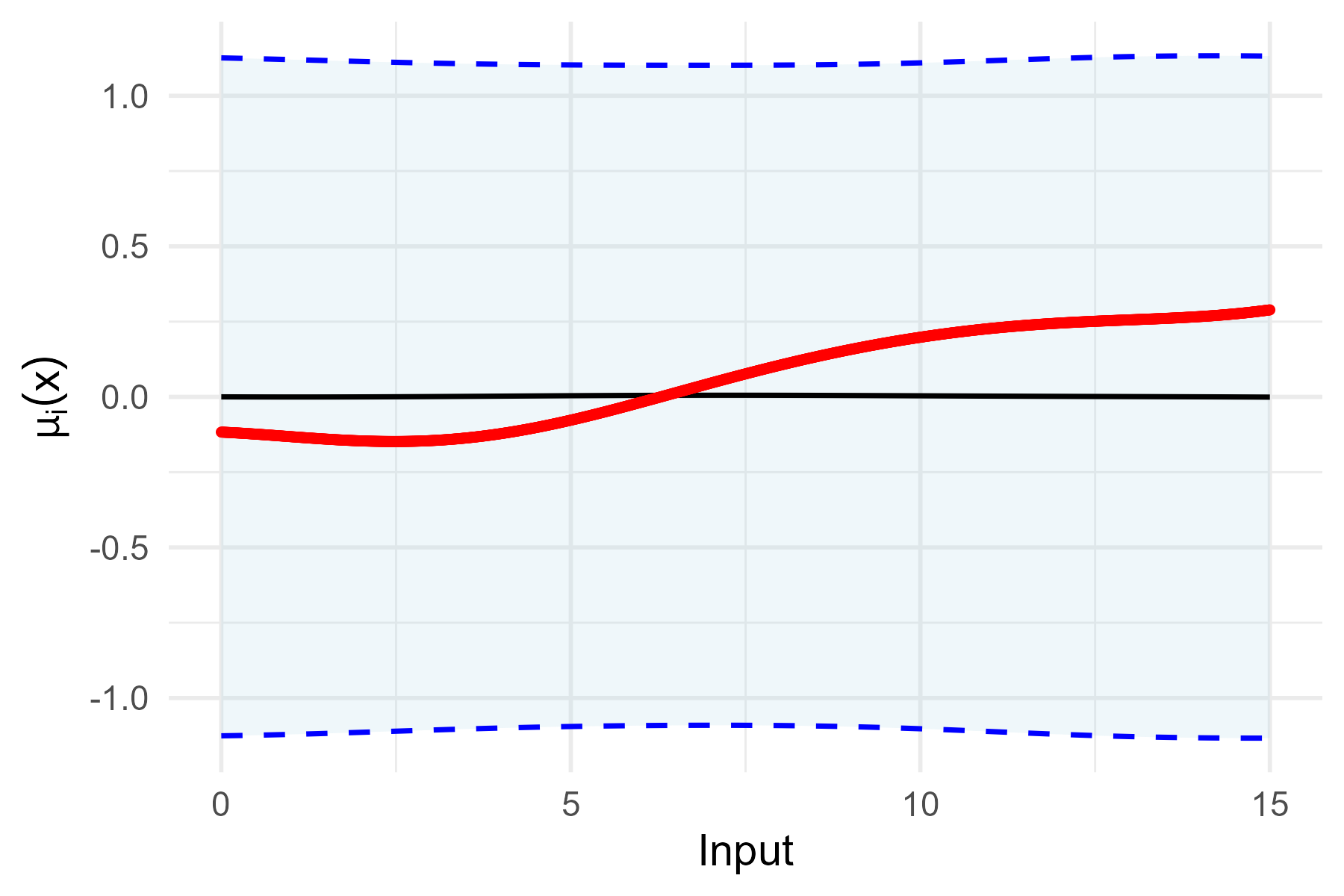}
    \caption{Illustration of the global scaled MAD envelope test under the null hypothesis. The red curve shows a representative observed difference $T_{\mathrm{obs}}(x)$ between two realizations generated from the same mean boundary $f_0(x)$ (see  Figure~\ref{fig:alternative_curves}). The dashed blue curves 
    are the upper and lower limits of the 95\% global envelope, and the solid black curve is the ensemble mean $\mu_{\mathrm{en}}(x)$, which is approximately zero across the domain.}
    \label{fig:GET_Null}
\end{figure}

Overall, the empirical size remains close to the nominal 5\% level across all combinations of grid densities and numbers of iterations, indicating that the MAD global envelope test is well calibrated for boundary comparison in this setting. To visualize the behavior of the test under the null, Figure~\ref{fig:GET_Null} displays a representative global envelope constructed from the null differences, with the observed curve $T_{\mathrm{obs}}(x)$ and the $95\%$ envelope derived from the critical MAD value (here corresponding to the upper $5$th percentile of the ordered $\{R_i\}$). As expected under the null, the observed curve remains well within the envelope over the entire domain.

\subsection{Power Computations}

To assess the power of the proposed test, we generate boundary curves under controlled departures from the null model. Whereas the null hypothesis assumes both curves arise from the same latent mean boundary $f_0(x)$, the alternative hypothesis introduces a localized perturbation to this mean function. This allows us to evaluate the ability of the test to detect boundary shifts of varying magnitude.

\begin{figure}[!ht]
    \centering
    \includegraphics[width=0.5\textwidth]{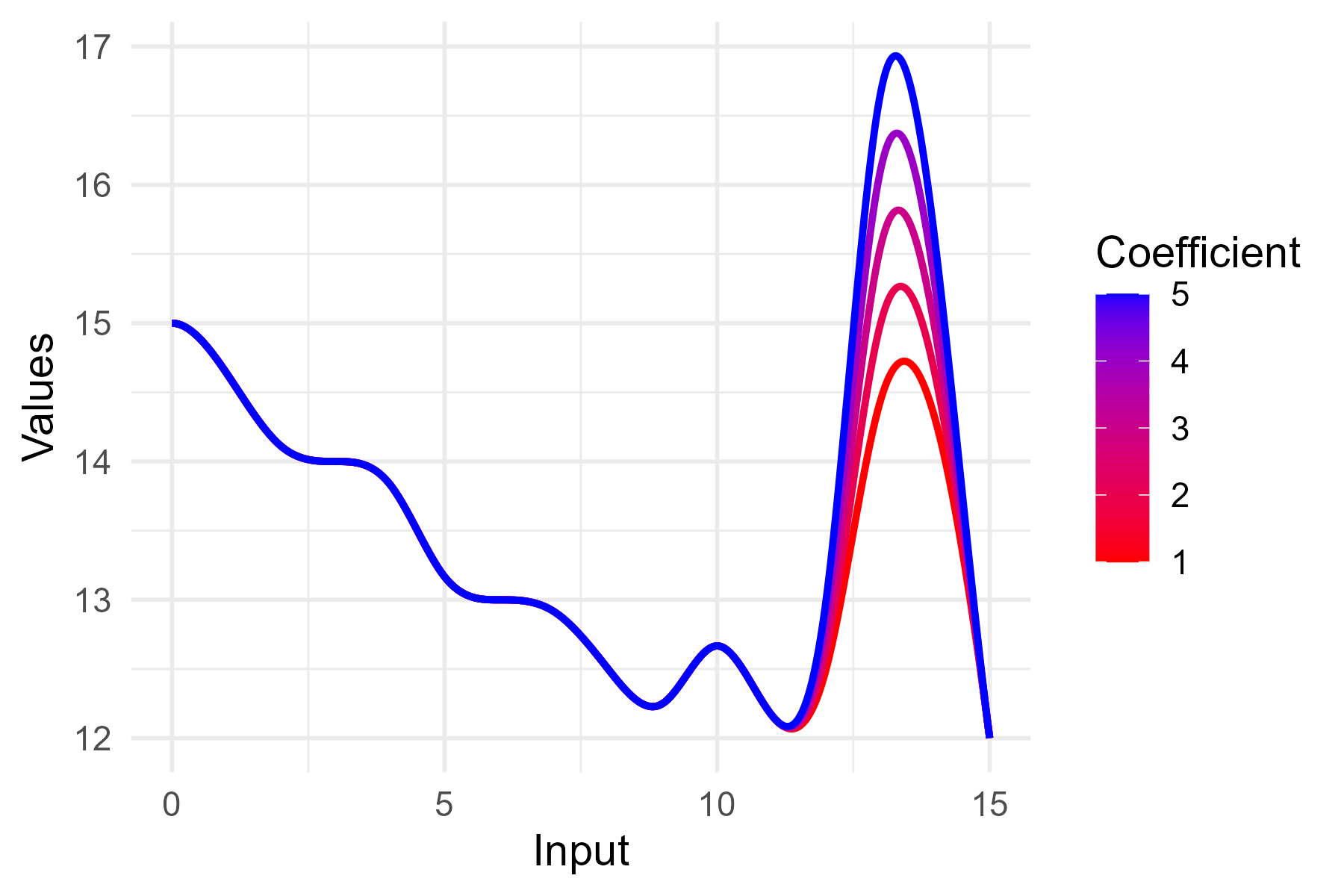}
    \caption{Mean boundary curves obtained by increasing the B-spline coefficient $\beta_{13}$ from its null value of 1 to 5 in increments of 0.5, producing progressively larger localized deviations used to define the alternative hypothesis.}
    \label{fig:Null_Hypothesis}
\end{figure}

We construct the alternative mean boundary $f_a(x)$ by modifying a single B-spline coefficient in $\bm{\beta}$. Specifically, we vary the fourteenth coefficient $\beta_{13}$ while keeping all other coefficients identical to those used to define $f_0(x)$. Because B-spline basis functions possess compact support, adjusting $\beta_{13}$ affects the shape of the boundary only over a localized region, thereby producing a controlled and interpretable deviation. The coefficient $\beta_{13}$ is increased in steps of 0.5 from its null value of 1 up to 5, yielding increasingly pronounced local displacements in the mean boundary (see Figure~\ref{fig:Null_Hypothesis}). This setup mimics the kinds of local climatic boundary shifts we aim to detect in the real data.

To generate independent realizations from under the alternative, we now generate from the perturbed mean function 
$$
f_{a; r}(x) = f_a(x) + \varepsilon_r(x), 
$$
where $\varepsilon_r(x)$ is the same GP error process described earlier. In each Monte Carlo iteration, we construct an ensemble of $M = 2500$ null difference curves $\{ T_i(x)\}_{i = 1}^{2500}$ using the unperturbed model $f_0(x)$, exactly as in the empirical size study. We then generate an ``observed" difference curve by pairing one realization from the alternative model with one from the null model 
$$
T_{a; obs}(x) = f_{a; r}(x) - f_{r}(x). 
$$
The null ensemble $\{T_i(x)\}$ is standardized using its Monte Carlo mean and standard deviation, and the MAD statistic $R_{a; obs}$ for the observed curve is also computed. A rejection occurs if $R_{a; obs}$ exceeds the corresponding critical value obtained from the null ensemble. This procedure is repeated for $N_{\mathrm{sim}}$ number of iterations, and the empirical power is estimated as the proportion of repetitions in which the null hypothesis is rejected.

\begin{figure}[!htbp]
    \centering
    \begin{minipage}[b]{0.48\textwidth}
        \centering
        \includegraphics[width=\textwidth]{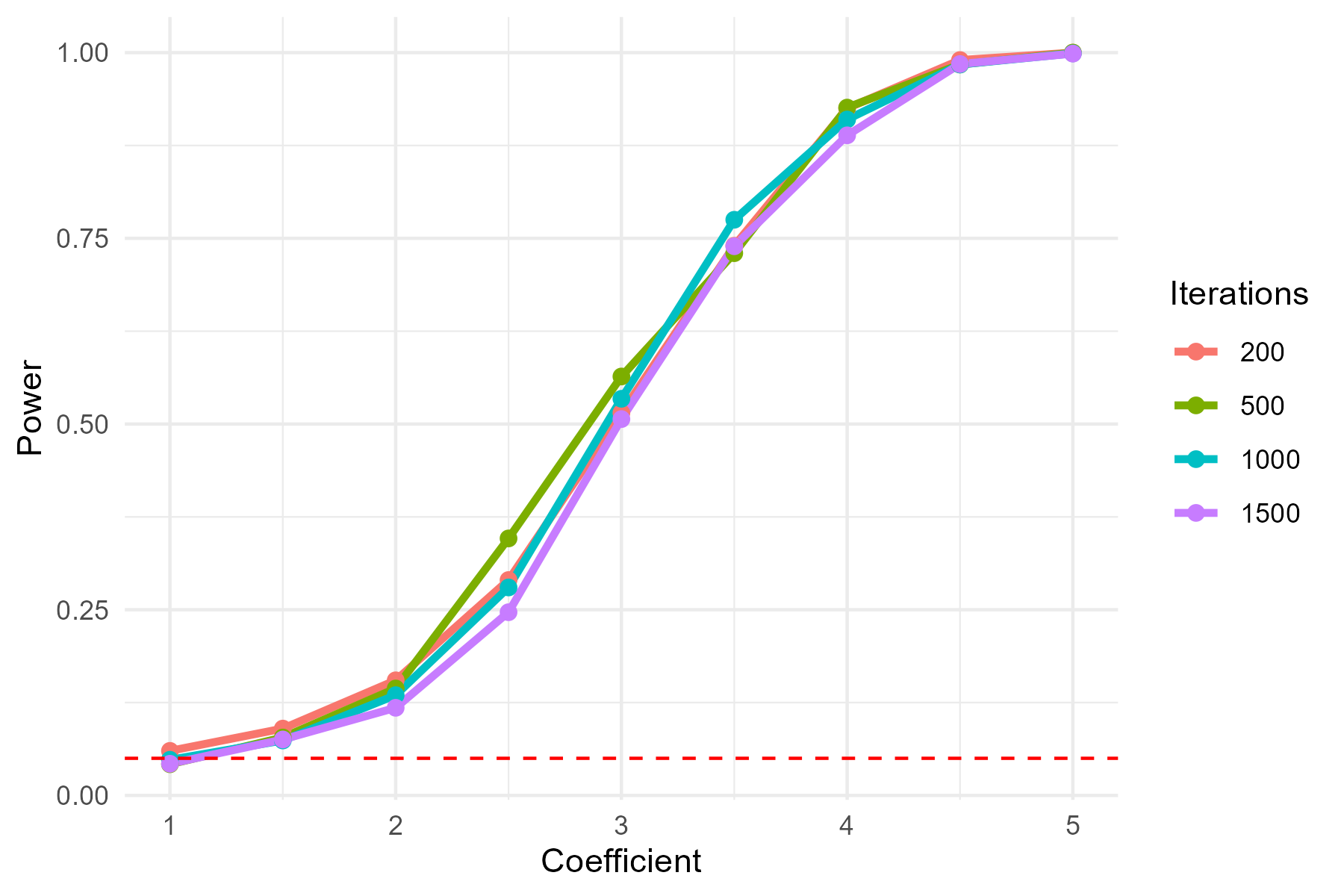}
        \caption*{(a) Varying numbers of iterations.}
    \end{minipage}
    \hfill
    \begin{minipage}[b]{0.48\textwidth}
        \centering
        \includegraphics[width=\textwidth]{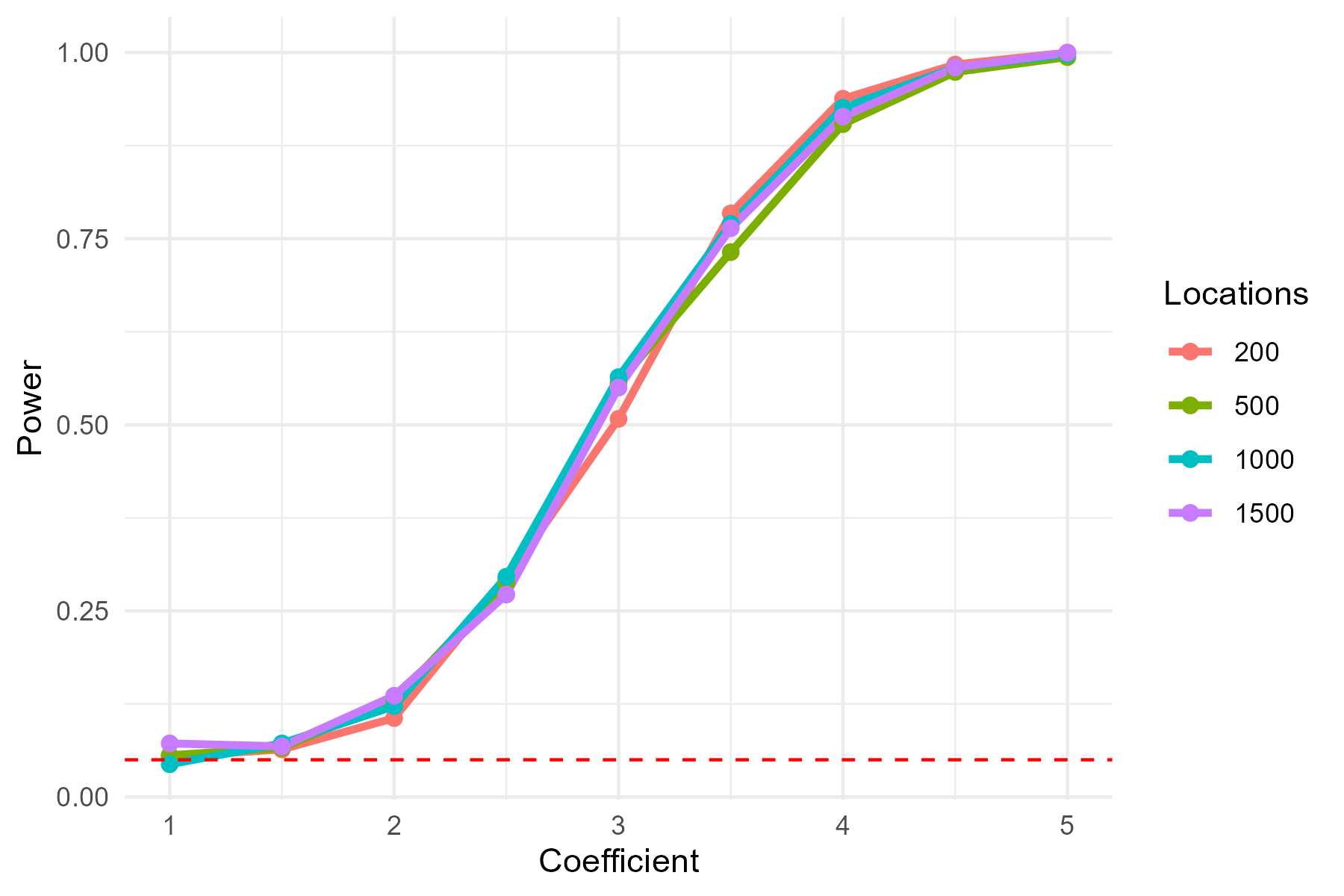}
        \caption*{(b) Varying spatial resolutions.}
    \end{minipage}
    \caption{Empirical power of the scaled MAD global envelope test under increasing perturbations of the B-spline coefficient $\beta_{13}$. Panel (a) shows power curves for fixed spatial resolution ($n = 1000$ grid points) with varying numbers of Monte Carlo repetitions. Panel (b) shows power curves for fixed simulation size ($N_{\mathrm{sim}} = 500$) with varying numbers of spatial grid points. The dotted red line marks the 5\% significance threshold. In both panels, power increases with the magnitude of the localized boundary perturbation, illustrating the sensitivity of the test to even moderate changes in boundary structure.}
    \label{fig:Power_Curve}
\end{figure}

Figure~\ref{fig:Power_Curve}(a) shows the power curves obtained from our simulation when we vary the number of simulation replicates over $N_{\mathrm{sim}} = \{200, 500, 1000, 1500\}$ with the number of grid points fixed at $n = 1000$. In addition, Figure~\ref{fig:Power_Curve}(b) shows the power curves when we vary the number of grid points over $n = \{200, 500, 1000, 1500\}$ keeping the number of simulation replications fixed at $N_{\mathrm{sim}} = 500$. In both cases, power increases monotonically with the magnitude of the coefficient perturbation, reflecting the ability of the scaled MAD statistic to detect even moderate local deviations in boundary structure.

\begin{figure}[!ht]
  \centering
  % Row 1
  \begin{subfigure}[b]{0.30\textwidth}
    \includegraphics[width=\textwidth]{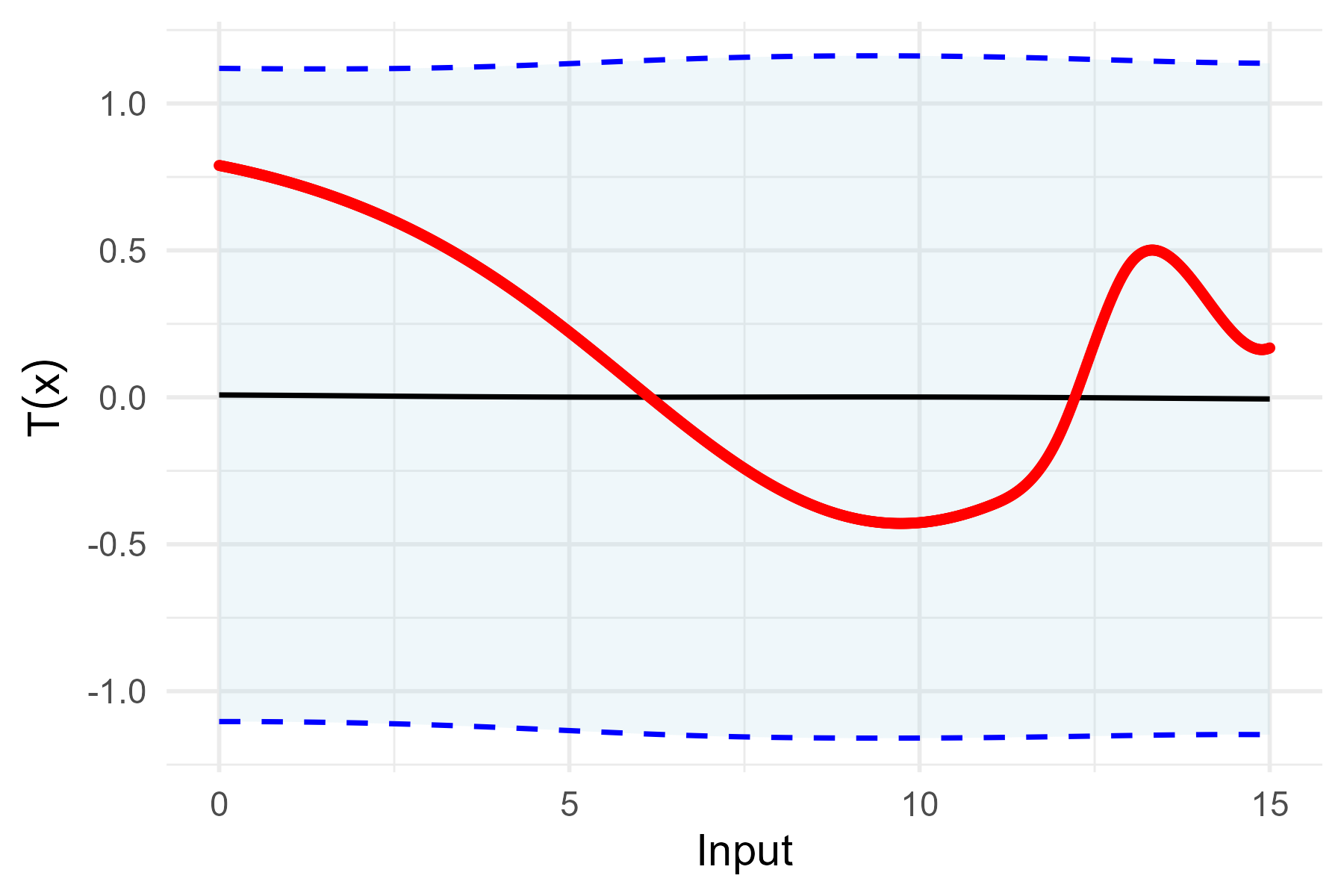}
    \caption{$\beta_{13} = 2$}
  \end{subfigure}
  \hfill
  \begin{subfigure}[b]{0.30\textwidth}
    \includegraphics[width=\textwidth]{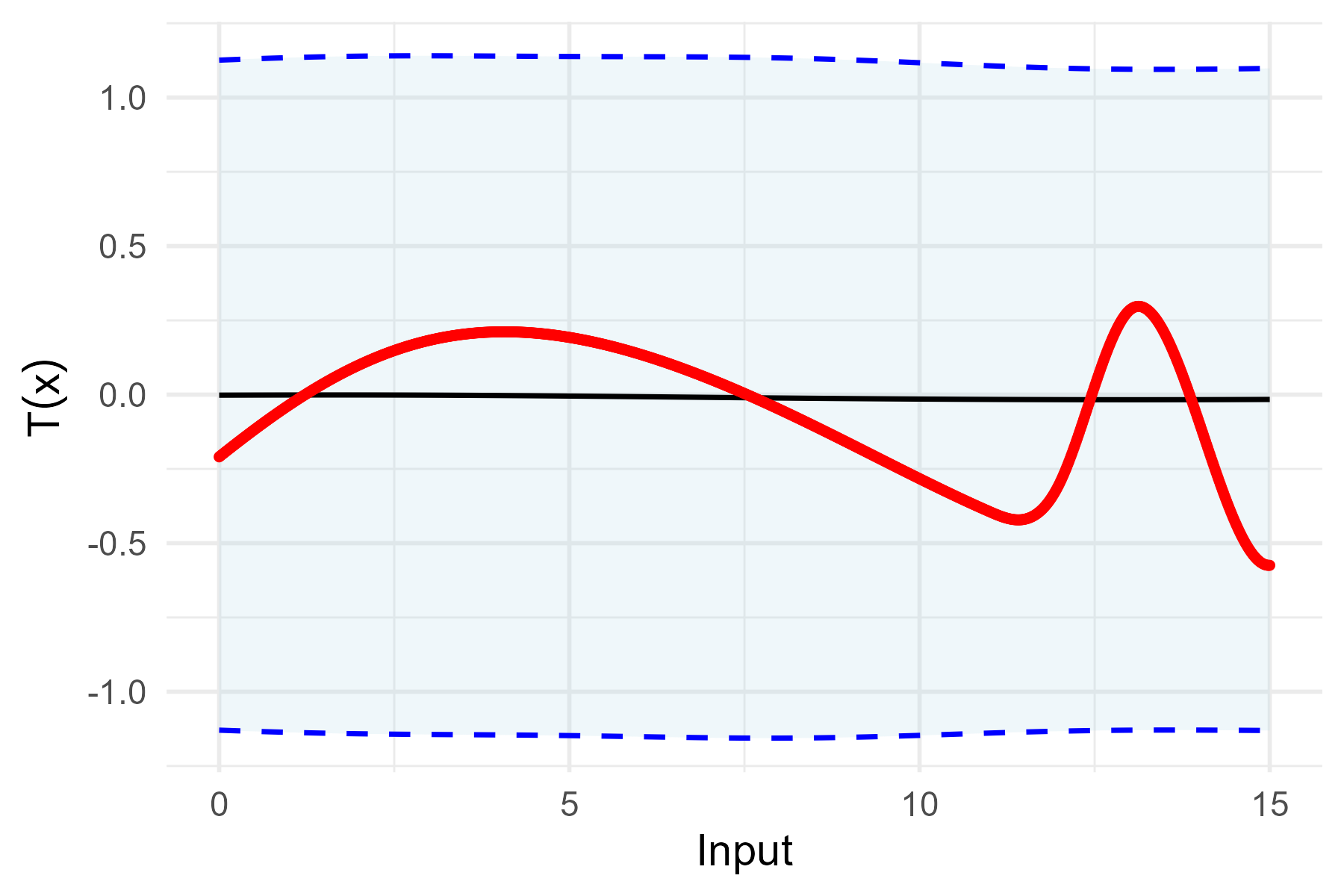}
    \caption{$\beta_{13} = 2.5$}
  \end{subfigure}
  \hfill
  \begin{subfigure}[b]{0.30\textwidth}
    \includegraphics[width=\textwidth]{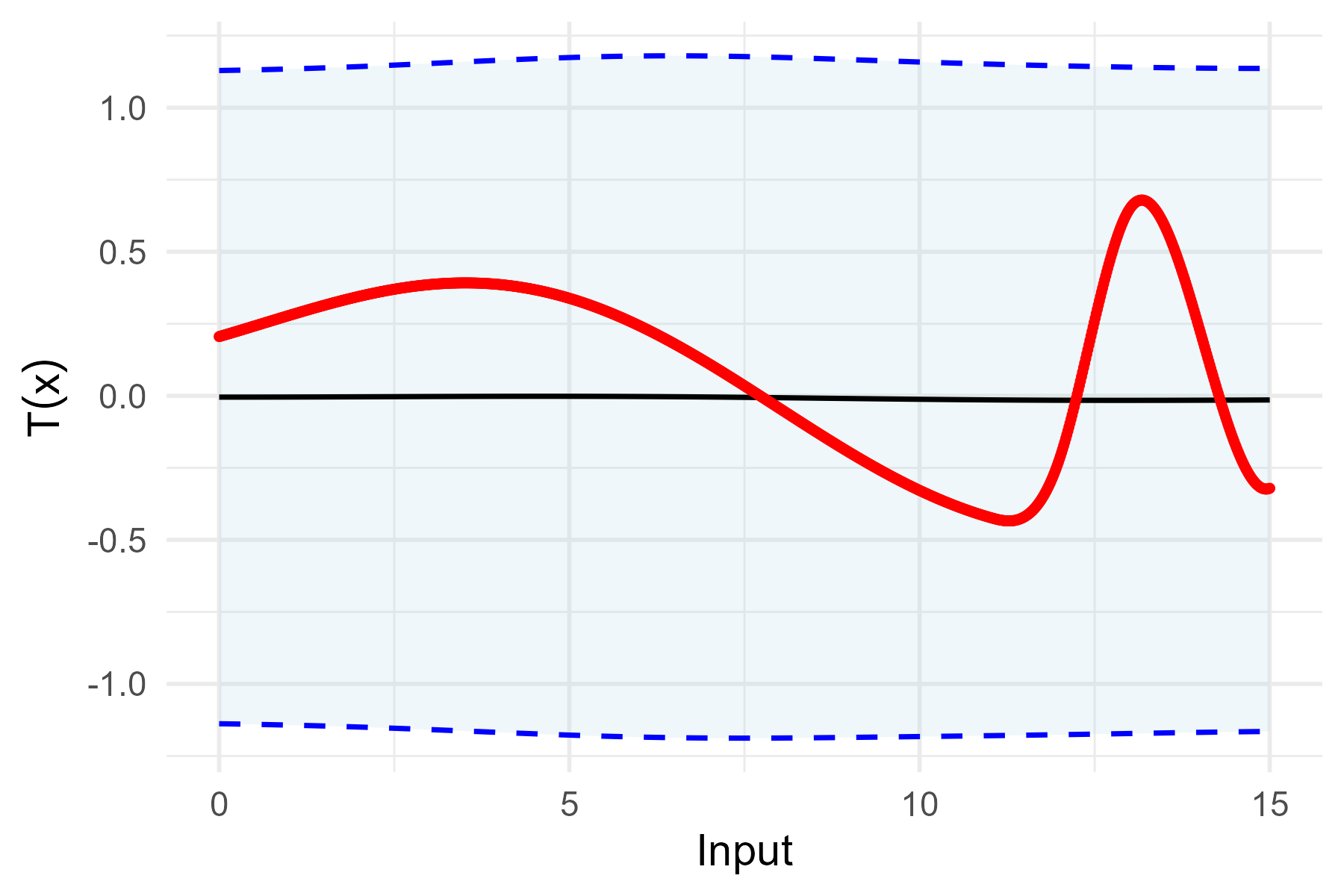}
    \caption{$\beta_{13} = 3$}
  \end{subfigure}

  \vspace{1em}

  % Row 2
  \begin{subfigure}[b]{0.30\textwidth}
    \includegraphics[width=\textwidth]{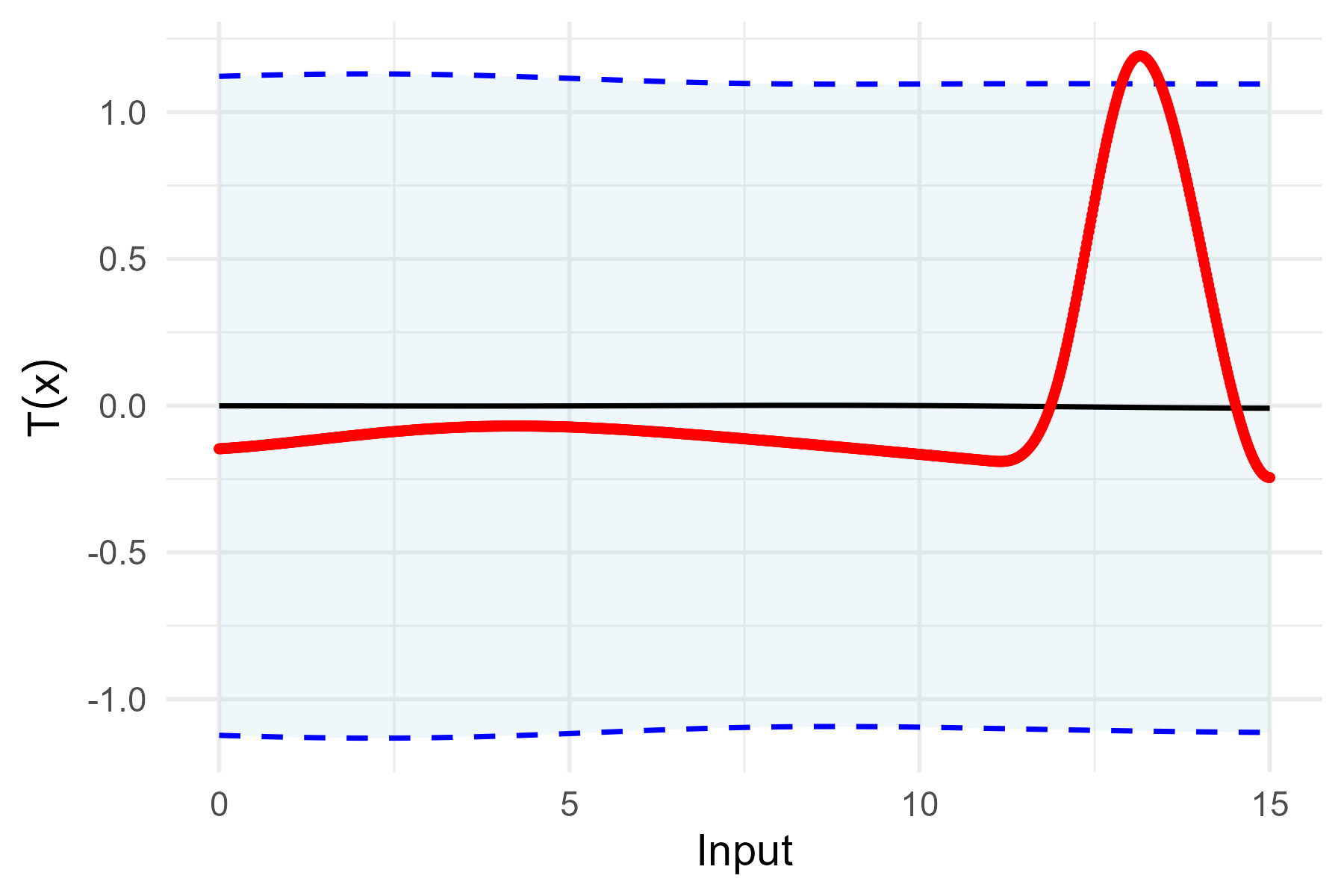}
    \caption{$\beta_{13} = 3.5$}
  \end{subfigure}
  \hfill
  \begin{subfigure}[b]{0.30\textwidth}
    \includegraphics[width=\textwidth]{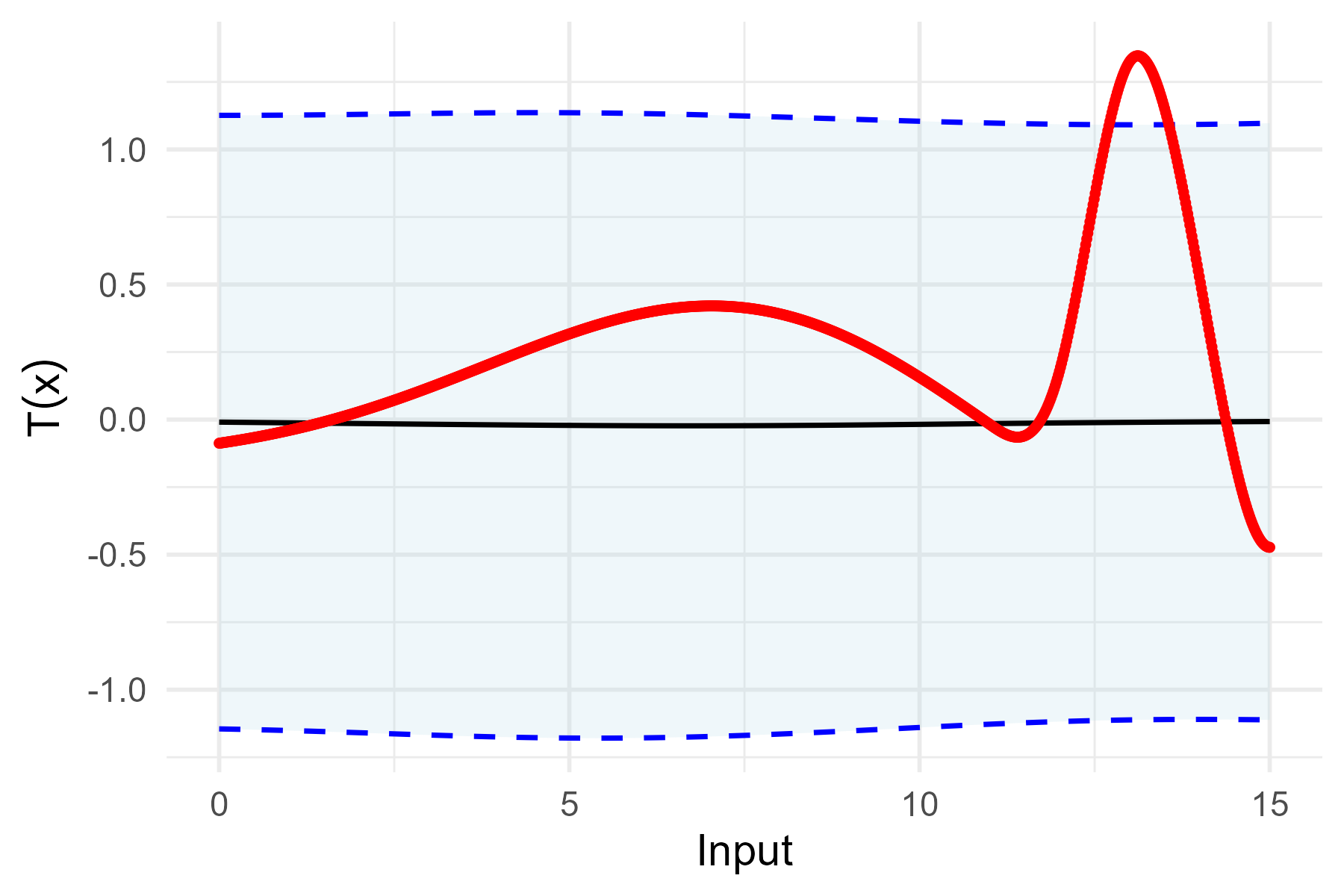}
    \caption{$\beta_{13} = 4$}
  \end{subfigure}
  \hfill
  \begin{subfigure}[b]{0.30\textwidth}
    \includegraphics[width=\textwidth]{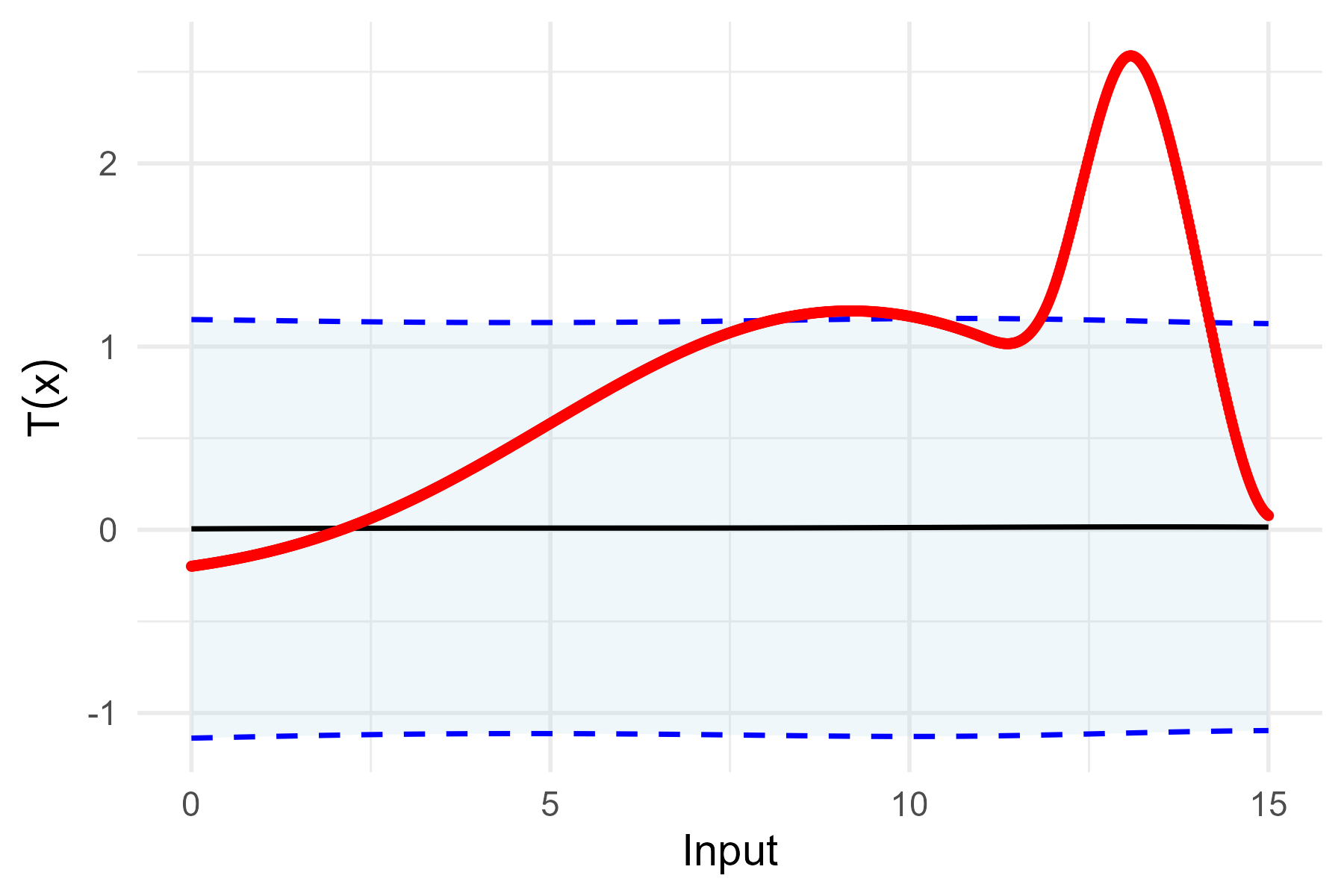}
    \caption{$\beta_{13} = 4.5$}
  \end{subfigure}

  \vspace{1em}

  % Center last plot
  \begin{subfigure}[b]{0.30\textwidth}
    \centering
    \includegraphics[width=\textwidth]{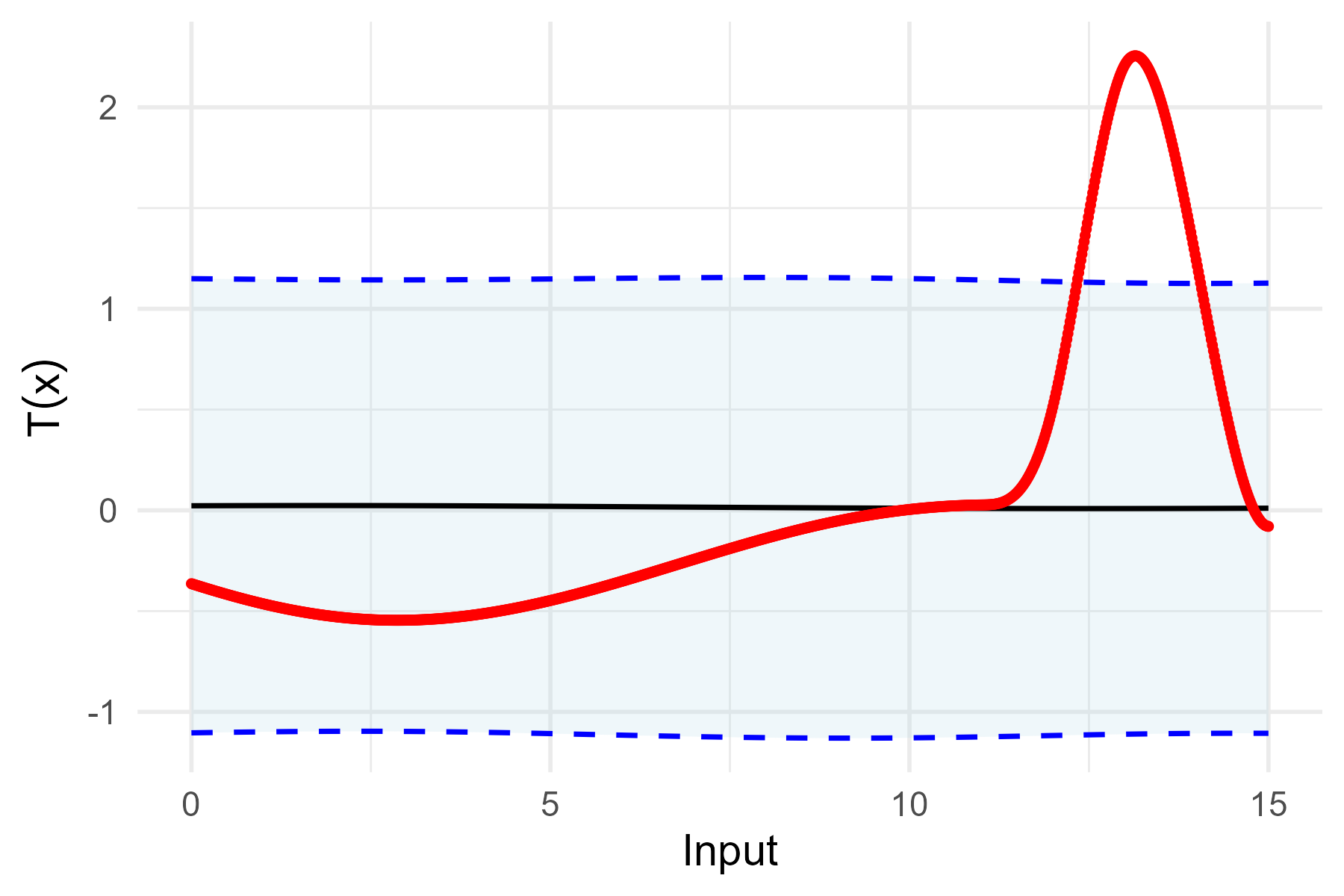}
    \caption{$\beta_{13} = 5$}
  \end{subfigure}

  \caption{Global scaled MAD envelopes for seven perturbation magnitudes. All figures use $n = 1000$ spatial locations. The observed difference curve exceeds the envelope precisely in the region influenced by the modified coefficient $\beta_{13}$.}
  \label{fig:get_plots}
\end{figure}

Finally, to visualize how departures from the null hypothesis reflect in the global envelope, Figure~\ref{fig:get_plots} displays the global envelopes for several values of $\beta_{13}$ under the alternative hypothesis. In each case, the deviation of the observed curve outside the envelope aligns precisely with the spatial region influenced by the perturbed coefficient. As the magnitude of the perturbation increases, these departures become progressively larger, demonstrating the ability of the global scaled MAD test to detect and localize even modest boundary shifts. It is important to note that a significant result from the test indicates that there is significant difference between each tested boundary line, however, by visualizing the GET we can confirm there are local  deviations and continuities between the compared lines as well. Then, it would be fair to say there that  the changes in specific regions resulted in changes of line as compared to distinguishing that there are two entirely different lines.

\section{Application to the Climate Classification Data} \label{sec:results}

In this section we apply the heteroskedastic Gaussian process (GP) regression model and the scaled MAD global envelope test to the K\"oppen–Trewartha climate (KTC) boundary dataset introduced in Section~\ref{sec:data_desc}. The analysis focuses on detecting temporal changes in two key climate interfaces across the Sahel–Sahara region, namely, the arid/semi-arid interface, referred to hereafter as the `primary' boundary, and the semi-arid/non-arid interface, referred to as the ``secondary' boundary. Our objectives are twofold. First, we assess whether the mean climatic boundaries corresponding to the 1960s, 1970s, and 1980s show statistically detectable shifts. Second, we investigate whether specific individual years, especially those associated with well-documented drought episodes, exhibit deviations from their decade-level mean boundaries. All predictions, uncertainty quantification, and hypothesis tests follow the methodology described in Section~\ref{sec:methods}.

\subsection{Decadal Boundary Comparisons}

To examine decade-scale changes in the climatic interfaces, the entire dataset is partitioned into three disjoint temporal subsets representing the 1960s (1960–1969), 1970s (1970–1979), and 1980s (1980–1989). For each year $t$, the temporal structure is encoded via the Fourier basis vector 
$\bm{r}_t$ as defined in Section~\ref{sec:methods}. To obtain a decade-level temporal covariate, we average these basis vectors over the corresponding ten years. For a decade $D$, this yields, 
$$
\bar{\bm{r}}_D = \frac{1}{|D|} \sum_{t \in D} \bm{r}_t, \qquad D \in \{1960s, 1970s, 1980s\}, 
$$
which is then used as the temporal covariate for predicting each decade’s mean boundary curve.

\begin{figure}[!htbp]
    \centering
    \begin{minipage}{0.49\textwidth}
        \centering
        \includegraphics[width=\textwidth]{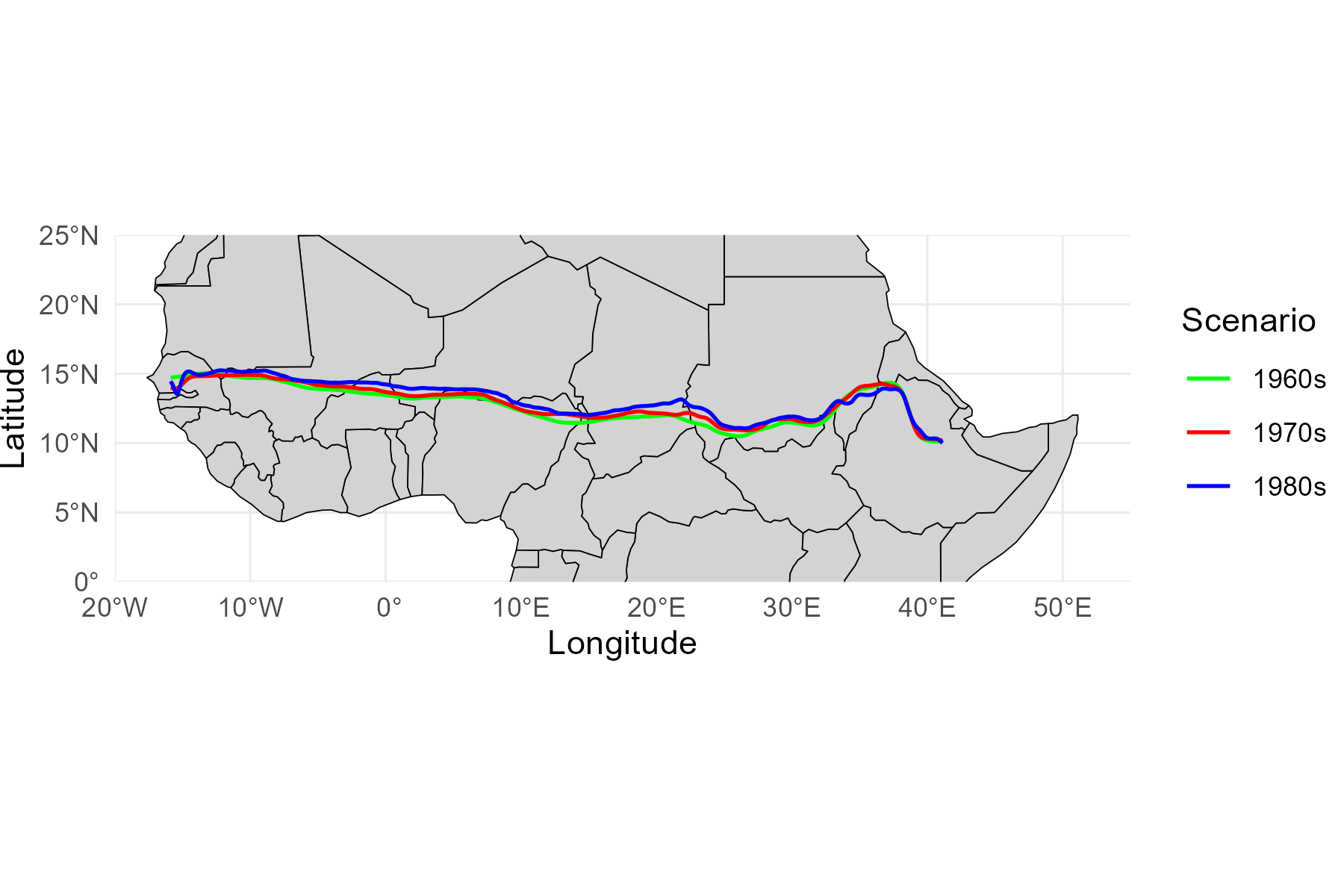}
        \vspace{-1.8cm}
        \caption*{(a) Primary boundary: predicted mean curves}
    \end{minipage}
    \hfill
    \begin{minipage}{0.49\textwidth}
        \centering
        \includegraphics[width=\textwidth]{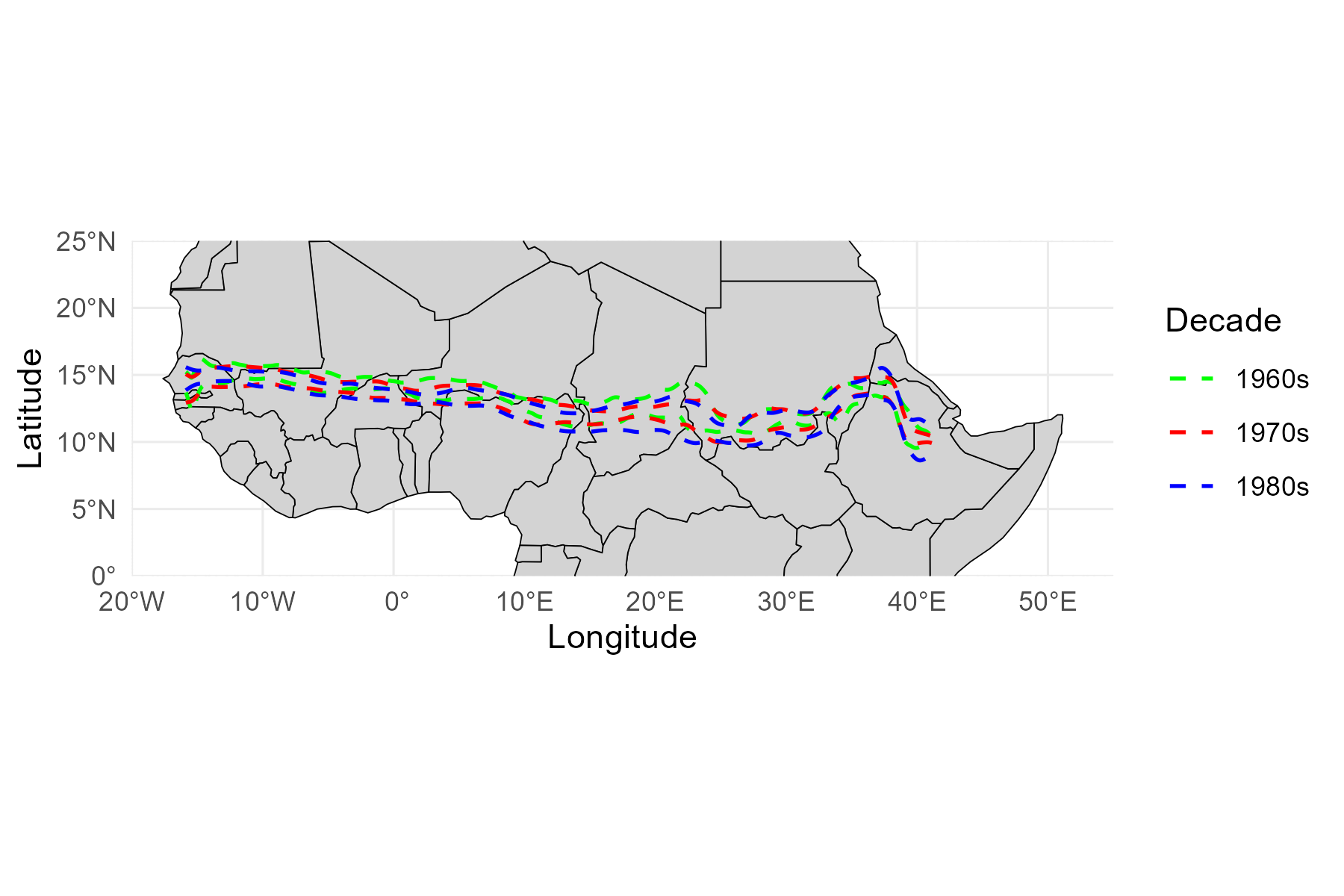}
        \vspace{-1.8cm}
        \caption*{(b) Primary boundary: prediction intervals}
    \end{minipage}
    \centering
    \begin{minipage}{0.49\textwidth}
        \centering
        \includegraphics[width=\textwidth]{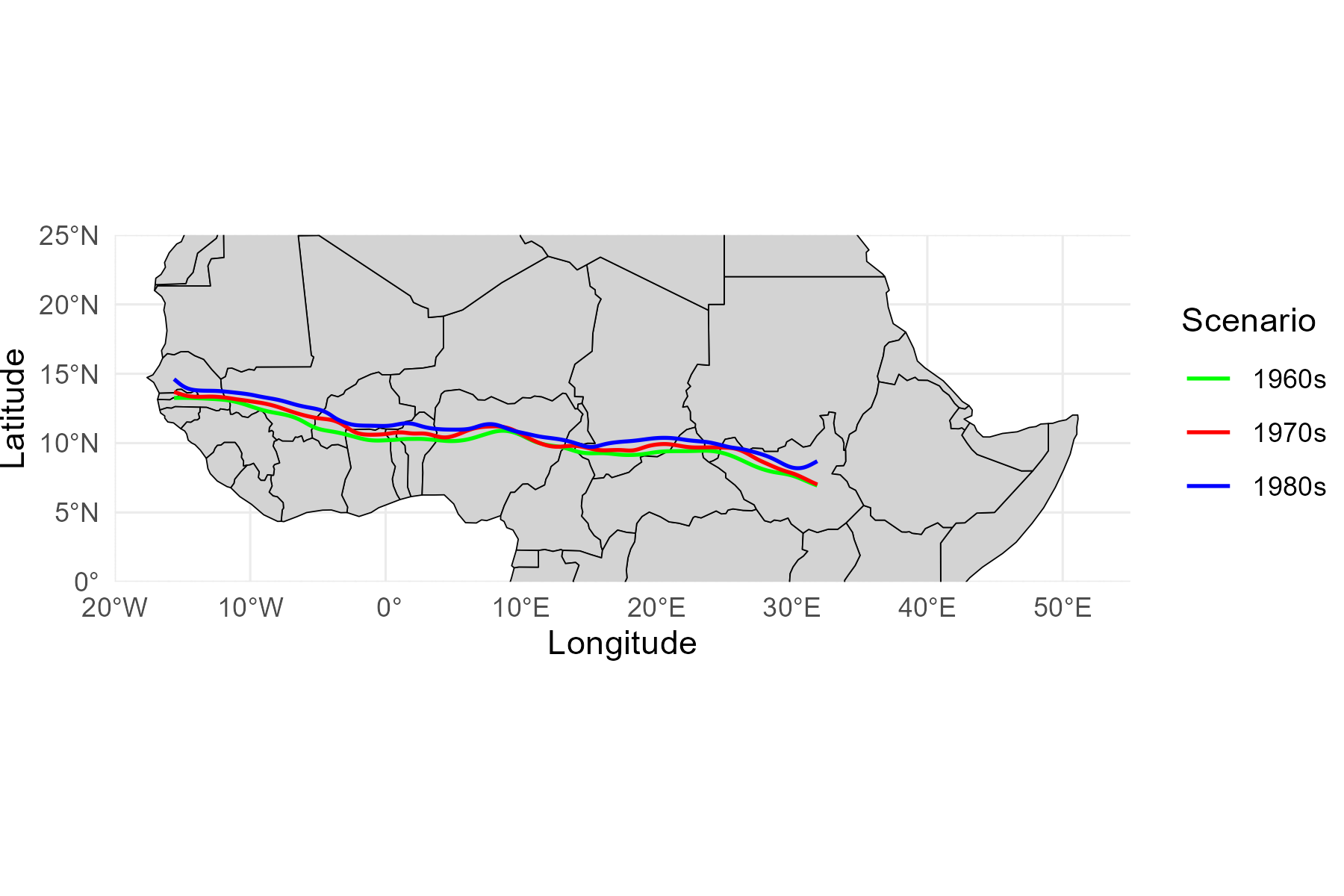}
        \vspace{-1.8cm}
        \caption*{(a) Secondary boundary: predicted mean curves}
    \end{minipage}
    \hfill
    \begin{minipage}{0.49\textwidth}
        \centering
        \includegraphics[width=\textwidth]{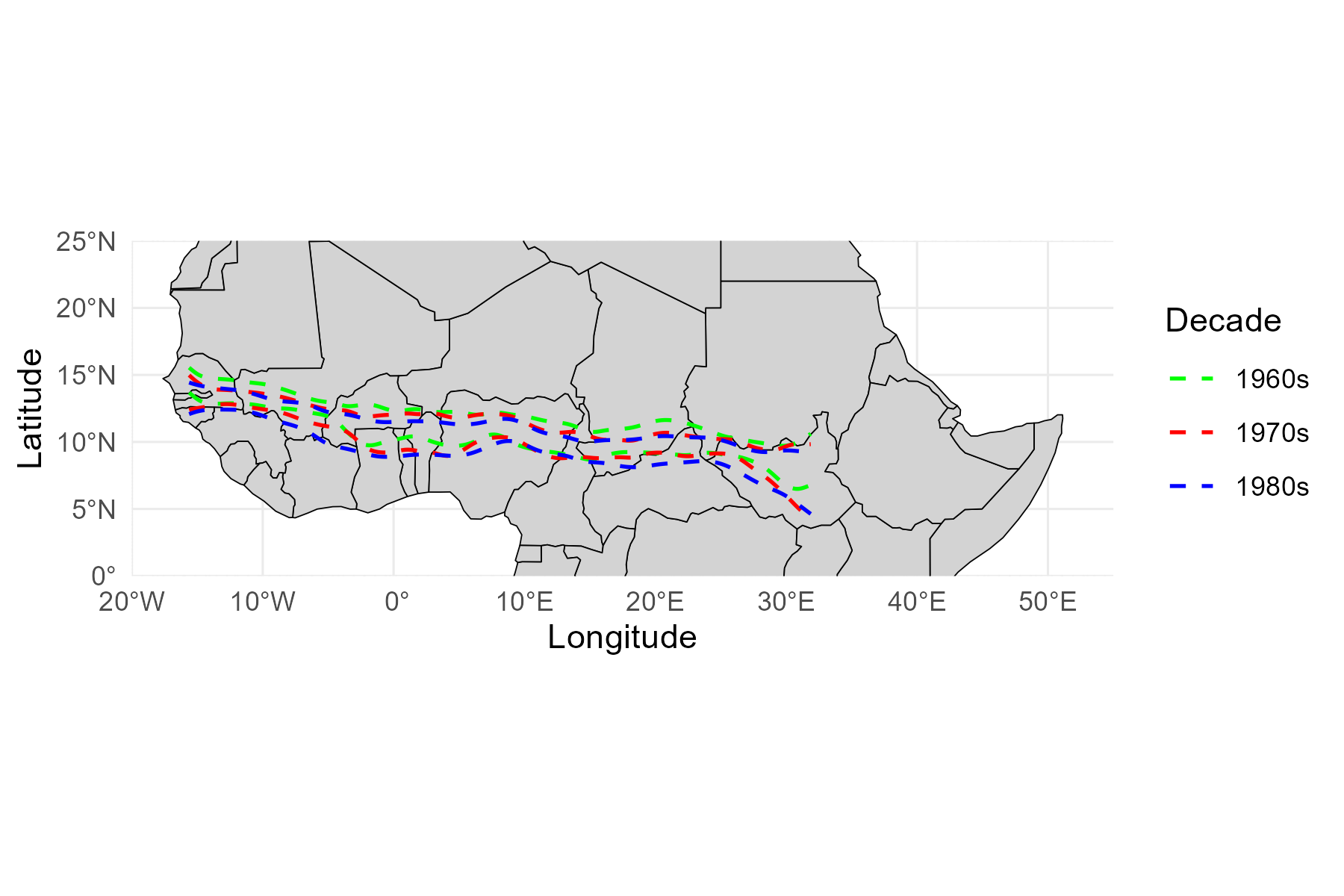}
        \vspace{-1.8cm}
        \caption*{(b) Secondary boundary: prediction intervals}
    \end{minipage}
    \caption{Decadal predicted mean boundary curves and corresponding prediction intervals for the primary (top row) and secondary (bottom row) climatic interfaces. Green, red, and blue curves represent the 1960s, 1970s, and 1980s, respectively.}
    \label{fig:Boundary_Lines}
\end{figure}

A heteroskedastic GP model is fitted separately to each decade-specific subset of boundary points, generating predicted mean boundary curves $\widehat{y}^{(D)}(x^*)$ and associated predictive variances on a dense grid of 1000 equally spaced longitudes spanning the observed spatial domain. Figure~\ref{fig:Boundary_Lines} displays these predicted decade-level boundaries along with their prediction intervals for both the primary (arid/semi-arid) and secondary (semi-arid/non-arid) interfaces. Across all decades, the estimated mean boundaries remain smooth and broadly consistent in their spatial patterns, with no visually pronounced shifts between the 1960s, 1970s, and 1980s. The differing widths of uncertainty intervals highlight the spatially varying heteroskedasticity captured by the GP.

% Put figures side by side 

To formally test whether the decade-level mean curves differ beyond natural climatic variability, we apply the scaled MAD global envelope test using the predictive distribution of each decade-specific GP model to simulate difference ensembles under the null. For any two decades $A$ and $B$, the observed functional statistic is the pointwise difference
$$
T_{obs}(x^*) = \widehat{y}^{(A)}(x^*) - \widehat{y}^{(B)}(x^*),
$$
which is then compared against a null ensemble of synthetic differences generated from the appropriate decade-level GP, following the framework of Case~1 in Section~\ref{sec:methods}. For each comparison, $95\%$ global envelopes are also constructed, and significance was assessed based on p-values and whether $T_{obs}(x^*)$ exited the envelope.

\begin{figure}[!htbp]
    \centering
    % --- Row 1 ---
    \begin{subfigure}[b]{0.3\textwidth}
        \includegraphics[width=\textwidth]{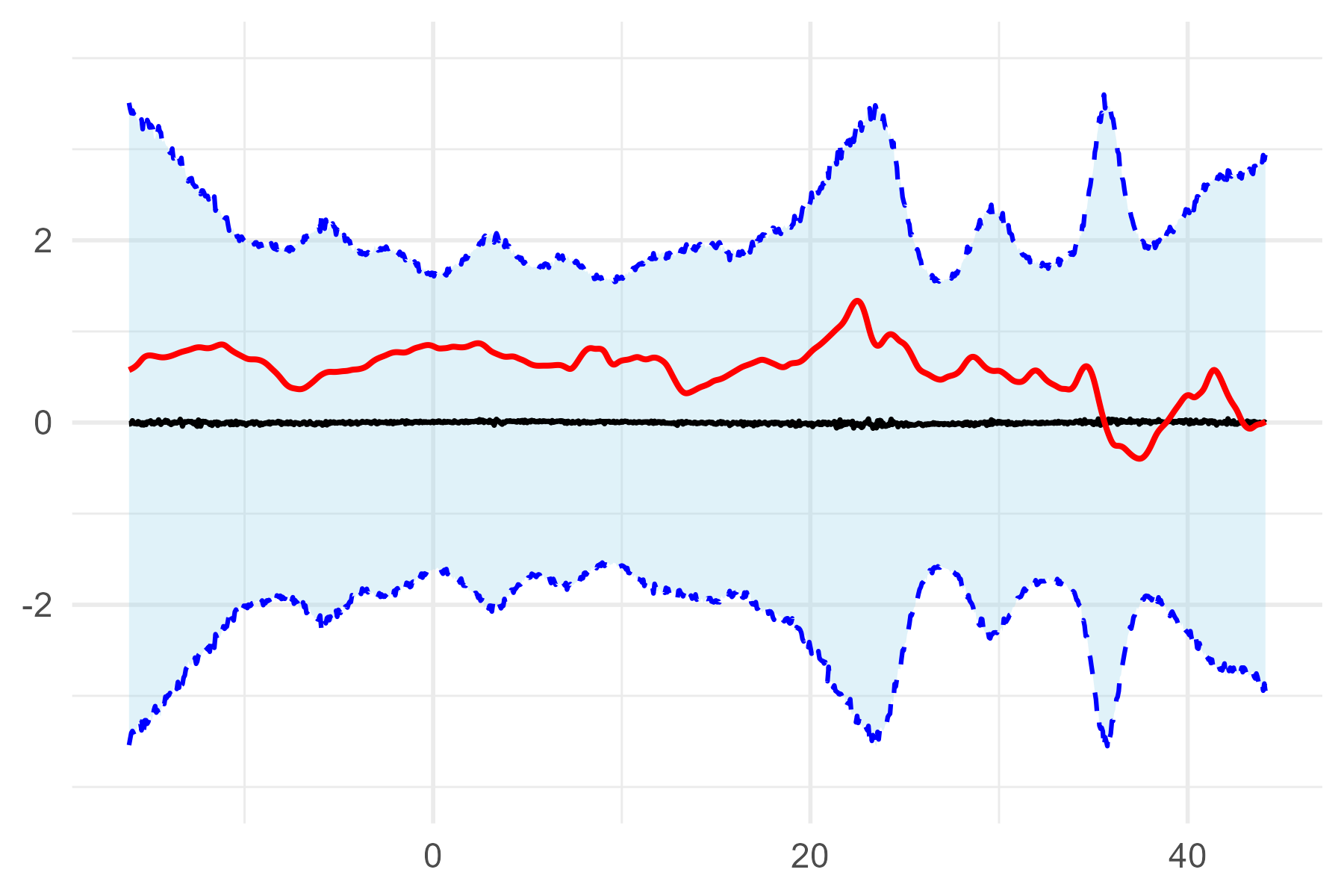}
        \caption{1960 vs 1970}
    \end{subfigure}
    \hfill
    \begin{subfigure}[b]{0.3\textwidth}
        \includegraphics[width=\textwidth]{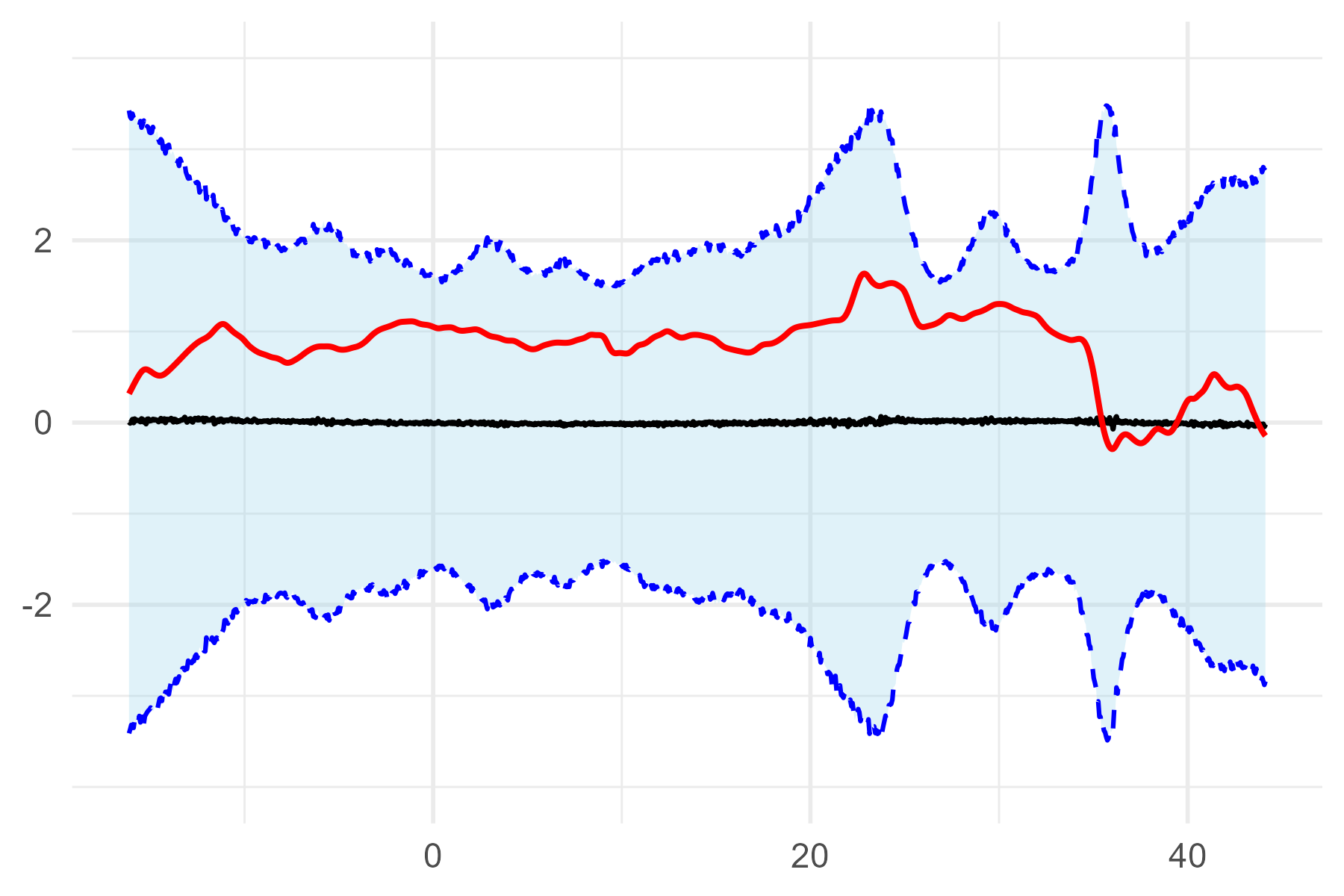}
        \caption{1960 vs 1980}
    \end{subfigure}
    \hfill
    \begin{subfigure}[b]{0.3\textwidth}
        \includegraphics[width=\textwidth]{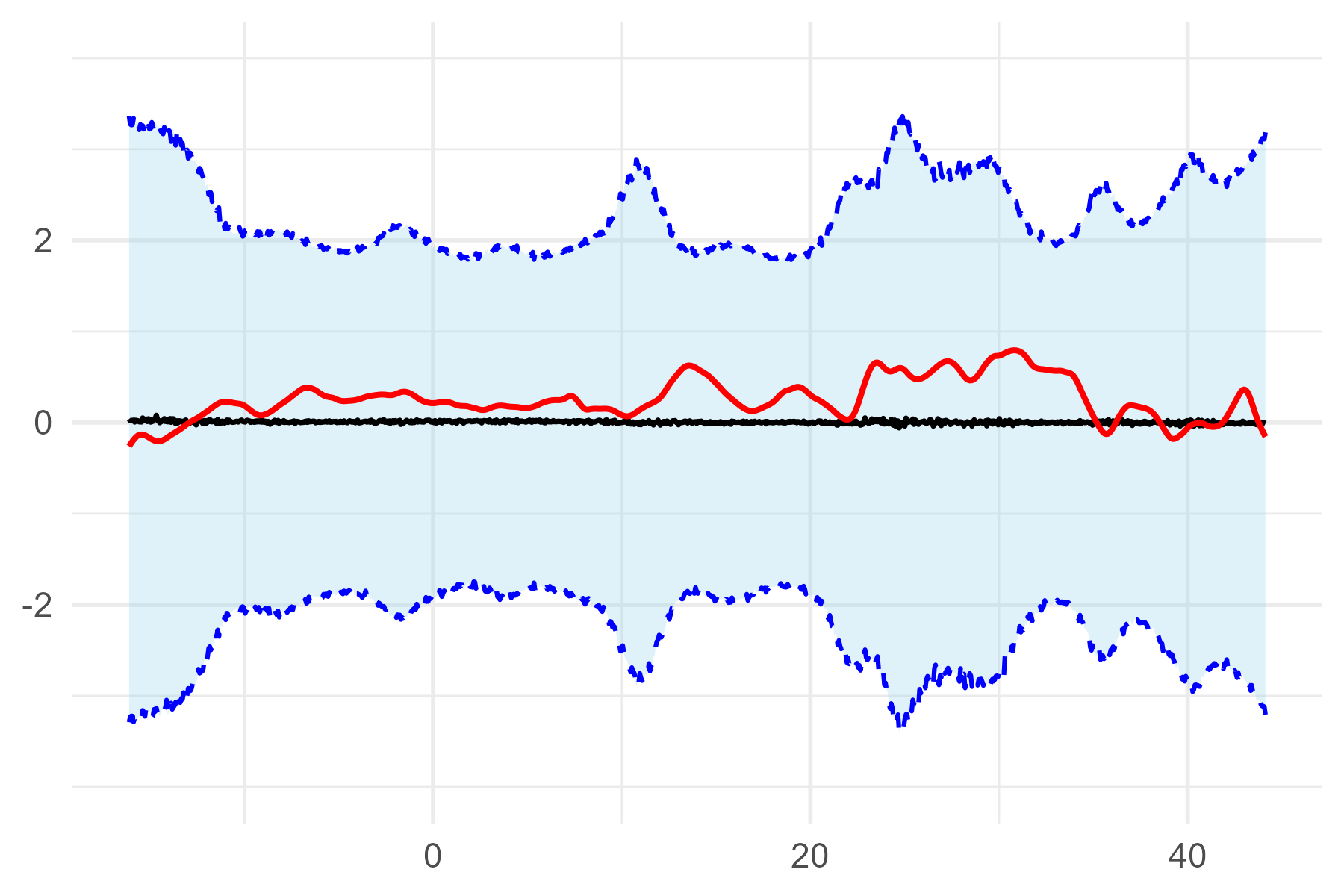}
        \caption{1970 vs 1980}
    \end{subfigure}

    % --- Row 2 ---
    \vspace{0.5cm}
    \begin{subfigure}[b]{0.3\textwidth}
        \includegraphics[width=\textwidth]{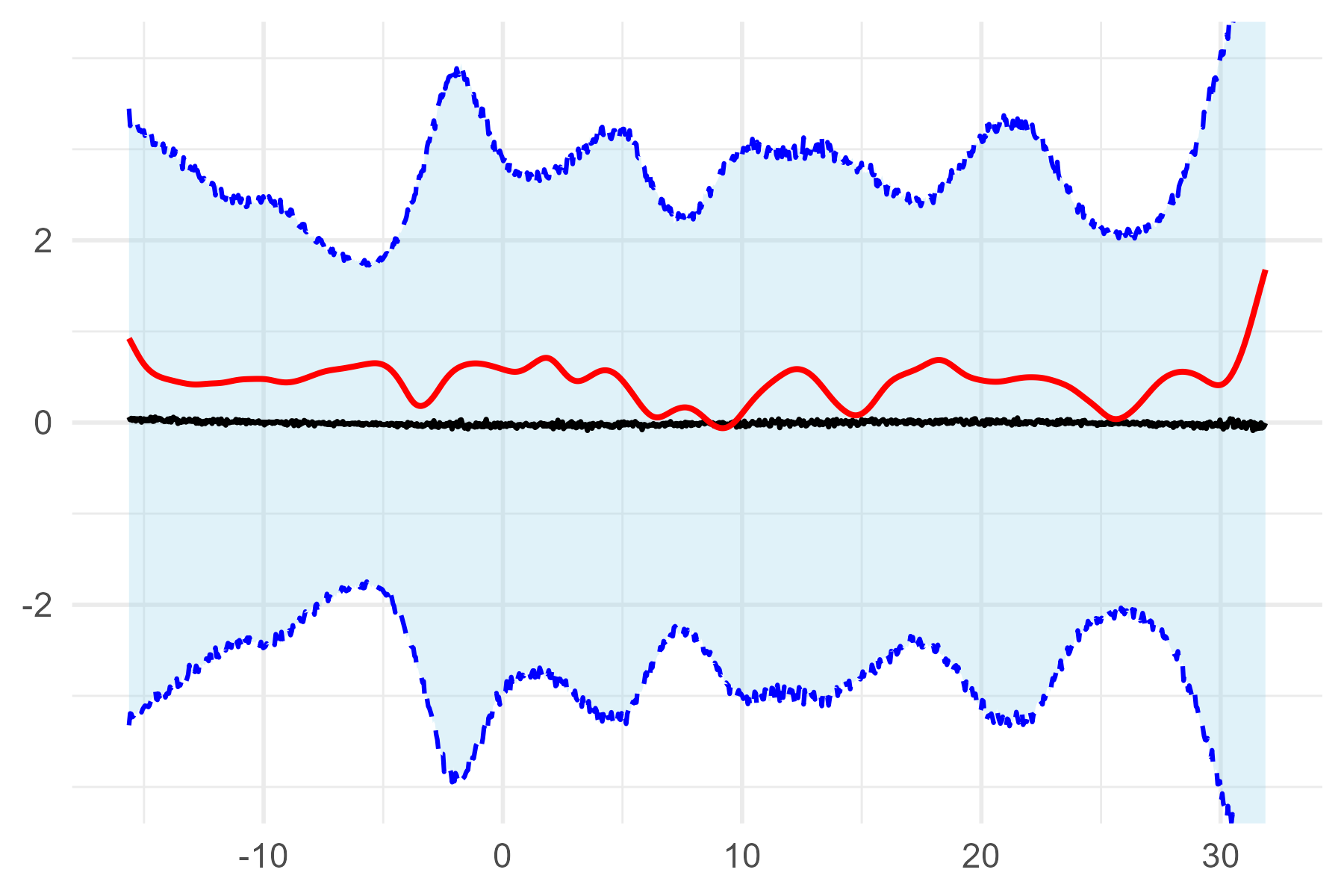}
        \caption{1960 vs 1970}
    \end{subfigure}
    \hfill
    \begin{subfigure}[b]{0.3\textwidth}
        \includegraphics[width=\textwidth]{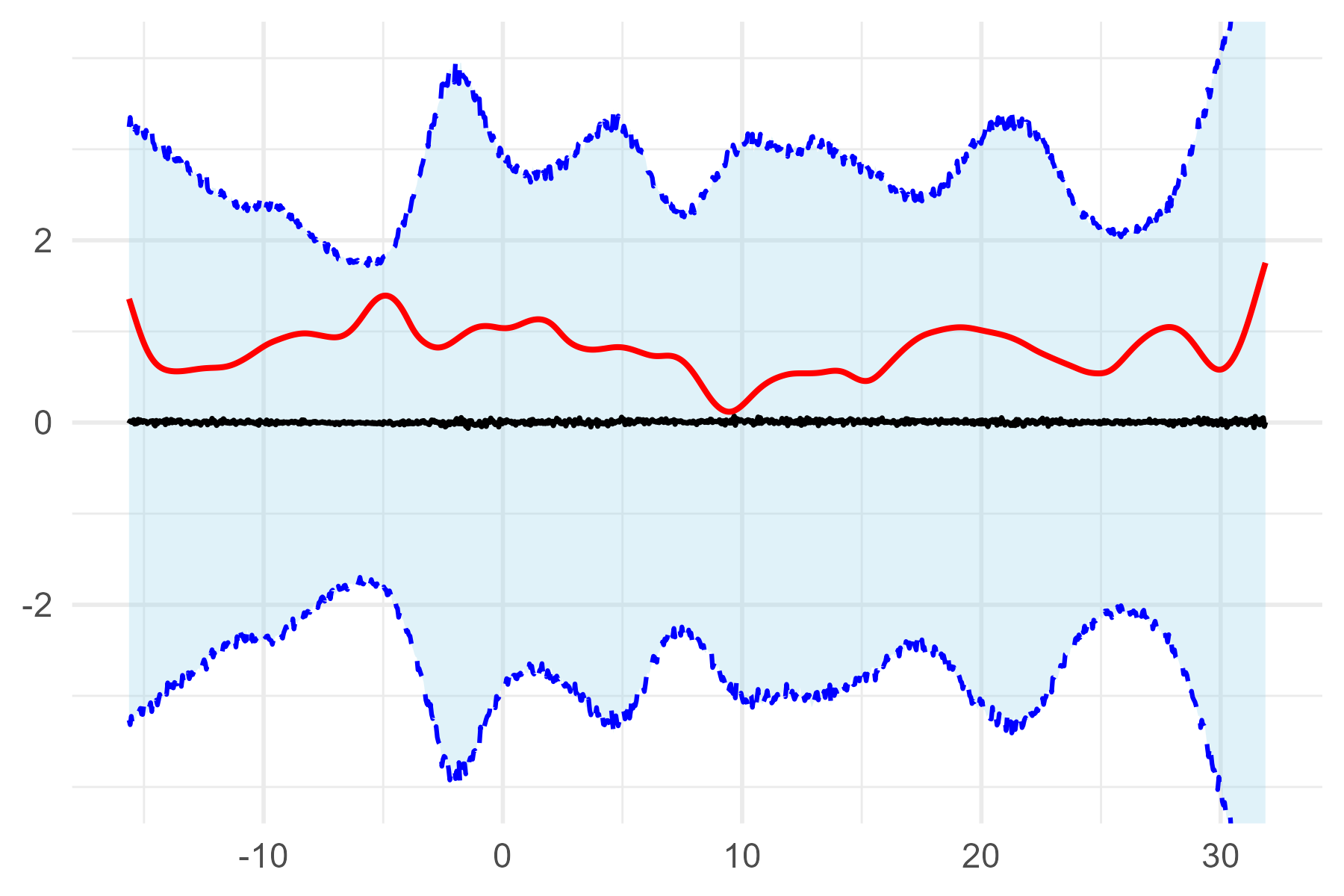}
        \caption{1960 vs 1980}
    \end{subfigure}
    \hfill
    \begin{subfigure}[b]{0.3\textwidth}
        \includegraphics[width=\textwidth]{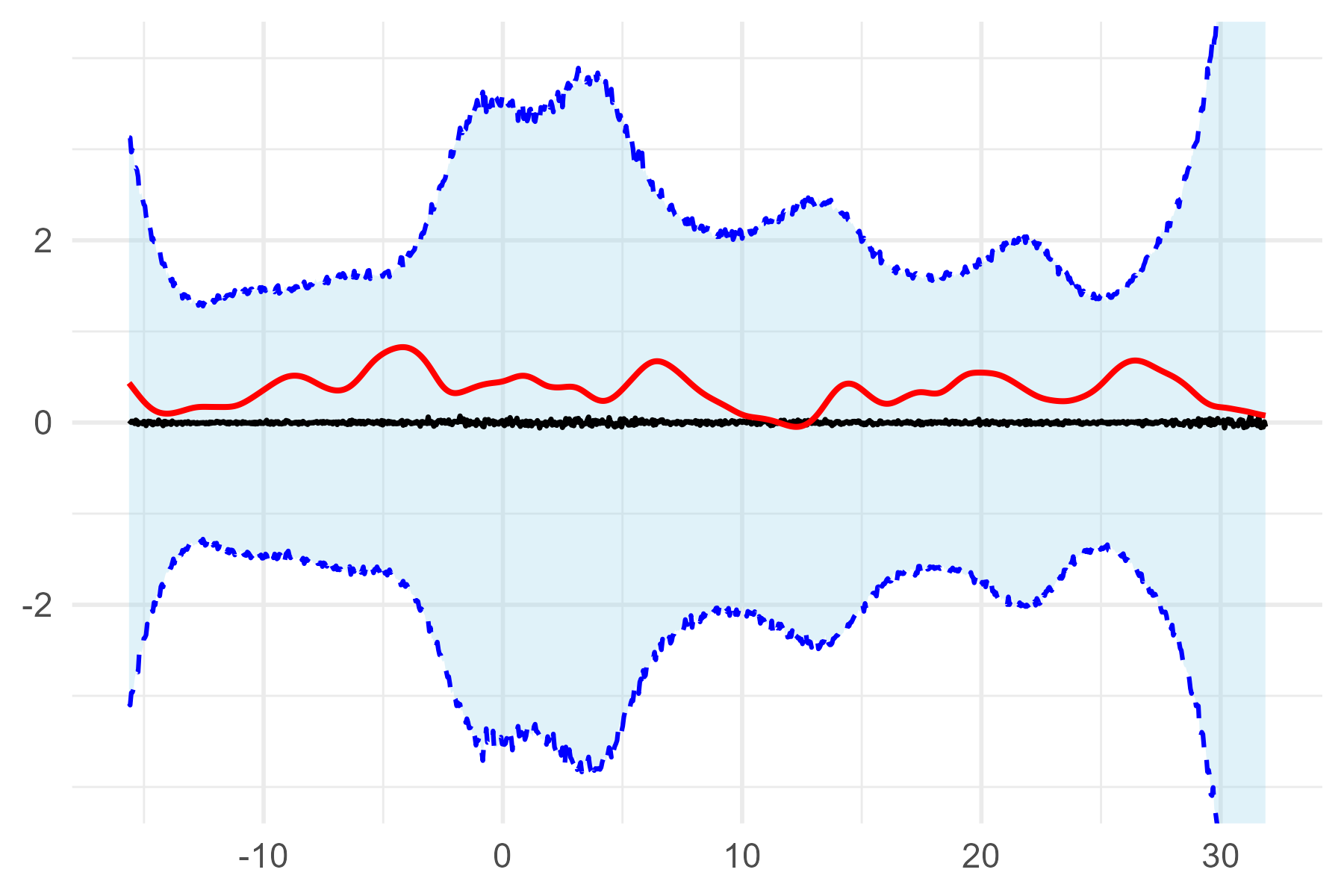}
        \caption{1970 vs 1980}
    \end{subfigure}

    \caption{Global envelope tests comparing decadal boundary differences. The dashed blue curves denote the 95\% global envelope, the solid black curve is the envelope mean $\mu(x)$, and the red curve is the observed difference $T(x)$ for each decade pair. The top row corresponds to the primary (arid/semi-arid) boundary, and the bottom row to the secondary (semi-arid/non-arid) boundary. In all cases, the observed curves remain entirely within the envelopes, and the associated p-values equal 1, indicating no statistically significant decadal differences.}
    \label{fig:global_envelope_comparison}
\end{figure}

The global envelope test results, shown in Figure~\ref{fig:global_envelope_comparison}, indicate no statistically significant differences among the decade-level boundaries for either climatic interface. In all pairwise comparisons, the observed difference curves remain entirely within their corresponding envelopes across the full spatial domain. Although the contrast between the 1960s and 1980s exhibits the largest visible departures, these deviations never exceed the envelope limits required to reject the null hypothesis. This conclusion is further supported by the test p-values, all of which equal 1, confirming the absence of statistically significant decade-to-decade changes. 

%\FloatBarrier

\subsection{Assessing Individual-Year Departures}

To complement the decadal comparison, we next assess whether particular years exhibit statistically meaningful departures from the climatological baseline represented by the 1960s. This decade is widely recognized as a relatively wet period across the Sahel–Sahara transition zone, preceding the severe and persistent droughts that characterized the 1970s and 1980s \citep{nicholson1994recent, LebelThierry2009Rtit, GangneronFabrice2022Paso}. Since the K\"oppen–Trewartha classification is driven by annual precipitation and temperature thresholds, pronounced drought years may produce localized displacements in the primary and secondary climatic boundaries. Our objective in this subsection is to determine whether any individual years between 1970 and 1989 exhibit boundary shifts that exceed the range of the natural interannual variability characterized by the 1960s baseline.

\begin{figure}[!ht]
    \centering
    %====================== Top Row ======================%
    \begin{minipage}{0.49\textwidth}
        \centering
        \includegraphics[width=\textwidth]{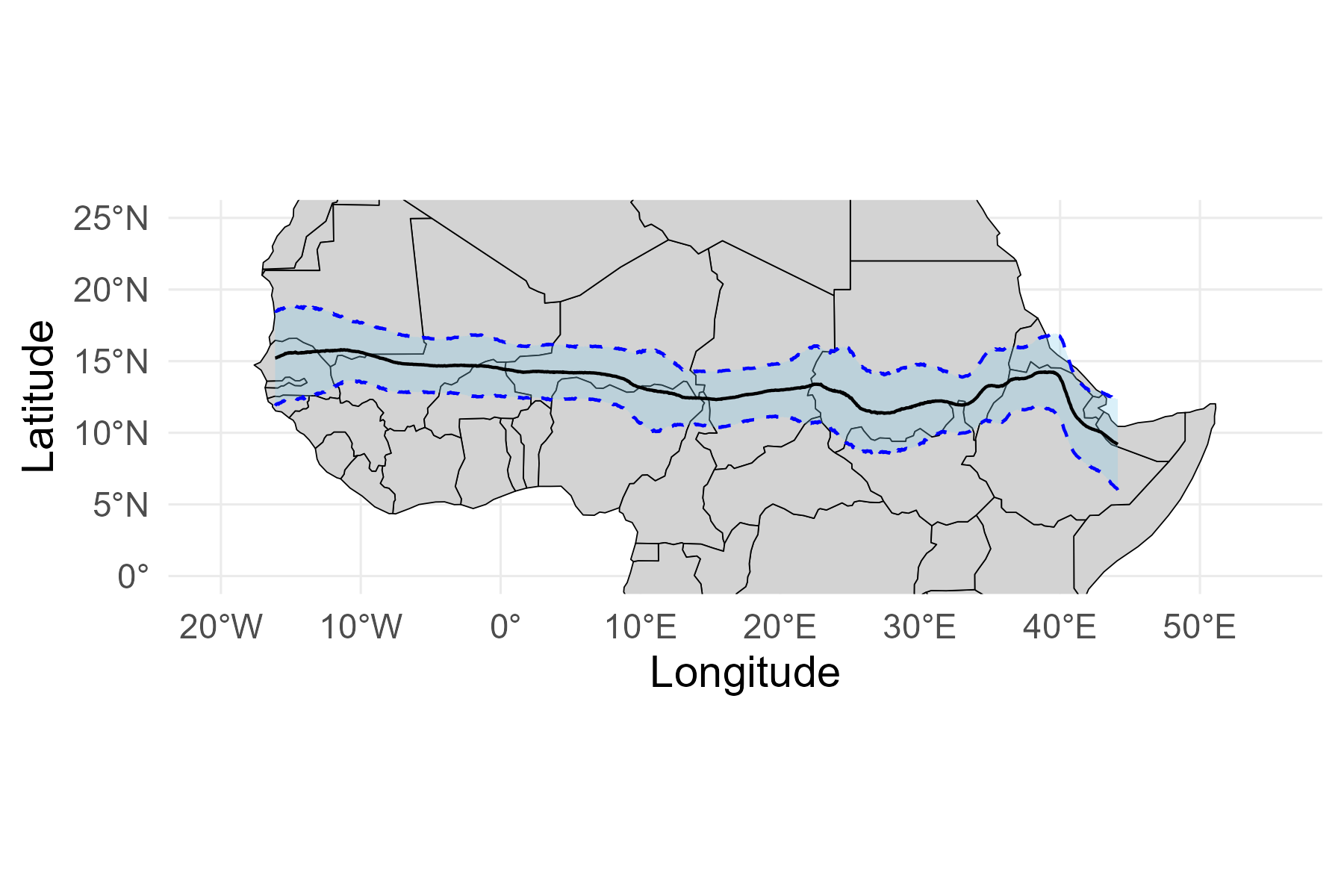}
        %\vspace{0.3cm}
        \includegraphics[width=\textwidth]{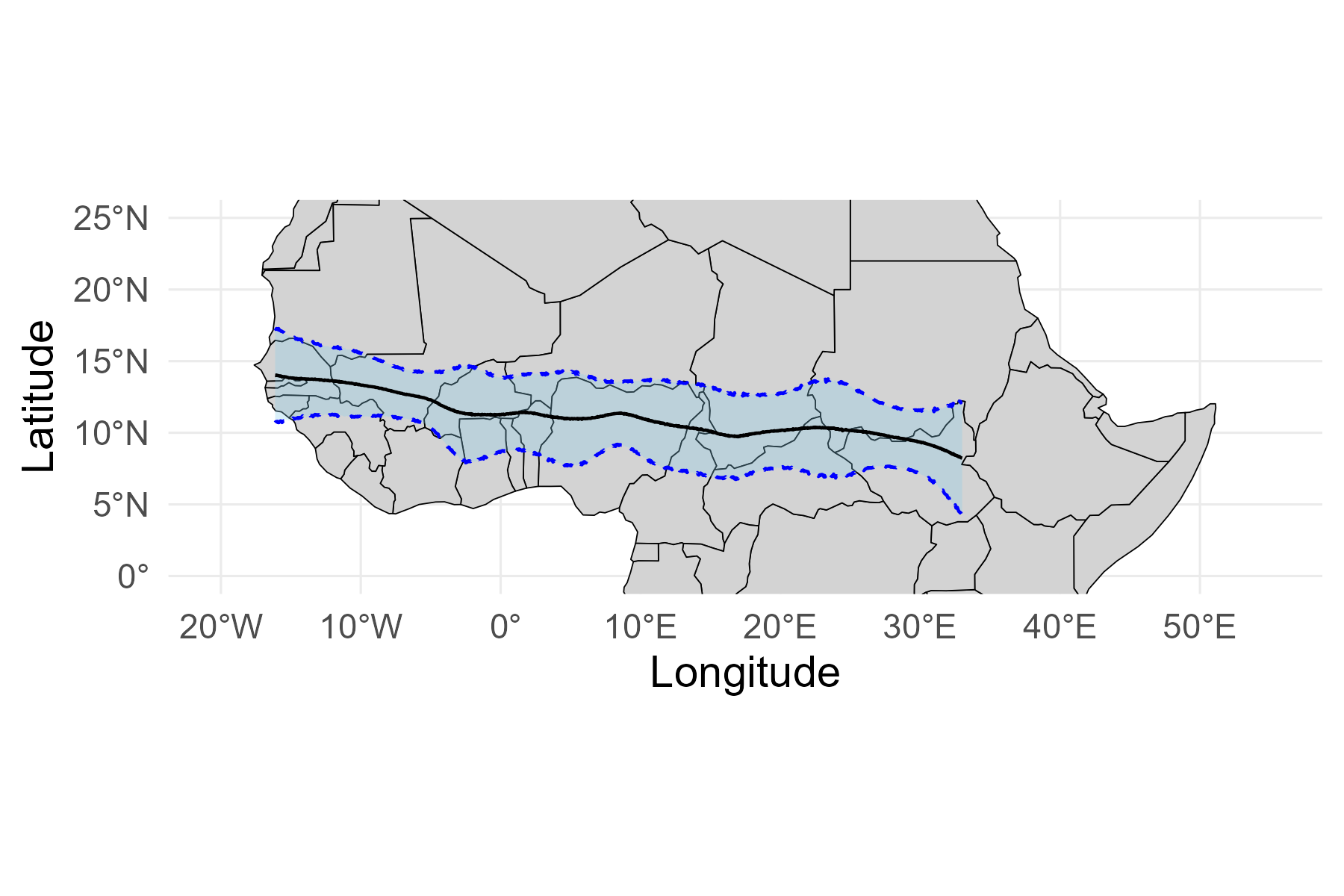}
    \end{minipage}
    %\hfill
    \begin{minipage}{0.49\textwidth}
        \centering
        \includegraphics[width=\textwidth]{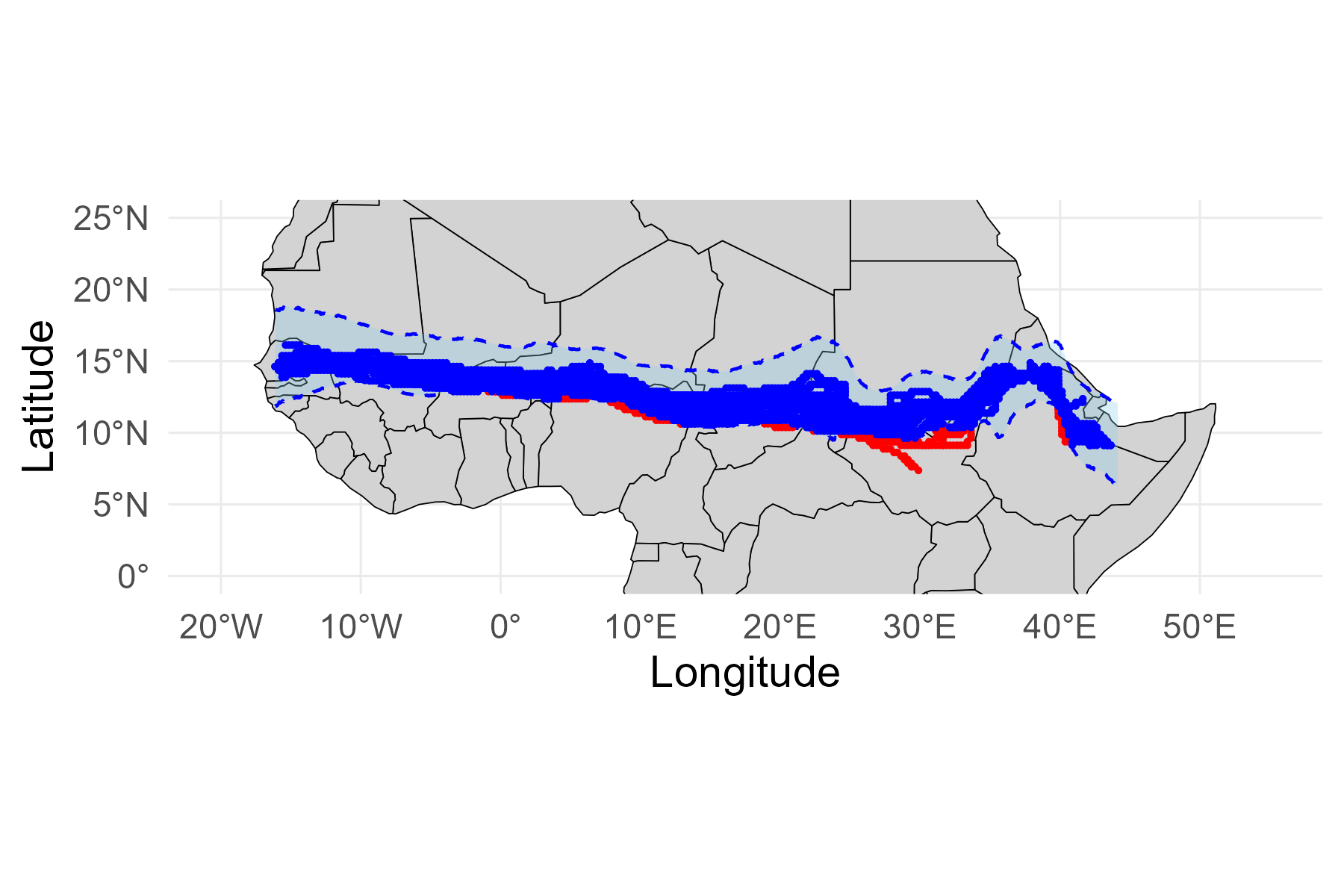}
        %\vspace{0.3cm}
        \includegraphics[width=\textwidth]{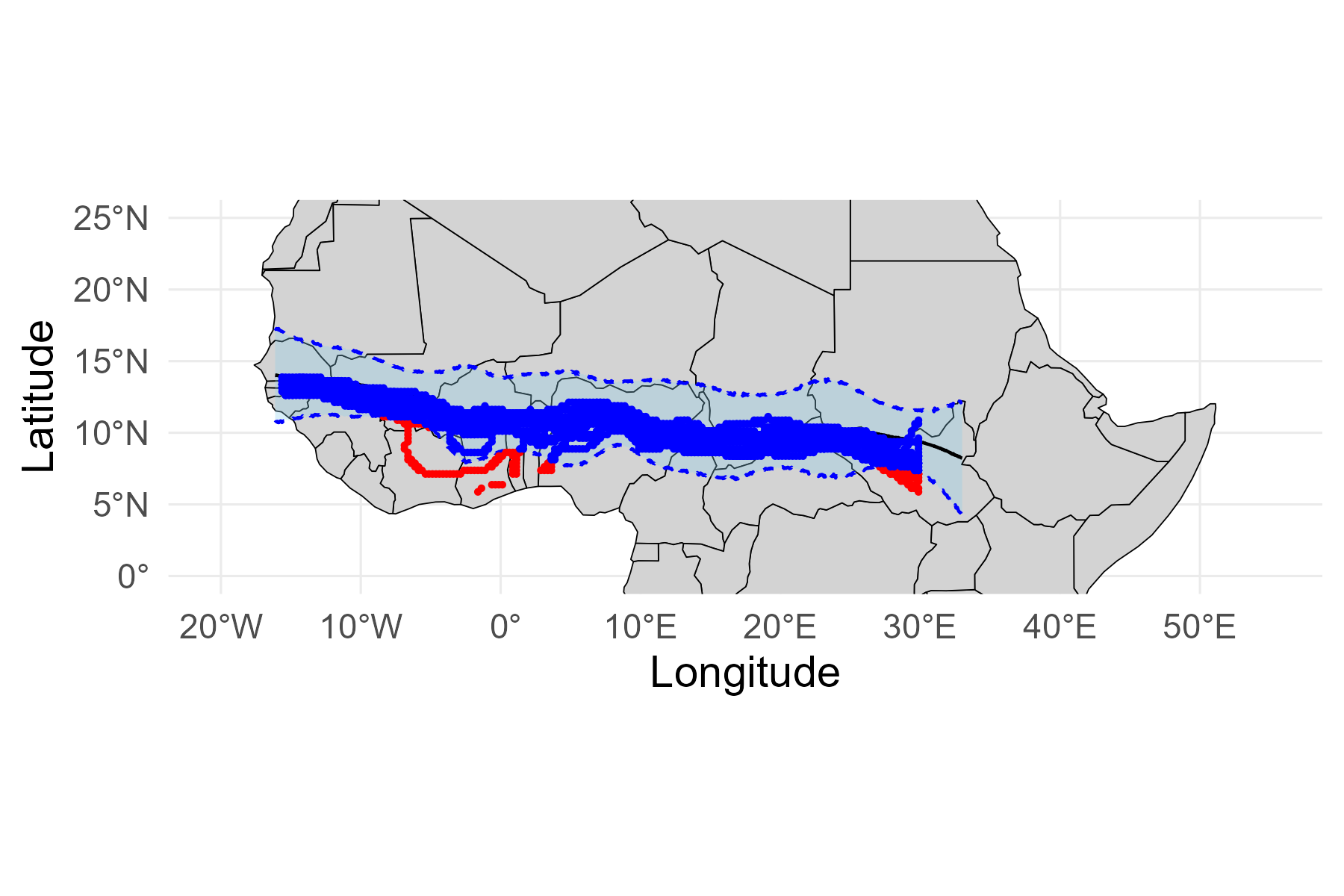}
    \end{minipage}
    \caption{Shifted scaled MAD global envelope test (GET) for the 1960s baseline boundary. Left panels show the decade-level mean primary (top) and secondary (bottom) boundaries with their global envelopes. Right panels overlay the observed boundary points from individual years (1960–1989), with blue points falling within the envelope and red points indicating potential departures.}
\label{fig:get_plots_1960}
\end{figure}

We begin by constructing a shifted, decade-level global envelope test based on the heteroskedastic GP fitted to the 1960–1969 data. Let $f^{(1960)}(x)$ denote the decade-level predicted mean boundary for the 1960s on a grid of 1000 longitude locations, and let $\phi_{1960s}$ denote the corresponding global envelope obtained using the scaled MAD procedure described in Section~\ref{sec:methods}. For each year $t = 1970, \ldots, 1989$, we overlay the observed boundary points on this envelope to assess whether the annual boundaries fall within the range of natural variability estimated by the 1960s model. Figure~\ref{fig:get_plots_1960} shows the 1960s shifted envelopes for both boundaries together with the observed points from subsequent decades. Most annual boundary points fall within the envelope region, but several areas of departures emerge. For the primary boundary, departures occur predominantly over northern South Sudan, southern Chad, northern Niger, and parts of Ethiopia. For the secondary boundary, departures appear along the West African coast, particularly in C\^ote d'Ivoire, southern Mali, Ghana, Togo, and Benin, as well as in parts of South Sudan. These localized departures motivate a more targeted examination of individual years.

\begin{figure}[!ht]
    \centering
    % First histogram
    \begin{subfigure}[b]{0.49\textwidth}
        \centering
        \includegraphics[width=\textwidth]{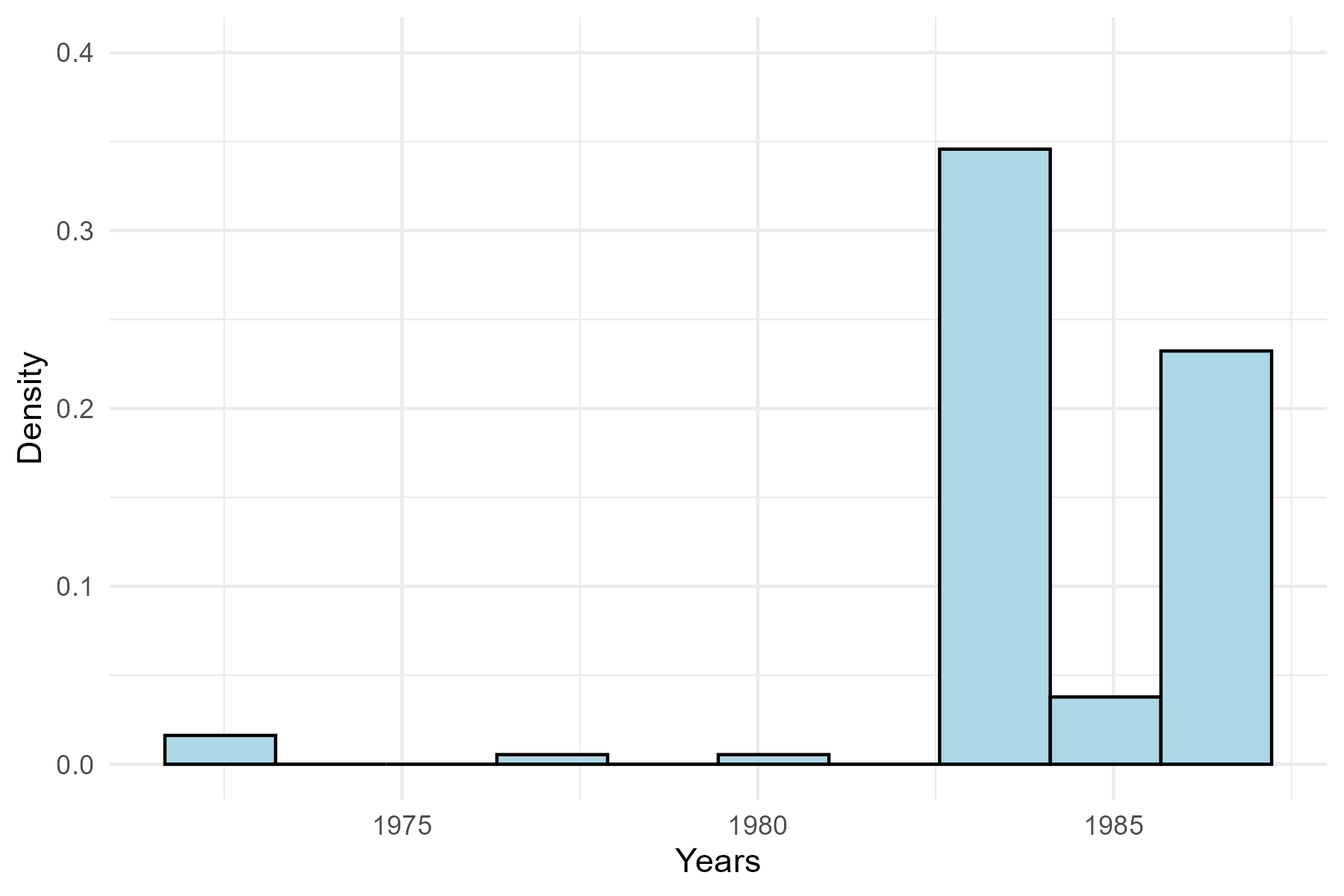}
        \caption{Proportion of exceedance (primary boundary).}
        \label{fig:hist1}
    \end{subfigure}
    %\hfill
    % Second histogram
    \begin{subfigure}[b]{0.49\textwidth}
        \centering
        \includegraphics[width=\textwidth]{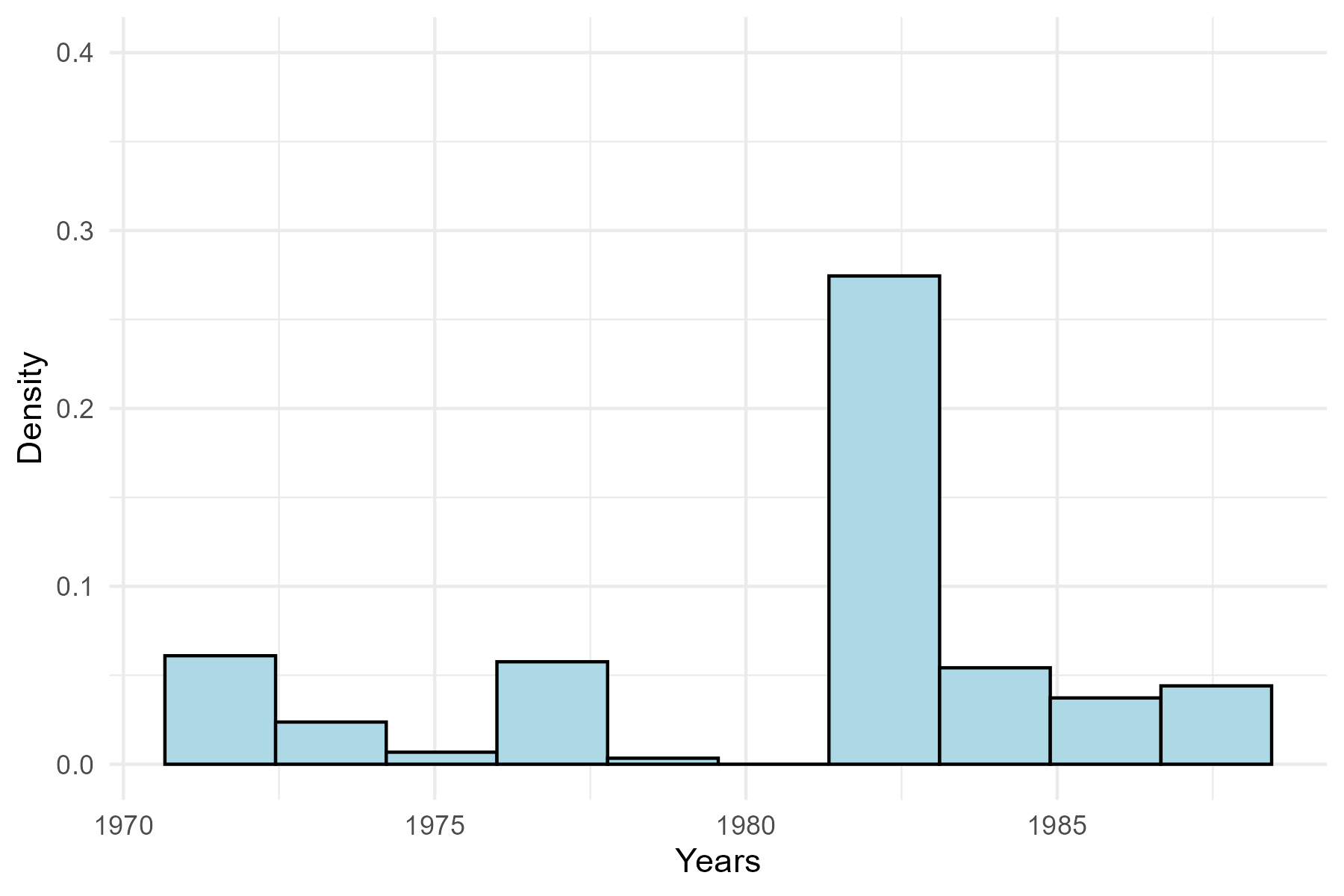}
        \caption{Proportion of exceedance (secondary boundary).}
        \label{fig:hist2}
    \end{subfigure}
    \caption{Year-specific exceedance probabilities of the 1960s global envelope for the primary (left) and secondary (right) climatic boundaries.}
    \label{fig:histograms}
\end{figure}

To identify the years that contribute most strongly to deviations from the 1960s baseline, we compute, for each year, the coverage probability of the 1960s global envelope, that is, the proportion of observed boundary points that fall within the 1960s envelope for that year. Across all years, the primary and secondary boundaries envelopes exhibit high overall coverages of 95.65\% and 95.84\%, respectively, indicating that fewer than 5\% of all observations fall outside the 1960s envelope. Figure~\ref{fig:histograms} shows the histogram of the year-specific proportion of boundary points that fall outside the 1960s envelope for both boundaries. The largest proportion of exceedance occurs in 1983 and 1984, two years that have been widely recognized as severe drought years in the region \citep{biasutti2019rainfall}.
Knowing that 1983 and 1984 had significant deviations from the decade boundaries, we wanted to check if these boundaries had significant deviations from preceding years as well. For each of these candidate years, we fit the heteroskedastic GP model using a rolling training window of six preceding years. Specifically, to predict the boundary for 1983, we train the GP on data from 1976–1981 (excluding 1982 as a gap year), and for 1984, the model is trained on 1977–1982. The temporal basis functions corresponding to the target year are used to compute the predictive mean and variance at the observed longitudes. For each year $t = \{1983, 1984\}$, the observed difference function is defined as
$$
T_{obs; ~t}(x) = \widehat{y}^{(t)}(x) - y_t(x), 
$$
and the difference ensemble under the null distribution is calculated by repeatedly sampling paired realizations from the multivariate normal predictive distribution obtained at time $t$, following the procedure in Section~\ref{sec:methods}. Global envelopes are then constructed using using the upper 5\% threshold of the MAD statistics values, simulated from the difference ensemble.

\begin{figure}[!ht]
    \centering
    % Row 1: Two images (slightly smaller)
    \begin{minipage}{0.49\textwidth}
        \centering
        \includegraphics[width=\textwidth]{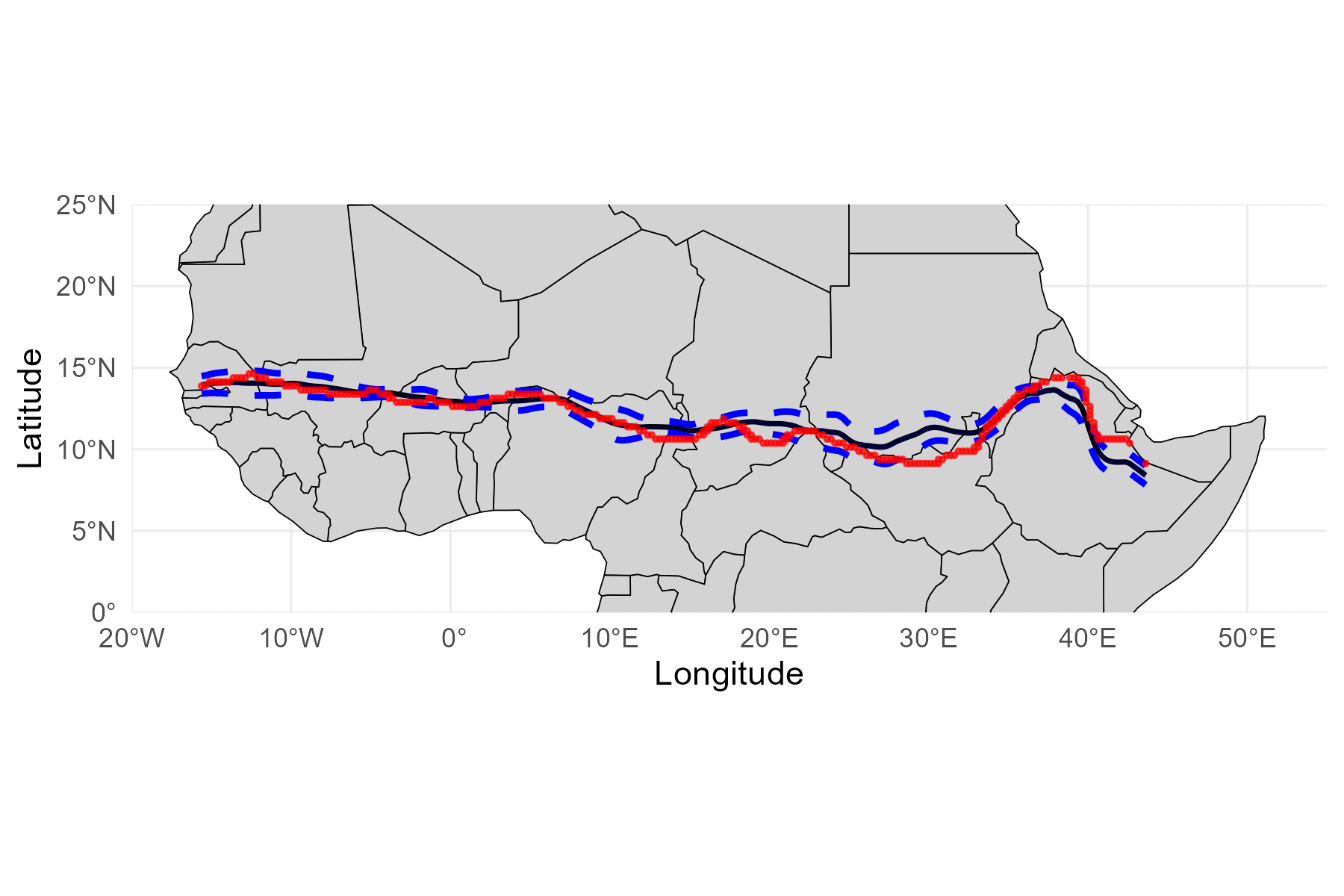}
        %\caption*{(a) Predicted mean line and interval for the semi-arid/non-arid boundary for 1983.}
    \end{minipage}
    \begin{minipage}{0.49\textwidth}
        \centering
        \includegraphics[width=\textwidth]{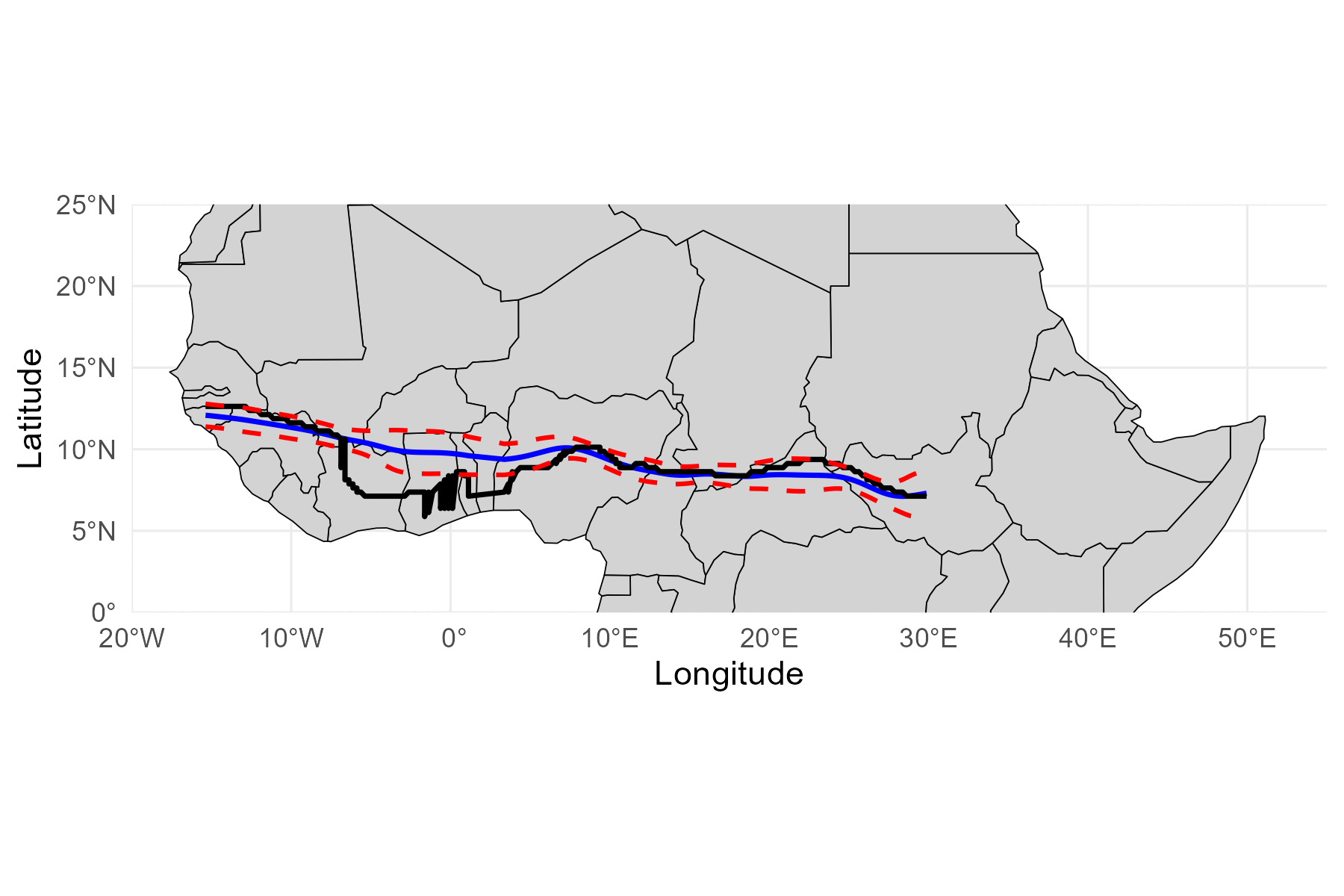}
        %\caption*{Prediction mean line and intervals for the semi-arid/non-arid boundary for 1983.}
    \end{minipage}
    %\vspace{0.5cm} % Reduced vertical spacing
    % Row 2: Two smaller images
    \begin{minipage}{0.49\textwidth}
        \centering
        \includegraphics[width=\textwidth]{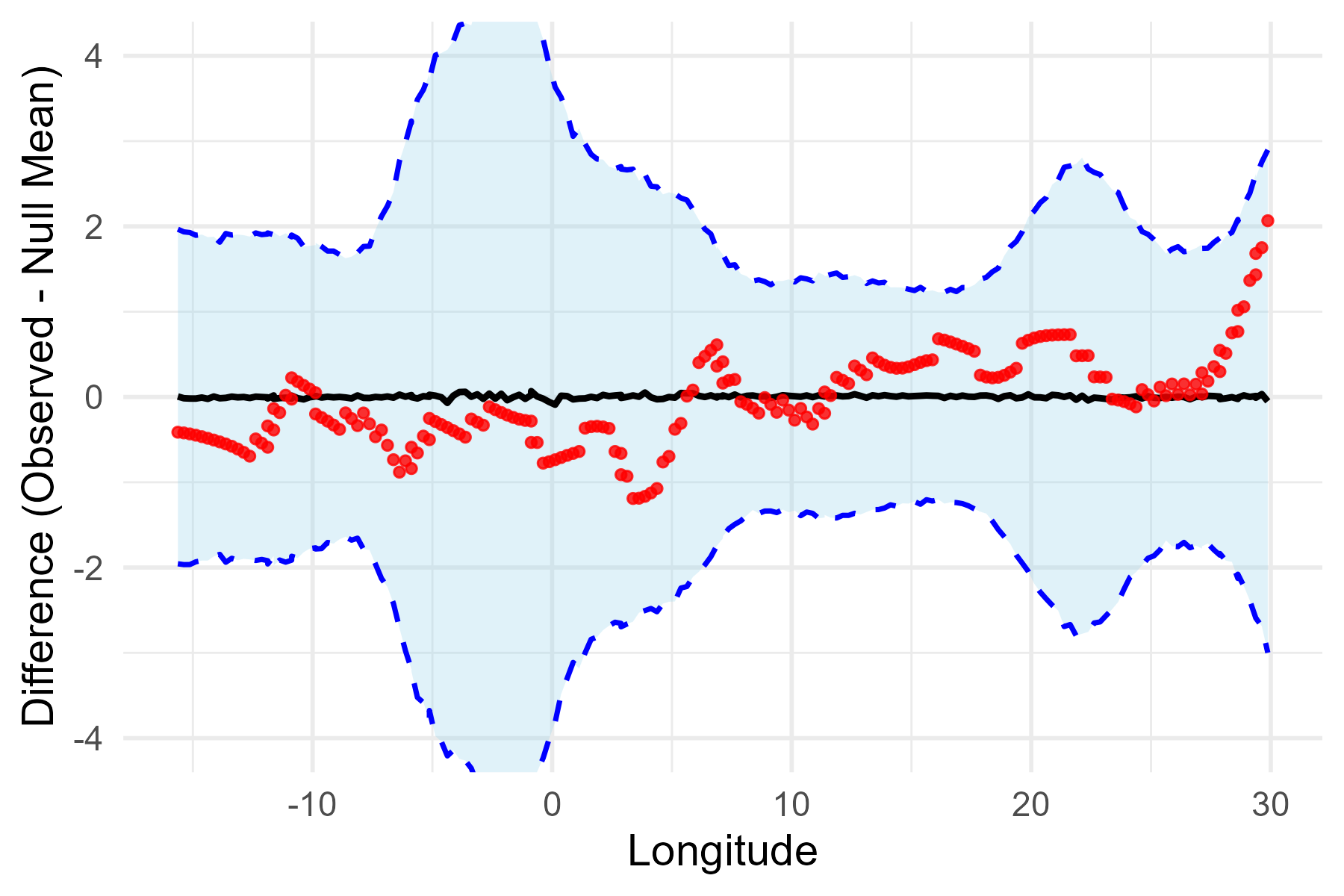}
        %\caption*{GET for the arid and semi-arid boundary in 1983. }
    \end{minipage}
    \hfill
    \begin{minipage}{0.49\textwidth}
        \centering
        \includegraphics[width=\textwidth]{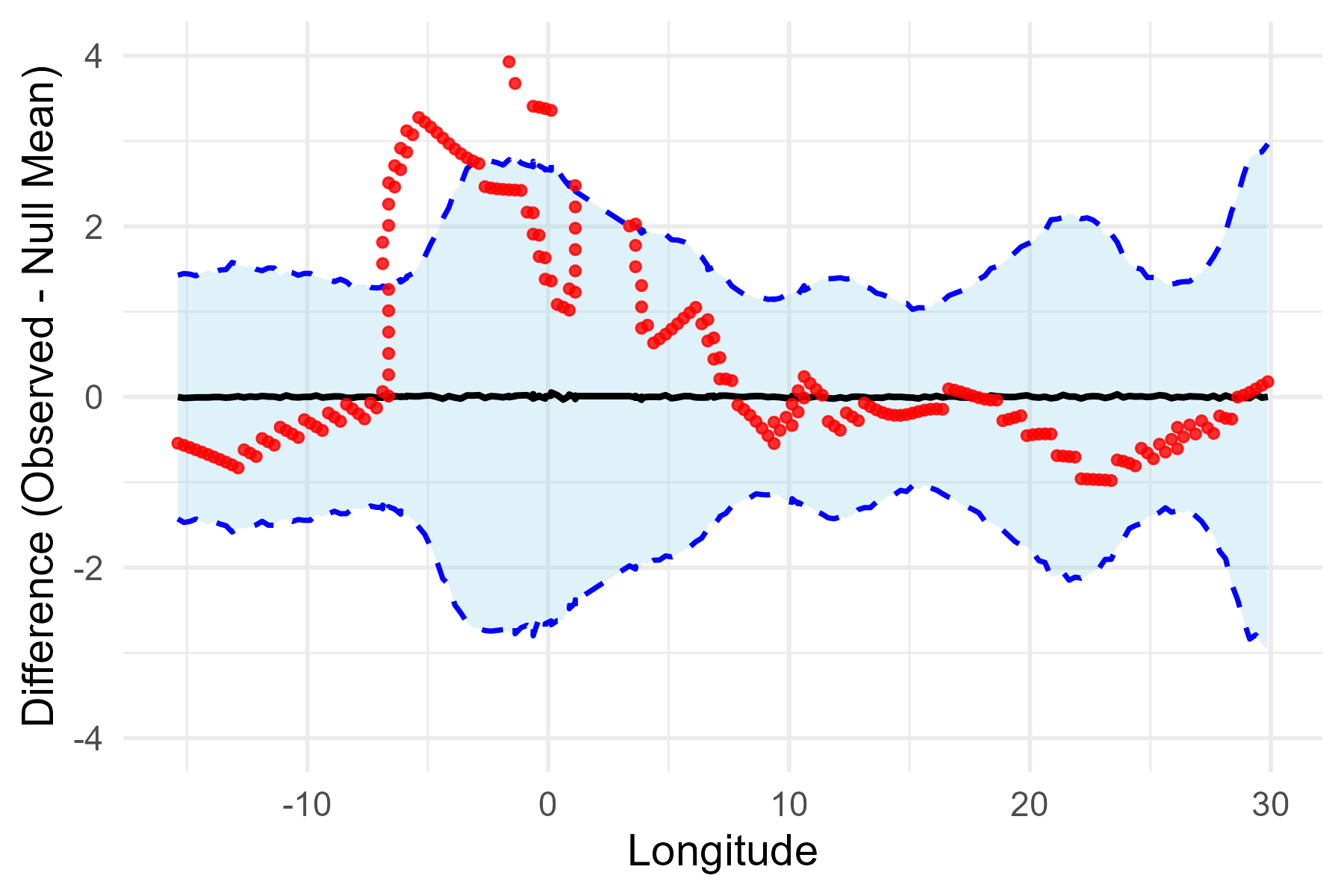}
        %\caption*{GET for the semi-arid and non-arid boundary in 1983. }
    \end{minipage}
    \caption{Predicted boundary lines, uncertainty intervals, and global envelope test results for the year 1983. Top panels show the predicted mean climatic boundary for 1983 (solid black) with 95\% prediction intervals (dashed blue), overlaid on the observed boundary points (red) for the primary (left) and secondary (right) boundaries. Bottom panels display the scaled MAD global envelope test for the primary (left) and secondary (right) boundaries.}
    \label{fig:1983_Boundary_Lines}
\end{figure}

\begin{figure}[!ht]
    \centering
    % Row 1: Two images (slightly smaller)
    \begin{minipage}{0.49\textwidth}
        \centering
        \includegraphics[width=\textwidth]{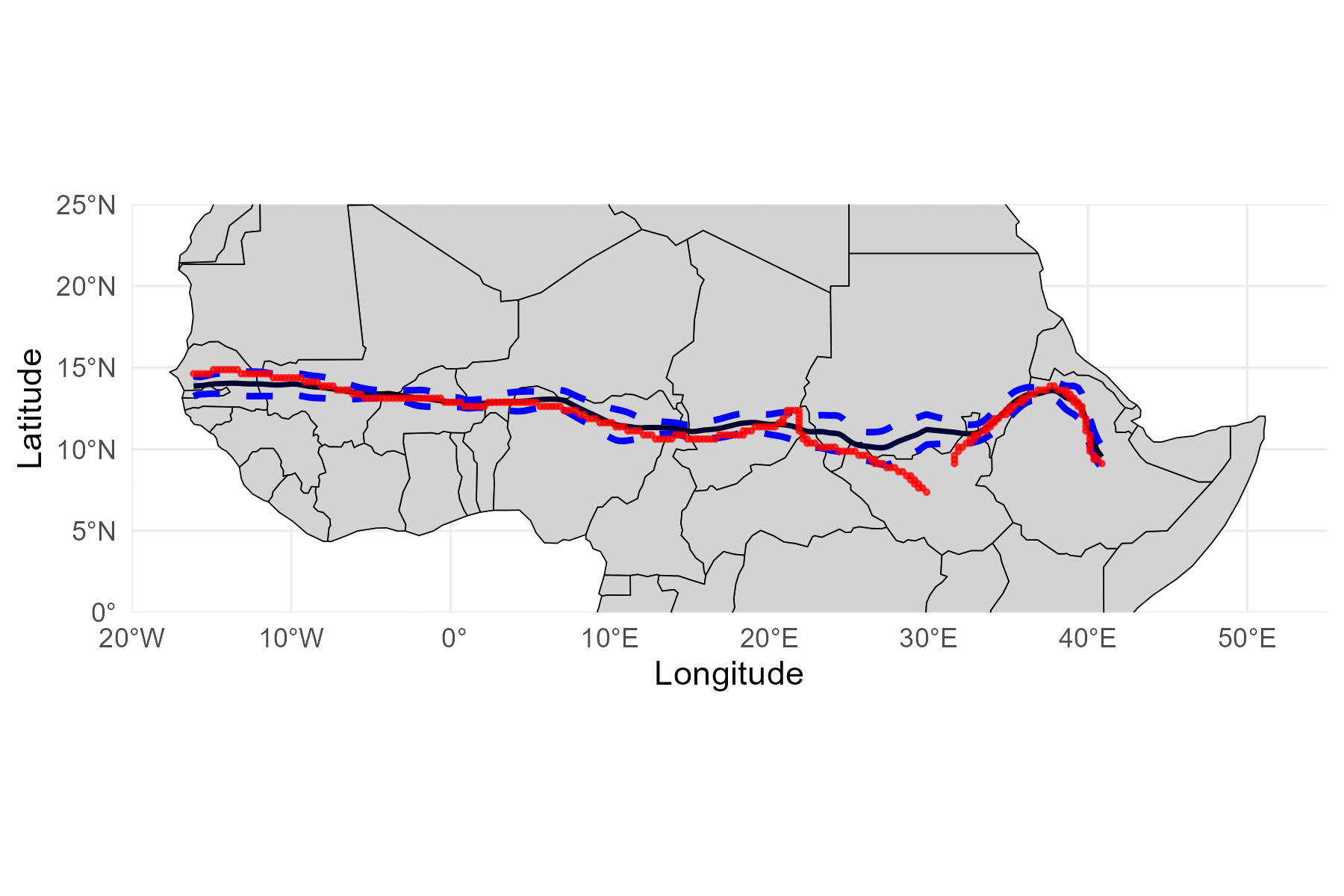}
        %\caption*{(a) Predicted mean line and interval for the semi-arid/non-arid boundary for 1984.}
    \end{minipage}
    \hfill
    \begin{minipage}{0.49\textwidth}
        \centering
        \includegraphics[width=\textwidth]{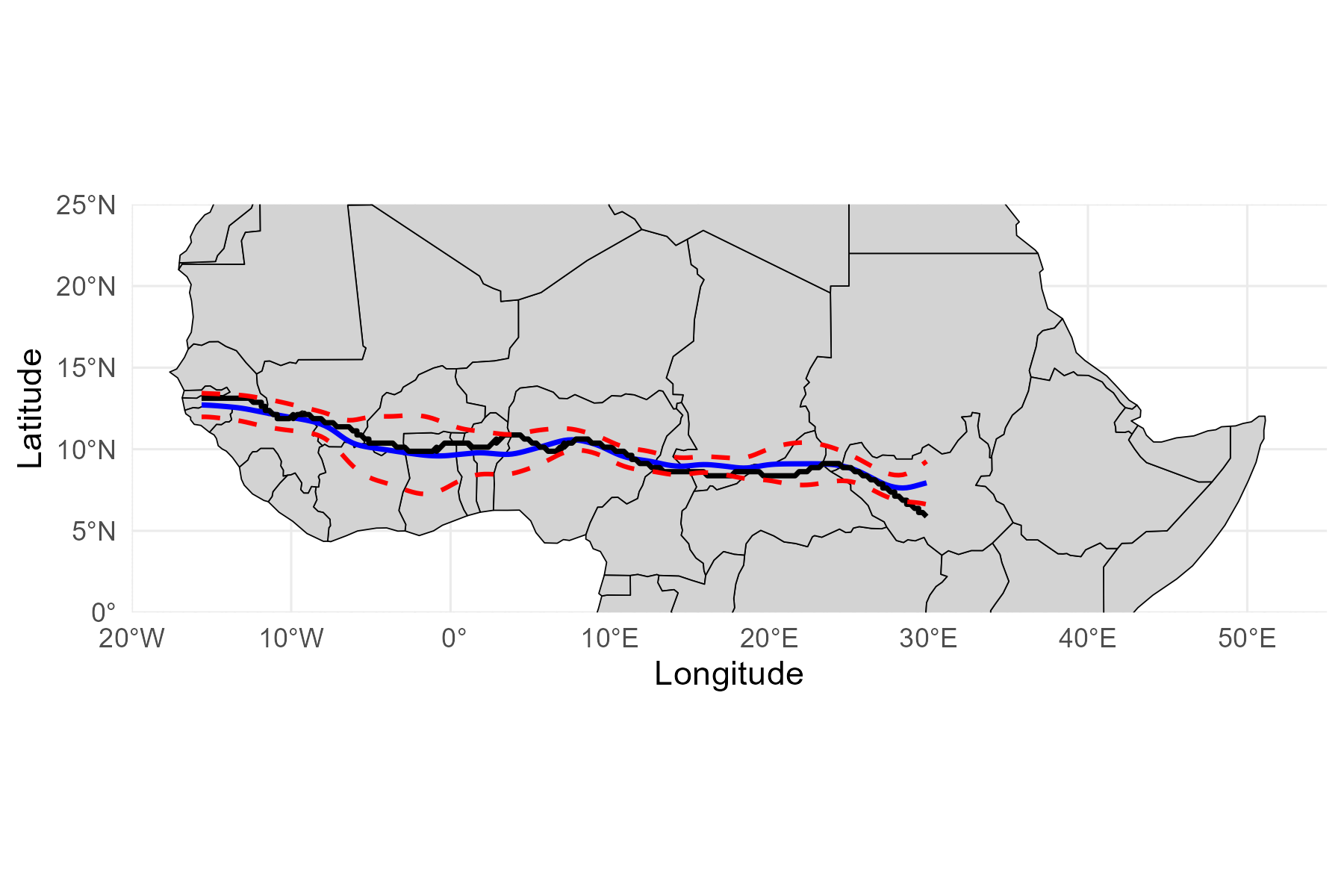}
        %\caption*{Prediction mean line and intervals for the semi-arid/non-arid boundary for 1984.}
    \end{minipage}
    %\vspace{0.5cm} % Reduced vertical spacing
    % Row 2: Two smaller images
    \begin{minipage}{0.49\textwidth}
        \centering
        \includegraphics[width=\textwidth]{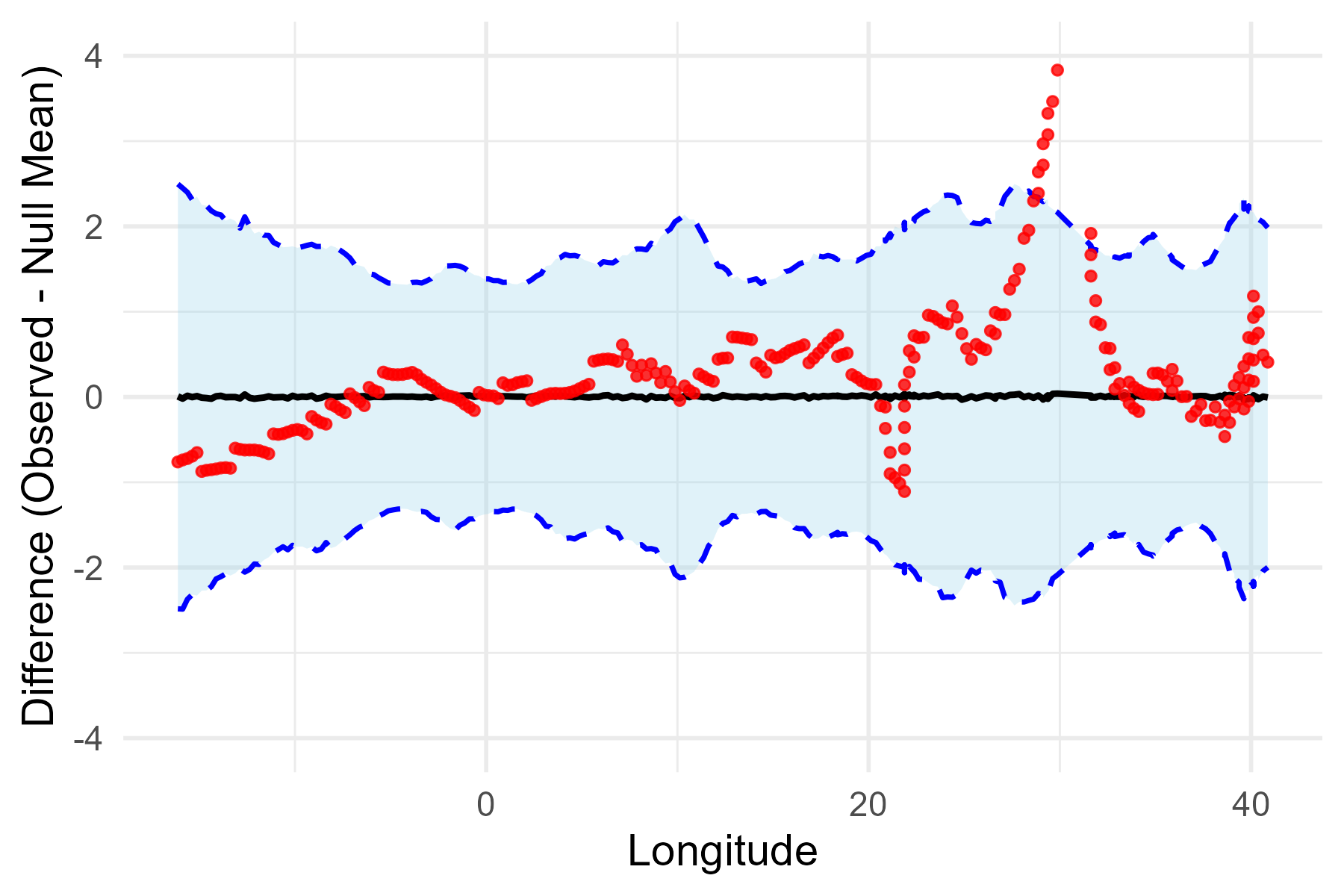}
        %\caption*{GET for the arid and semi-arid boundary in 1984. }
    \end{minipage}
    \hfill
    \begin{minipage}{0.49\textwidth}
        \centering
        \includegraphics[width=\textwidth]{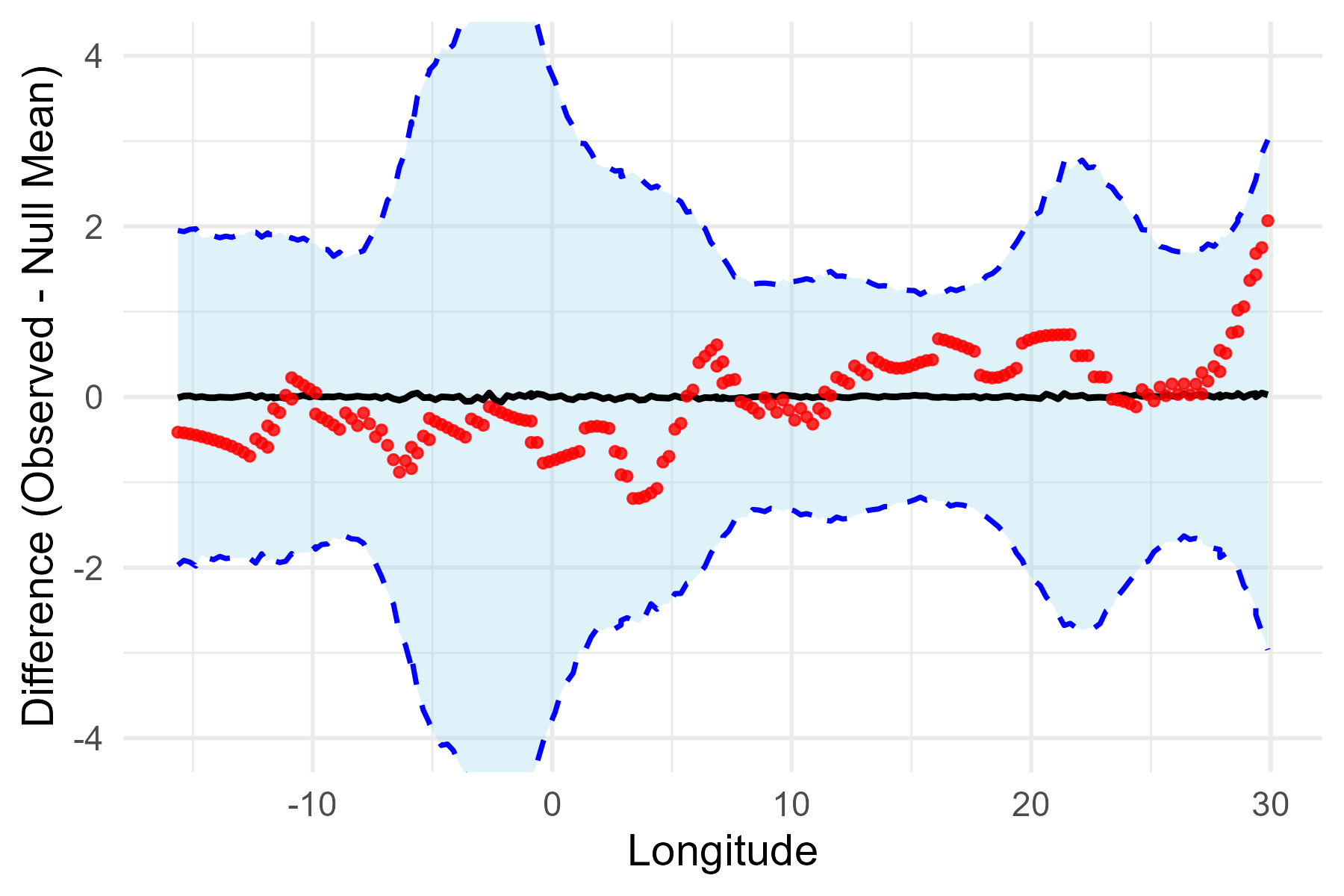}
        %\caption*{GET for the semi-arid and non-arid boundary in 1984. }
    \end{minipage}
    \caption{Predicted boundary lines, uncertainty intervals, and global envelope test results for the year 1984. Top panels show the predicted mean climatic boundary for 1984 (solid black) with 95\% prediction intervals (dashed blue), overlaid on the observed boundary points (red) for the primary (left) and secondary (right) boundaries. Bottom panels display the scaled MAD global envelope test for the primary (left) and secondary (right) boundaries.}
    \label{fig:1984_Boundary_Lines}
\end{figure}

The results for 1983 are presented in Figure~\ref{fig:1983_Boundary_Lines}. For the primary boundary, the observed difference curve remains entirely within the envelope, yielding a p-value of 0.78 and indicating no statistically significant departure from the 1976–1981 model's prediction. In contrast, the secondary boundary exhibits a highly significant deviation, with a p-value of 0.0. The exceedance occurs predominantly along the West African coastal sector, coinciding with the areas identified earlier as showing reduced envelope coverage. The results for 1984, shown in Figure~\ref{fig:1984_Boundary_Lines}), show a complementary pattern. Here, the primary boundary exhibits statistically significant deviations, with a p-value of 0.018, indicating that the 1984 arid/semi-arid interface lies outside the range expected under the 1977–1982 baseline climatic regime. Conversely, the secondary boundary for 1984 shows no significant departure (p = 1), with observed differences well contained within the global envelope. Since the simulation results under the alternative hypothesis demonstrated that deviations from the global envelope occur precisely in the regions where the boundaries shift, we can confirm these local shifts in the climatic interfaces during drought years and conclude that the scaled MAD global envelope test is capable of detecting localized annual boundary anomalies that are also visible in decade-level comparisons.

%  Mention that it is very scalable test,
%  Talk about the flaws, it is very fast and reduce  

% Send it 

% Plot variance, add coverage, add discussion, and conclusion summary 
% Adjust scale Mad, reframe, 
%\FloatBarrier
\section{Discussion} \label{sec:disc}
Our analysis reveals that decade‐to‐decade differences in the arid/semi-arid and semi-arid/non-arid K\"oppen–Trewartha climatic boundaries are not statistically distinguishable, and this aligns with earlier climate science studies showing that long-term shifts in the Sahel–Sahara ecological transition zone are subtle when aggregated at multi-decadal scales \citep{nicholson1994recent}. However, the global-scale MAD test identifies meaningful year-specific boundary shifts during major drought years, highlighting the short-term impact of low rainfall on these interfaces. In 1983, we detect a significant displacement of the secondary boundary (semi-arid/non-arid) along the West African coast, particularly across C\^ote d’Ivoire, Ghana, Togo, Benin, and southern Nigeria. This pattern is consistent with the documented 1983 West African drought, which is widely attributed to anomalously warm sea-surface temperatures in the tropical Atlantic and Pacific oceans \citep{bader20111983, nicholson2013west}. 

In 1984, the primary boundary shows a statistically significant north–south displacement, with exceedances concentrated over South Sudan. This finding is strongly supported by previous work describing 1984 as one of the most severe East African drought years in the 20th century \citep{williams2012recent, omoj2016temporal}. The boundary shifts detected for 1983 and 1984 also align with broader literature linking drought severity to socio-economic vulnerability. Extreme precipitation deficits across the Sahel have been shown to worsen poverty gaps, reduce household welfare, and destabilize agricultural production systems \citep{gascoigne2024welfare, tefera2025rainfall, esd-12-17-2021, kotir2011climate}. The regions where we detect specific anomalous boundary displacement, mainly West African coastal states in 1983, and South Sudan in 1984, correspond closely to areas repeatedly identified as drought hotspots with high exposure to climate-induced livelihood stress.

Importantly, our findings align with the well-established understanding in the climate science literature that, while major environmental events can trigger sharp, localized shifts in ecological and climatic boundaries, these anomalies often dissipate when averaged over longer temporal scales. Numerous studies have shown that, despite the succession of severe droughts affecting the Sahel during the 1970s and 1980s, long-term boundary migration has been relatively mild, with pronounced year-to-year fluctuations occurring against a largely stable multi-decadal backdrop \citep{LebelThierry2009Rtit, GangneronFabrice2022Paso}. Our results support this claim; although the decade-level KTC boundaries remain statistically unchanged, the scaled MAD test reveals clear, short-lived—but climatologically meaningful—boundary shifts during the extreme drought years of 1983 and 1984, highlighting the importance of methods capable of capturing fine-scale interannual variability.

\section{Conclusion} \label{sec:conc}

This study introduces a flexible for detecting spatial and temporal shifts in climatic boundaries by combining heteroskedastic Gaussian process regression with the scaled MAD Global Envelope test. Applied to the K\"oppen–Trewartha arid and semi-arid classification interfaces across the Sahel–Sahara region from 1960–1989, the method shows that while decade-scale boundary positions remained statistically stable, year-specific anomalies during well-documented drought years were significant. 

A key strength of the approach lies in its computational scalability. Gaussian process regression is typically expensive for large datasets, but by modeling the boundary as a one-dimensional spatial curve with heteroskedastic structure, inference and simulation become extremely fast, allowing for global envelope tests to be performed within seconds. This makes the method well suited for emerging high-resolution climate model data, remote-sensing products, or long-term reanalysis data where annual or seasonal boundary reconstructions may be of interest. The prooposed framework also offers clear advantages over existing boundary-change detection tools. The scaled MAD envelope test provides a formal statistical basis for evaluating whether observed spatial differences exceed natural variability, while simultaneously offering interpretable visual diagnostics. 

However, The current study is not aloof from limitations. The initial extraction of boundary points from gridded climate data relies on Canny edge detection, which can occasionally introduce discontinuities or jagged segments that influence the GP fit. These discontinuities could slightly shift local estimates of the mean boundary or inflate localized variance. Moreover, although the one-dimensional boundary representation captures the primary latitudinal structure of the Sahel–Sahara interface, it does not explicitly accommodate potential long-range spatial dependencies across the continent of Africa. 

Future work may address these limitations by incorporating spatially coherent edge-detection algorithms, leveraging multi-resolution representations of climate boundaries, or extending the proposed framework to two-dimensional manifold-based boundary modeling and testing. Another promising direction is the integration of physical covariates, such as soil moisture, vegetation indices, or sea-surface temperature anomalies in the analysis, to better understand the drivers of local boundary shifts. As drought frequency and aridity intensification are projected to increase across the region, developing scalable statistical tools capable of detecting emerging boundary shifts in real time will be essential for monitoring climate impacts and informing adaptation strategies.

\subsection*{Data availability statement}
The data \cite{beaudoing_rodell_2020} collected is from the Goddard Earth Sciences Data and Information Services Center and can be accessed by account users from: \url{https://disc.gsfc.nasa.gov/datasets/GLDAS_NOAH025_M_2.1/summary?keywords=NLDAS}.
\subsection*{Funding statement}

\subsection*{Conflict of interest}
The authors declare no conflicts of interest.

% Add the local geography differences between the western Africa and eastern africa, we are limited in  geographies 

\bibliographystyle{plainnat}
\bibliography{Reference}

@article{HuangJianping2020Gdvt,
  title={Global desertification vulnerability to climate change and human activities},
  author={Huang, Jianping and Zhang, Guolong and Zhang, Yanting and Guan, Xiaodan and Wei, Yun and Guo, Ruixia},
  journal={Land Degradation \& Development},
  volume={31},
  number={11},
  pages={1380--1391},
  year={2020},
  publisher={Wiley Online Library}
}

@article{GangneronFabrice2022Paso,
author = {Gangneron, Fabrice and Pierre, Caroline and Robert, Elodie and Kergoat, Laurent and Grippa, Manuela and Guichard, Françoise and Hiernaux, Pierre and Leauthaud, Crystele},
address = {Berlin/Heidelberg},
copyright = {The Author(s), under exclusive licence to Springer-Verlag GmbH Germany, part of Springer Nature 2022. Springer Nature or its licensor holds exclusive rights to this article under a publishing agreement with the author(s) or other rightsholder(s); author self-archiving of the accepted manuscript version of this article is solely governed by the terms of such publishing agreement and applicable law.},
issn = {1436-3798},
journal = {Regional environmental change},
language = {eng},
number = {4},
publisher = {Springer Berlin Heidelberg},
title = {Persistence and success of the Sahel desertification narrative},
volume = {22},
year = {2022},
}

@article{LebelThierry2009Rtit,
author = {Lebel, Thierry and Ali, Abdou},
copyright = {2008 Elsevier B.V.},
issn = {0022-1694},
journal = {Journal of hydrology (Amsterdam)},
language = {eng},
number = {1},
pages = {52-64},
publisher = {Elsevier B.V},
title = {Recent trends in the Central and Western Sahel rainfall regime (1990–2007)},
volume = {375},
year = {2009},
}

@article{Binois_2018,
   title={Practical Heteroscedastic Gaussian Process Modeling for Large Simulation Experiments},
   volume={27},
   ISSN={1537-2715},
   url={http://dx.doi.org/10.1080/10618600.2018.1458625},
   DOI={10.1080/10618600.2018.1458625},
   number={4},
   journal={Journal of Computational and Graphical Statistics},
   publisher={Informa UK Limited},
   author={Binois, Mickaël and Gramacy, Robert B. and Ludkovski, Mike},
   year={2018},
   month=jul, pages={808–821} }

@article{singh1989review,
  title={Review article digital change detection techniques using remotely-sensed data},
  author={Singh, Ashbindu},
  journal={International journal of remote sensing},
  volume={10},
  number={6},
  pages={989--1003},
  year={1989},
  publisher={Taylor \& Francis}
}

@article{lei2022boundary,
  title={Boundary extraction constrained siamese network for remote sensing image change detection},
  author={Lei, Jie and Gu, Yijie and Xie, Weiying and Li, Yunsong and Du, Qian},
  journal={IEEE Transactions on Geoscience and Remote Sensing},
  volume={60},
  pages={1--13},
  year={2022},
  publisher={IEEE}
}

@Article{Cheng_etal_2024,
AUTHOR = {Cheng, Guangliang and Huang, Yunmeng and Li, Xiangtai and Lyu, Shuchang and Xu, Zhaoyang and Zhao, Hongbo and Zhao, Qi and Xiang, Shiming},
TITLE = {Change Detection Methods for Remote Sensing in the Last Decade: A Comprehensive Review},
JOURNAL = {Remote Sensing},
VOLUME = {16},
YEAR = {2024},
NUMBER = {13},
ARTICLE-NUMBER = {2355},
URL = {https://www.mdpi.com/2072-4292/16/13/2355},
ISSN = {2072-4292},
DOI = {10.3390/rs16132355}
}

@article{jenkins2010high,
  title={High-resolution remote sensing of upland swamp boundaries and vegetation for baseline mapping and monitoring},
  author={Jenkins, Ross B and Frazier, Paul S},
  journal={Wetlands},
  volume={30},
  number={3},
  pages={531--540},
  year={2010},
  publisher={Springer}
}

@article{nicholson1994recent,
  title={Recent rainfall fluctuations in Africa and their relationship to past conditions over the continent},
  author={Nicholson, Sharon E},
  journal={The Holocene},
  volume={4},
  number={2},
  pages={121--131},
  year={1994},
  publisher={Sage Publications Sage CA: Thousand Oaks, CA}
}

@article{opencv_library,
    author = {Bradski, G.},
    journal = {Dr. Dobb's Journal of Software Tools},
    title = {{The OpenCV Library}},
    year = {2000}
}

@article{biasutti2019rainfall,
  title={Rainfall trends in the African Sahel: Characteristics, processes, and causes},
  author={Biasutti, Michela},
  journal={Wiley Interdisciplinary Reviews: Climate Change},
  volume={10},
  number={4},
  pages={e591},
  year={2019},
  publisher={Wiley Online Library}
}

@article{Koppen1918,
  author    = {Köppen, Wladimir},
  title     = {Classification of Climates According to Temperature, Precipitation and Annual Run},
  journal   = {Petermanns Mitteilungen},
  volume    = {64},
  pages     = {193--203},
  year      = {1918}
}

@book{Trewartha1966,
  author    = {Trewartha, Glenn T.},
  title     = {An Introduction to Climate},
  publisher = {McGraw-Hill},
  year      = {1966},
  edition   = {4},
  address   = {New York}
}

@article{tivenan26,
  author    = {Tivenan, Stephen and Sahoo, Indranil and Qian, Yanjun},
  title     = {Probabilistic Classification and Uncertainty Quantification of Sahara Desert Climate Using Feedforward Neural Networks},
  journal   = {Data Science in Science},
  volume    = {},
  pages     = {},
  year      = {2026}
}

@article{myllymaki2017global,
  title={Global envelope tests for spatial processes},
  author={Myllym{\"a}ki, Mari and Mrkvi{\v{c}}ka, Tom{\'a}{\v{s}} and Grabarnik, Pavel and Seijo, Henri and Hahn, Ute},
  journal={Journal of the Royal Statistical Society Series B: Statistical Methodology},
  volume={79},
  number={2},
  pages={381--404},
  year={2017},
  publisher={Oxford University Press}
}

@article{Rethore2013,
  author = {Julien Rethore and Marc Francois},
  title = {Curve and boundaries measurement using B-splines and virtual images},
  journal = {Optics and Lasers in Engineering},
  year = {2013},
  note = {Preprint},
  url = {https://www.researchgate.net/publication/260867286\_Curve\_and\_boundaries\_measurement\_using\_B-splines\_and\_virtual\_images}
}

@phdthesis{Pan2011,
  author = {Huijun Pan},
  title = {Bivariate B-splines and its applications in spatial data analysis},
  school = {Texas A\&M University},
  year = {2011},
  url = {https://oaktrust.library.tamu.edu/items/af6afb88-17f6-4d74-8bc2-9d19c589cebb/full}
}

@article{Liu2024,
  author = {Lei Liu and Jun Dai},
  title = {Estimation of partially linear single-index spatial autoregressive model using B-splines},
  journal = {Electronic Research Archive},
  volume = {32},
  number = {12},
  pages = {6822--6846},
  year = {2024},
  doi = {10.3934/era.2024319}
}

@Article{hess-18-3635-2014,
AUTHOR = {Masih, I. and Maskey, S. and Muss\'a, F. E. F. and Trambauer, P.},
TITLE = {A review of droughts on the African continent:  a geospatial and long-term perspective},
JOURNAL = {Hydrology and Earth System Sciences},
VOLUME = {18},
YEAR = {2014},
NUMBER = {9},
PAGES = {3635--3649},
URL = {https://hess.copernicus.org/articles/18/3635/2014/},
DOI = {10.5194/hess-18-3635-2014}
}

@Article{esd-12-17-2021,
AUTHOR = {Kew, S. F. and Philip, S. Y. and Hauser, M. and Hobbins, M. and Wanders, N. and van Oldenborgh, G. J. and van der Wiel, K. and Veldkamp, T. I. E. and Kimutai, J. and Funk, C. and Otto, F. E. L.},
TITLE = {Impact of precipitation and increasing temperatures on drought trends in eastern Africa},
JOURNAL = {Earth System Dynamics},
VOLUME = {12},
YEAR = {2021},
NUMBER = {1},
PAGES = {17--35},
URL = {https://esd.copernicus.org/articles/12/17/2021/},
DOI = {10.5194/esd-12-17-2021}
}

@article{tefera2025rainfall,
  title={Rainfall variability and drought in West Africa: challenges and implications for rainfed agriculture},
  author={Tefera, Meron Lakew and Seddaiu, Giovanna and Carletti, Alberto and Awada, Hassan},
  journal={Theoretical and Applied Climatology},
  volume={156},
  number={1},
  pages={41},
  year={2025},
  publisher={Springer}
}

@article{kotir2011climate,
  title={Climate change and variability in Sub-Saharan Africa: a review of current and future trends and impacts on agriculture and food security},
  author={Kotir, Julius H},
  journal={Environment, Development and Sustainability},
  volume={13},
  number={3},
  pages={587--605},
  year={2011},
  publisher={Springer}
}

@article{gascoigne2024welfare,
  title={The welfare cost of drought in sub-saharan africa},
  author={Gascoigne, Jon and Baquie, Sandra and Vinha, Katja Pauliina and Skoufias, Emmanuel and Calcutt, Evie and Kshirsagar, Varun and Meenan, Conor and Hill, Ruth},
  journal={Policy Research Working Paper},
  volume={10683},
  year={2024}
}

@Article{IECG2020-08544,
AUTHOR = {Ahangarha, Marjan and Shah-Hosseini, Reza and Saadatseresht, Mohammad},
TITLE = {Deep Learning-Based Change Detection Method for Environmental Change Monitoring Using Sentinel-2 Datasets},
JOURNAL = {Environmental Sciences Proceedings},
VOLUME = {5},
YEAR = {2021},
NUMBER = {1},
ARTICLE-NUMBER = {15},
URL = {https://www.mdpi.com/2673-4931/5/1/15},
ISSN = {2673-4931},
DOI = {10.3390/IECG2020-08544}
}

@article{WombleWilliamH.1951DS,
language = {eng},
number = {2961},
pages = {315-322},
publisher = {The American Association for the Advancement of Science},
title = {Differential Systematics},
volume = {114},
year = {1951},
author = {Womble, William H.},
address = {United States},
issn = {0036-8075},
journal = {Science (American Association for the Advancement of Science)},
}

@article{Banerjee01122006,
author = {Sudipto Banerjee and Alan E Gelfand},
title = {Bayesian Wombling},
journal = {Journal of the American Statistical Association},
volume = {101},
number = {476},
pages = {1487--1501},
year = {2006},
publisher = {ASA Website},
doi = {10.1198/016214506000000041},

    note ={PMID: 20221318},


URL = { 
    
        https://doi.org/10.1198/016214506000000041
    
    

},
eprint = { 
    
        https://doi.org/10.1198/016214506000000041
    
    

}

}

@article{lu2005bayesian,
  title={Bayesian areal wombling for geographical boundary analysis},
  author={Lu, Haolan and Carlin, Bradley P},
  journal={Geographical Analysis},
  volume={37},
  number={3},
  pages={265--285},
  year={2005},
  publisher={Wiley Online Library}
}

@article{gelfand2015bayesian,
  title={Bayesian wombling: finding rapid change in spatial maps},
  author={Gelfand, Alan E and Banerjee, Sudipto},
  journal={Wiley Interdisciplinary Reviews: Computational Statistics},
  volume={7},
  number={5},
  pages={307--315},
  year={2015},
  publisher={Wiley Online Library}
}

@article{fitzpatrick2010ecological,
  title={Ecological boundary detection using Bayesian areal wombling},
  author={Fitzpatrick, Matthew C and Preisser, Evan L and Porter, Adam and Elkinton, Joseph and Waller, Lance A and Carlin, Bradley P and Ellison, Aaron M},
  journal={Ecology},
  volume={91},
  number={12},
  pages={3448--3455},
  year={2010},
  publisher={Wiley Online Library}
}

@article{halder2024bayesian,
  title={Bayesian modeling with spatial curvature processes},
  author={Halder, Aritra and Banerjee, Sudipto and Dey, Dipak K},
  journal={Journal of the American Statistical Association},
  volume={119},
  number={546},
  pages={1155--1167},
  year={2024},
  publisher={Taylor \& Francis}
}

@article{tian2023northward,
  title={Northward shifts of the sahara desert in response to twenty-first-century climate change},
  author={Tian, Chuyin and Huang, Guohe and Lu, Chen and Song, Tangnyu and Wu, Yinghui and Duan, Ruixin},
  journal={Journal of Climate},
  volume={36},
  number={10},
  pages={3417--3435},
  year={2023}
}

@incollection{salgado2004sahel,
  title={Sahel: The end of the road},
  author={Salgado, Sebasti{\~a}o},
  booktitle={Sahel},
  year={2004},
  publisher={University of California Press}
}

@article{patton1962note,
  title={A note on the classification of dry climates in the Koppen system},
  author={Patton, Clyde P},
  year={1962},
  publisher={California Council of Geography Teachers}
}

@misc{beaudoing_rodell_2020,
  author = {Beaudoing, H. and Rodell, M.},
  title = {{GLDAS Noah Land Surface Model L4 monthly 0.25 x 0.25 degree V2.1}},
  year = {2020},
  publisher = {NASA/GSFC/HSL},
  address = {Greenbelt, Maryland, USA},
  howpublished = {Goddard Earth Sciences Data and Information Services Center (GES DISC)},
  note = {Accessed: 11/24/2024},
  doi = {10.5067/SXAVCZFAQLNO}
}

@article{NicholsonSharonE.2018RotA,
author = {Nicholson, Sharon E. and Funk, Chris and Fink, Andreas H.},
copyright = {2018},
issn = {0921-8181},
journal = {Global and Planetary Change},
language = {eng},
pages = {114-127},
publisher = {Elsevier B.V},
title = {Rainfall over the African continent from the 19th through the 21st century},
volume = {165},
year = {2018},
}

@article{glantz1987drought,
  title={Drought in Africa},
  author={Glantz, Michael H},
  journal={Scientific American},
  volume={256},
  number={6},
  pages={34--41},
  year={1987},
  publisher={JSTOR}
}

@article{kassas1995desertification,
  title={Desertification: a general review},
  author={Kassas, Mohammad},
  journal={Journal of Arid Environments},
  volume={30},
  number={2},
  pages={115--128},
  year={1995},
  publisher={Elsevier}
}

@article{VogtJ.V.2011Maao,
author = {Vogt, J. V. and Safriel, U. and Von Maltitz, G. and Sokona, Y. and Zougmore, R. and Bastin, G. and Hill, J.},
address = {Chichester, UK},
copyright = {Copyright © 2011 John Wiley & Sons, Ltd.},
issn = {1085-3278},
journal = {Land Degradation \& Development},
language = {eng},
number = {2},
pages = {150-165},
publisher = {John Wiley & Sons, Ltd},
title = {Monitoring and assessment of land degradation and desertification: Towards new conceptual and integrated approaches},
volume = {22},
year = {2011},
}

@article{goldberg1997regression,
  title={Regression with input-dependent noise: A Gaussian process treatment},
  author={Goldberg, Paul and Williams, Christopher and Bishop, Christopher},
  journal={Advances in neural information processing systems},
  volume={10},
  year={1997}
}

@article{bader20111983,
  title={The 1983 drought in the West Sahel: a case study},
  author={Bader, J{\"u}rgen and Latif, Mojib},
  journal={Climate dynamics},
  volume={36},
  number={3},
  pages={463--472},
  year={2011},
  publisher={Springer}
}

@article{nicholson2013west,
  title={The West African Sahel: A review of recent studies on the rainfall regime and its interannual variability},
  author={Nicholson, Sharon E},
  journal={International Scholarly Research Notices},
  volume={2013},
  number={1},
  pages={453521},
  year={2013},
  publisher={Wiley Online Library}
}

@article{williams2012recent,
  title={Recent summer precipitation trends in the Greater Horn of Africa and the emerging role of Indian Ocean sea surface temperature},
  author={Williams, A Park and Funk, Chris and Michaelsen, Joel and Rauscher, Sara A and Robertson, Iain and Wils, Tommy HG and Koprowski, Marcin and Eshetu, Zewdu and Loader, Neil J},
  journal={Climate dynamics},
  volume={39},
  number={9},
  pages={2307--2328},
  year={2012},
  publisher={Springer}
}

@article{omoj2016temporal,
  title={Temporal and spatial characteristics of the June-August seasonal rainfall and temperature over South Sudan},
  author={Omoj, P and Ogallo, L and Oludhe, C and Gitau, W},
  journal={J. Meteorol},
  volume={9},
  number={5},
  year={2016}
}

@article{cui2022fast,
  title={Fast univariate inference for longitudinal functional models},
  author={Cui, Erjia and Leroux, Andrew and Smirnova, Ekaterina and Crainiceanu, Ciprian M},
  journal={Journal of Computational and Graphical Statistics},
  volume={31},
  number={1},
  pages={219--230},
  year={2022},
  publisher={Taylor \& Francis}
}

@article{crainiceanu2012bootstrap,
  title={Bootstrap-based inference on the difference in the means of two correlated functional processes},
  author={Crainiceanu, Ciprian M and Staicu, Ana-Maria and Ray, Shubankar and Punjabi, Naresh},
  journal={Statistics in medicine},
  volume={31},
  number={26},
  pages={3223--3240},
  year={2012},
  publisher={Wiley Online Library}
}

@article{zhu1997algorithm,
  title={Algorithm 778: L-BFGS-B: Fortran subroutines for large-scale bound-constrained optimization},
  author={Zhu, Ciyou and Byrd, Richard H and Lu, Peihuang and Nocedal, Jorge},
  journal={ACM Transactions on Mathematical Software},
  volume={23},
  number={4},
  pages={550--560},
  year={1997},
  publisher={ACM New York, NY, USA}
}
\end{document}